\documentclass[aps,twocolumn,prd,showpacs,showkeys,preprintnumbers,superscriptaddress,nobibnotes,floatfix,longbibliography,notitlepage,nofootinbib]{revtex4-2}

\usepackage{amsmath}
\usepackage{amsfonts}
\usepackage{amssymb}
\usepackage{graphicx}
\usepackage{tabularx}
\usepackage{afterpage}
\usepackage{float}
\usepackage{chngcntr}
\usepackage[colorlinks=true,allcolors=blue]{hyperref}
\usepackage[capitalize]{cleveref}
\usepackage[dvipsnames,table]{xcolor}

\begin{document}

\title{Cross Sections and Inelasticity Distributions of \\High-Energy Neutrino Deep Inelastic Scattering}
\author{Philip L. R. Weigel}
\email{pweigel@mit.edu}
\affiliation{Department of Physics, Massachusetts Institute of Technology, Cambridge, MA 02139, USA}
\author{Janet M. Conrad}
\affiliation{Department of Physics, Massachusetts Institute of Technology, Cambridge, MA 02139, USA}
\author{Alfonso Garcia-Soto}
\affiliation{Instituto de F{\'\i}sica Corpuscular (IFIC), CSIC and Universitat de Val{\`e}ncia, 46980 Paterna, Val{\`e}ncia, Spain}

\date{\today}

\begin{abstract}
This study presents a comprehensive model for neutrino deep inelastic scattering (DIS) cross sections spanning energies from 50 GeV to 5$\times10^{12}$ GeV with an emphasis on applications to neutrino telescopes.
We provide calculations of the total charged-current DIS cross sections and inelasticity distributions up to NNLO for isoscalar nucleon targets and up to NLO order for nuclear targets.
Several modifications to the structure functions are applied to improve the modeling of the cross sections at low energies where perturbative QCD is less accurate and at high energies where there is non-negligible top quark production, and small-$x$ logarithms need to be resumed.
Using the FONLL general-mass variable-flavor number scheme, we account for heavy quark mass effects and separate the heavy flavor components of the structure functions, obtaining predictions of their relative contributions to the cross sections and the uncertainties arising from the parton distribution functions.
Additionally, the effects of final state radiation are implemented in the calculation of the double-differential cross section and discussed in terms of their impact on measurements at neutrino telescopes.

\end{abstract}
\maketitle
\section{Introduction} \label{sec:intro}
Existing neutrino telescopes can detect neutrino interactions with energies from a few GeV to hundreds of PeV \cite{Ackermann:2022rqc, IceCube:2020acn, IceCube:2020wum, IceCube:2011ucd, ANTARES:2013iuz, Baikal-GVD:2022fis}.
This range of energies provided by the atmospheric and astrophysical neutrino fluxes opens many opportunities for precision measurements and to search for new physics \cite{Ackermann:2022rqc, Valera:2022ylt}. 
The IceCube Neutrino Telescope \cite{IceCube:2008qbc} has served as an incubator for techniques for reconstructing particle and shower energies, directions, and flavors.
IceCube has been recently joined by the KM3NeT \cite{KM3Net:2016zxf, Bozza:2023slv} and Baikal-GVD detectors \cite{Baikal-GVD:2020irv, Dvornicky:2024ujq}. 
Beyond these, a number of future experiments using a variety of detection techniques have been proposed that offer unique probes of high-energy astrophysics, astronomy, particle physics, and cosmology \cite{GRAND:2018iaj, Romero-Wolf:2020pzh, Aguilar:2019jay, IceCube-Gen2:2020qha, POEMMA:2020ykm, PUEO:2020bnn, RadarEchoTelescope:2021rca, IceCube-Gen2:2021rkf}.
The existing and proposed experiments aim to observe atmospheric and astrophysical potentially up to ZeV energies \cite{Ackermann:2022rqc}, and so an excellent understanding of the Standard Model neutrino cross sections across a large range of energies is key to many BSM searches using these detectors. 

Neutrino interactions at energies above 50 GeV are dominated by deep inelastic scattering (DIS), the case where neutrino interactions have a large 4-momentum exchange, leading to scattering from individual quarks, breaking up the nucleon target. 
DIS interactions may be neutral current (NC) or charged current (CC).   
The simplest diagram involving CC DIS is $\nu_\ell (k) + q_i (p)\rightarrow \ell (k') + q_j^\prime (p')$, where $\nu_\ell$ is an incoming neutrino that produces lepton $\ell$ in the interaction and a quark of type $q_i$ becomes type $q_j^\prime$ due to charge exchange. 
In such an interaction in the fixed target frame, with  
energy $\nu=E_\nu-E_\ell$ exchanged between the neutrino and the quark, the inelasticity is defined as $y = \nu/E_\nu$. The variable $Q^2$ is the negative squared four-momentum transfer, $Q^2=-q^2=-(k-k')^2$.
DIS is typically defined as the region where $Q^2\gtrsim1~\rm{GeV}^2$ and the invariant mass of the final state hadronic system $W > 2~\rm{GeV}$.
The variable $x$ is the fractional momentum carried by a struck quark in the scattering, where $x=Q^2/2 m_N E_\nu y$ given nucleon mass $m_N$. 
Taken together, these variables can be used to describe the charged-current DIS cross section,
\begin{equation}
\begin{split}
\label{eq:crosssec}
\frac{d^2\sigma^{\nu/\bar{\nu}}}{dxdy} = & \frac{G_{F}^2 m_{N} E}{\pi}
\left( \frac{M_{W}^2}{Q^2 + M_{W}^2} \right)^2[x y^2 F_{1}(x,Q^2) \\ & + (1-y) F_{2}(x,Q^2) \pm x y (1-\tfrac{y}{2}) F_{3}(x,Q^2) ]
\end{split}
\end{equation}
where $G_F$ is the Fermi weak coupling constant and $M_{W}$ is the mass of the W boson. The $\pm$ sign preceding the $F_{3}$ term corresponds to $+$ for neutrinos and $-$ for antineutrinos. Later, we will discuss modifications to this equation that account for the mass effects of the target nucleon and outgoing charged lepton.

This cross section depends on three structure functions, $F_1(x,Q^2)$, $F_2(x,Q^2)$, and $F_3(x,Q^2)$, that describe the parton content of the nucleon target. 
Other parameterizations of \cref{eq:crosssec} make use of the longitudinal structure function $F_{L}$ or $R$, which can be expressed in terms of $F_{1}$ and $F_{2}$ \cite{Formaggio:2012cpf}. 
The plus/minus sign preceding the $xF_3$ term on \cref{eq:crosssec} indicates the opposite sign contribution of this structure function for neutrinos vs. antineutrinos. 
The CC DIS cross section has been measured up to 300 GeV with experimental uncertainties $\lesssim 3\%$ \cite{Formaggio:2012cpf} by experiments such as CCFR/NuTeV \cite{Seligman:1997fe, NuTeV:2005wsg}, CDHS \cite{Berge:1987zw} and NOMAD \cite{NOMAD:2007krq}.
IceCube has measured the neutrino-nucleon cross sections using the attenuation of neutrinos by the earth \cite{IceCube:2017roe} and with measurements of starting events \cite{IceCube:2020rnc}.
Recently, the FASER collaboration measured the flux-averaged $\nu_{\mu}$ and $\nu_{e}$ cross sections using TeV neutrinos from $pp$ collisions at the LHC \cite{FASER:2023zcr, FASER:2024hoe}, a first step towards closing the so-called ``energy gap'' between the highest energy measurements from the previous accelerator experiments and the lowest energy measurements from IceCube ($E_{\nu}>6$ TeV) \cite{IceCube:2017roe, Reno:2023sdm}.

The cross-section measurements at energies above $300$ GeV remain imprecise, requiring the use of methods from perturbative QCD to compute the structure functions and extrapolate to higher energies.
These structure functions can then be used in \cref{eq:crosssec} to compute the cross section. The primary input for the structure function calculation is the parton distribution functions (PDFs), which come from global fits to a variety of experimental data \cite{Hou:2019efy,Bailey:2020ooq,NNPDF:2021njg}. 
More recently, substantial efforts have been made towards improved modeling of the nuclear parton distribution functions (nPDFs), which capture effects such as (anti-)shadowing and the EMC effect \cite{Duwentaster:2022kpv, Helenius:2022rmf, AbdulKhalek:2022fyi, Eskola:2021nhw}. 
For neutrino telescopes primarily composed of H$_2$O, only a fraction of the scattered events originate from the hydrogen. Therefore, it is critical to account for the nuclear effects associated with scattering off of oxygen. Furthermore, when considering Earth absorption effects, the modifications to the cross section from interacting with heavier nuclear targets, such as iron, must be considered \cite{Klein:2020nuk}.

In this paper, we provide calculations of the neutrino CC DIS cross sections\footnote{A neutral-current DIS cross section compatible with the charged-current cross section presented here will be released in the future.} from 50 GeV to 5$\times10^{12}$ GeV, emphasizing their application to neutrino telescopes. To briefly summarize this work, we highlight the following aspects of this work:

\begin{itemize}
\item Structure functions calculated at the Next-to-Next-to-Leading Order (NNLO) in QCD for nucleon targets
\item Description of heavy quark mass effects using the FONLL mass scheme with flavor-separated structure functions
\item Implementation of the CKMT+PCAC-NT method for low-$Q^2$ structure functions to include the shallow inelastic scattering (SIS) region
\item Next-to-Leading Order (NLO) calculations with nuclear PDF sets to capture nuclear effects
\item Inclusion of final state radiation (FSR) to the single- and double-differential cross sections
\item Evaluation of the (n)PDF uncertainties associated with observables related to the cross sections and inelasticity distributions
\end{itemize}

The files for the total and differential cross sections from this work will be publicly released. See \cref{sec:appendix_xsec} for more details.

\section{Structure Functions} \label{sec:structure_functions}

At leading order (LO), the neutrino CC DIS structure functions can be computed in terms of sums and differences of the PDFs. For example, $F_2$ and $xF_3$ at are given by,
\begin{equation}
    F_{2}^{\nu}(x, Q^2) = 2x(d + s + b + \bar{u} + \bar{c} + \bar{t})
\end{equation}
\begin{equation}
   xF_{3}^{\nu}(x, Q^2) = 2x(d + s + b - \bar{u} - \bar{c} - \bar{t})\,.
\end{equation}
At LO, $F_1$ is computed through the Callan-Gross relation, which states that $F_{L} = 2x F_1 - F_2 = 0$. Beyond LO, the structure functions are expressed as convolutions of the PDFs and the perturbative expansion of the coefficient functions up to a fixed order in $\alpha_{s}$.

The structure functions used for the cross-section calculations will be computed at NNLO for the proton, neutron, and isoscalar nucleon targets and NLO for the nuclear targets, determined by the availability of the PDF and nPDF sets.
The nucleon target cross sections will be based on the CT18ANNLO and CT18ANLO PDF sets \cite{Hou:2019efy}, and the nuclear targets with the EPPS21 nPDF sets \cite{Eskola:2021nhw}. 
The NLO $\nu{-}p$ cross section will be used with the NLO $\nu{-}^{16}\rm{O}$ cross section to construct the $\nu$-H$_2$O cross section, maintaining consistency in the perturbative order and PDF usage for the calculations.
In the following sections, we describe the method for constructing the structure functions used to compute the cross sections and inelasticity distributions.

\subsection{Heavy Quark Treatment and FONLL}\label{sec:hq_fonll}
At leading order, charm quarks are primarily produced in neutrino deep inelastic scattering through the $\nu_{\ell} s \rightarrow \ell^{-} c$ process. Contributions to $c$ quark production from the $d$ and $b$ quarks are Cabbibo-suppressed, and the charm $F_{2}$ structure function at leading order is given by,
\begin{equation}
    F_{2}^{c,\textrm{LO}} = 2x\left(\vert V_{cd} \vert^2 d + \vert V_{cs} \vert^2 s + \vert V_{cb} \vert^2 b \right)\,.
\end{equation}

Computing the structure functions beyond LO requires a more sophisticated treatment. In this work, we utilize the \texttt{APFEL} code \cite{Bertone:2013vaa} to compute the $F_{2}$, $F_{L}$, and $xF_{3}$ structure functions. 
We use the FONLL general-mass variable-flavor number scheme (GM-VFNS) \cite{Forte:2010ta, Ball:2011mu, Ball:2015dpa} to properly account for all heavy flavor thresholds. 
In the FONLL scheme, the massive coefficient functions (M) are used exclusively for $Q^2<m_{h}^2$, where $m_{h}$ is the mass of the heavy quark $h$. For $Q^2>m_{h}^2$, the massless coefficients (ZM) are added with a subtraction term corresponding to the $m_{h}^2\rightarrow 0$ limit of the massive term (M0) that prevents double counting of terms.
Given these components, the FONLL structure functions of heavy quark mass $m_h$ is given by,
\begin{equation}
    F_{i} = F_{i}^{\rm{M}} + D(Q^2, m_{h}^2)\left[ F_{i}^{\rm{ZM}} - F_{i}^{\rm{M}0}\right]
\end{equation}
where $D(Q^2, m_{h}^2)$ is a damping term expressed as $D(Q^2, m_{h}^2) = \theta(Q^2 - m_{h}^2)(1 - Q^2/m_{h}^2)^2$, where $\theta$ is the Heaviside function. 

In this work, we use the FONLL-C scheme as implemented in \texttt{APFEL} for NNLO calculations of the proton, neutron, and isoscalar nucleon cross sections and the FONLL-B scheme for NLO calculations of the nuclear targets.
We note that recent work has evaluated the impact of approximate N$^3$LO massless coefficients, which has an $\mathcal{O}(2)\%$ effect at $10^{12}$ GeV \cite{Xie:2023suk}, as well as the NNLO massive coefficient functions \cite{Berger:2016inr, Gao:2017kkx}. 
The NNLO contributions to the massive coefficient functions are negligible because of modifications to the low-$Q^2$ structure functions, which will be discussed in \cref{sec:lowq2}. 

We define the heavy structure functions in accordance with previous work \cite{Forte:2010ta, Ball:2011mu, Bertone:2013vaa}, where a given heavy flavor structure function is proportional to the terms containing CKM elements of the heaviest mixed quark. 
For example, the charm structure function does not contain terms proportional to $\vert V_{cb}\vert^2$ term as those are assigned to the bottom structure function.
At higher $Q^2$ and at lower $x$, this implies that the hadronic final state may not include a charm quark since there are also contributions from initial state charm quarks from the $cW \rightarrow s$ and $cW \rightarrow d$ processes.
This means that the cross section calculated with the charm structure function (the ``charm cross section'') defined here does not coincide exactly with charm production.
These additional contributions may be negligible at low energies due to the smallness of the $c/\bar{c}$ PDFs in the allowed phase space.
At NNLO, the light structure function will have contributions that contain heavy quarks from $g\rightarrow h\bar{h}$ originating from internal lines \cite{Forte:2010ta}.
We note that this also implies an often-used experimental definition of the charm structure function \cite{H1:2009uwa, ZEUS:2009lvu}, defined by selecting final states that contained charm hadrons, differs from the one considered here.
The definition used in the FONLL construction cancels divergent terms in the form of $\log(Q^2/m_{h}^2)$ \cite{Forte:2010ta, xFitterDevelopersTeam:2019ygc}.
These structure functions will later be used individually to calculate cross sections such that the light/charm/bottom/top cross sections correspond to the calculations with the light/charm/bottom/top structure functions.

The production of top quarks from neutrino DIS is very similar to that of charm quarks.
A detailed discussion about top quark production and its associated observables can be found in Ref. \cite{Barger:2016deu}. 
This work relies on publicly available PDF sets, which typically assume $n_f = 5$, which do not include the $t$ quark PDF for $Q^2 > m_t^2$. 
We re-evolve the PDF sets with a maximum number of active flavors $n_f=6$ to capture all heavy quark effects, including the top.
The primary channel for top production is $b W \rightarrow t$ requires $W^2 > m_{t}^2$ so the production of top quarks will only become relevant at the highest energies.
Introducing the $t$ PDF allows for the $t W \rightarrow b$ process to occur, though because of the smallness of the $t$ PDF, this process is suppressed.
It has been shown that neglecting the mass effects in the computation of the top structure function (i.e. using the zero-mass variable-flavor number scheme) can lead to an over-prediction of the top cross section \cite{Garcia:2020jwr}.

\subsection{Nuclear Effects} \label{sec:nuclearmod}
Until now, we have discussed the cross sections in the context of free nucleons. 
However, nearly every neutrino experiment uses nuclear targets. 
Nuclear modifications to the parton distribution functions are an active field of investigation \cite{AbdulKhalek:2022fyi, Duwentaster:2022kpv, Kusina:2020lyz, Eskola:2021nhw, Helenius:2022rmf, Klasen:2023uqj, Khanpour:2020zyu}. 
These nuclear effects are accounted for in the nuclear parton distribution functions.
The choice of the EPPS21 nPDF sets aligns with the choice of the CT18A PDF sets for the nucleon targets, as CT18ANLO is used for the free proton baseline of EPPS21. The effects of nuclear modifications can impact the absorption of neutrinos passing through the Earth, as they interact with nuclear targets like iron, silicon, oxygen, etc. \cite{Klein:2020nuk}. This may be an important effect for neutrino telescopes that study high-energy atmospheric and astrophysical neutrinos that pass through the Earth.

\subsection{Target Mass Corrections} \label{sec:tmc}

In the low-$Q^2$ regime close to the scale of the target mass $m_{N}^2$, corrections proportional to $m_{N}^2/Q^2$ are relevant \cite{Georgi:1976ve,Barbieri:1976rd,DeRujula:1976baf,Ellis:1982cd}. 
Corrections arise from accounting for the target mass in the light-cone momentum fraction, in which the Bjorken $x$ variable is replaced by the Nachtmann variable $\xi$ such that
\begin{equation}
    \xi = \frac{2x}{1+\sqrt{1+4x^2 m_{N}^2/Q^2}}
\end{equation}
in PDF evaluations. 
Target mass corrections (TMC) also introduce mixing that arises in the formation of the hadronic structure functions from partonic structure functions. 
Finally, additional corrections can be interpreted in the parton model framework due to transverse momentum effects up to the scale $m_{N}$ in the collinear parton model \cite{Ellis:1982cd}.
We implement the corrections due to the mass of the DIS target following Refs.~\cite{Kretzer:2003iu, Schienbein:2007gr}, where the corrected structure functions depend on additional functions $h_{2}(\xi,Q^2)$, $h_{3}(\xi,Q^2)$, and $g_{2}(\xi,Q^2)$ that are integrals over the uncorrected structure functions.
A recent detailed analysis of TMC for nuclear targets showed that it is straightforward to extend the TMC-corrected nucleon structure functions to TMC-corrected nuclear structure functions \cite{Ruiz:2023ozv}.

\subsection{Low-$Q^2$ Structure Functions} \label{sec:lowq2}
At the lowest energies accessible to neutrino telescopes, there are significant contributions to the cross sections from the low-$Q^2$ region of the structure function \cite{Candido:2023utz,Yang:1998zb,Bodek:2002ps,Bodek:2004pc,Bodek:2021bde,Reno:2006hj,Jeong:2023hwe}. 
At low $Q^2$, below the scale where the quark-parton model is reliable, extrapolations of the structure functions must be used. 
The low-$Q^2$ structure function $F_{i}^{\rm{low}}$ is used below a scale $Q_0^2$, chosen as a matching scale for the structure functions evaluated with PDFs and $F_{i}^{\rm{low}}$.
To ensure continuity of the structure functions for all $x$ at $Q^2=Q_0^2$,
\begin{equation}
    F_{i}(x, Q^2 < Q_0^2) = F_{i}^{\rm{low}}(x,Q^2) \times \frac{F_{i}(x,Q_{0}^2)}{F_{i}^{\rm{low}}(x,Q_{0}^2)}\,.
\end{equation}

The low-$Q^2$ structure function used here relies on the CKMT parameterization of the electromagnetic structure function $F_2$ \cite{Capella:1994cr,Kaidalov:1998pn} as a starting point. 
The electromagnetic structure functions are adapted to weak interaction scattering \cite{Reno:2006hj,Jeong:2023hwe}, where the weak structure function includes both the conserved vector current contribution plus a modification to account for the partially conserved axial current (PCAC) \cite{Kulagin:2007ju} so that \cite{Jeong:2023hwe},
\begin{equation}
    \label{eq:lowq_sf}
    F_i^{\rm{low}}(x,Q^2) = F_i^{\rm{CKMT}}(x,Q^2) + F_i^{\rm{PCAC}}(x, Q^2)\,.
\end{equation}
The transition scale is chosen to be $Q_{0}^{2} = 4~\rm{GeV}^2$, where the PCAC contribution to $F_2(x,Q_0^2)$ has been shown to be less than $0.5\%$ \cite{Jeong:2023hwe}. For $E_{\nu}=50$ GeV, with this approach, the contribution to the charged-current cross section from $Q^2<1$ GeV$^2$ is 7\%\ for $\nu_\mu$ and $14\%$ for $\bar\nu_\mu$, reducing to 4\%\ and 8\%\ respectively for $E_{\nu}=100$ GeV. 
The CKMT and PCAC modifications applied to the charm structure functions are additionally modified with the slow rescaling variable through the transformation $x\rightarrow \chi = x (1 + Q^2/m_{c}^2)$, which accounts for the effect of the charm quark mass.
The inclusion of these modifications to the structure functions will be used to include the SIS region of the kinematic space to the cross sections, which is defined as $Q^2 \lesssim 1~\rm{GeV}^2$ and $(m_{N} + m_{\pi})^2 < W^2 < 4~\rm{GeV}^2$ \cite{SajjadAthar:2020nvy}.

\subsection{Small-$x$ Resummation} \label{sec:smallx}
At the highest energies, terms that go as $\log(1/x)$ become important and must be resummed.
It has been shown that these effects can significantly alter the PDFs at low  $x$.
The resummation of these logarithms is done through the \texttt{HELL} (High-Energy Large Logarithms) code \cite{Bonvini:2016wki, Bonvini:2017ogt, Bonvini:2018xvt, Bonvini:2018iwt} which computes the correction factors for the coefficient functions up to next-to-leading log (NLL).
These terms are applied during the evolution of the PDFs and the computation of the structure functions.
Both the BGR \cite{Bertone:2018dse} and CT18 \cite{Xie:2023suk} neutrino cross-section models utilize these corrections in their calculations.
For this work, the PDFs are evolved with \texttt{APFEL} to include these corrections.

At the highest energies considered, there are non-negligible contributions with $x < 10^{-9}$ \cite{Xie:2023suk}.
This is beyond the limits of most PDF sets, so the PDFs below this limit must be extrapolated.
The publicly available grids for the coefficient functions from \texttt{HELL} only extend down to $x=10^{-9}$, so they were recomputed to $x=10^{-12}$ to cover the kinematic space completely.
These grids are then used to evolve the PDFs with small-$x$ resummation and compute the structure functions with NLL corrections.
We find that the contribution of these structure function corrections to the total cross sections can be as large as a few percent at $10^{12}$ GeV.

\section{Cross Sections} \label{sec:xsecs}
The structure functions described in \cref{sec:structure_functions} were formulated to allow for the cross sections to be evaluated across a wide range of $x$ and $Q^2$.
The low-$Q^2$ modifications and heavy quark treatment become increasingly important at lower energies, while the effects of top production and small-$x$ resummation arise at the highest energies.
To compute the cross sections, we utilize a complete version of \cref{eq:crosssec} that takes into account effects arising from $m_N$, $m_\ell$, $F_{4}$, and $F_{5}$ \cite{Albright:1974ts},

\begin{equation}
    \label{eq:complete_xs}
    \begin{split}
    \frac{d^2\sigma^{\nu/\bar{\nu}}}{dx dy} &= \frac{G_{F}^2 m_{N} E_{\nu}}{\pi \left(1+Q^2/M_{W}^2\right)^2} \Bigg[ \left( x y^2 + \frac{m_{\ell}^2 y}{2 E_{\nu} m_{N}} \right) F_{1} \\   
    &+\left( 1 - \frac{m_{\ell}^2}{4E_{\nu}^2} - \left( 1 + \frac{x m_N}{2 E_{\nu}}\right)y\right)F_{2}\\
    &\pm\left(xy\left(1 - \frac{y}{2}\right)-\frac{m_{\ell}^2 y}{4 E_{\nu} m_{N}}\right) F_{3}\\
    &+ \frac{m_{\ell}^2(m_{\ell}^2 + Q^2)}{4 E_{\nu}^2 m_{N}^2 x} F_{4} - \frac{m_{\ell}^2}{E_{\nu} m_{N}} F_{5}
    \Bigg]\,.
    \end{split}
\end{equation}

Due to the size of $m_\ell^2$, $F_{4}$ and $F_{5}$ are only important in the consideration of $\nu_\tau$ cross sections as the terms scale with $m_\ell^2$ for charged lepton $\ell$.
We include them here for completeness and assume the Albright-Jarlskog relations: $F_{4} = 0$ and $F_{5} = 2xF_{2}$ \cite{Albright:1974ts}.
While this only holds at LO, the contributions from these terms are small in the context of $\nu_{\mu}$ cross sections.

\subsection{Final State Radiation} 
\label{sec:fsr}
We include the effects of final state radiation (FSR) on the outgoing lepton leg in CC processes.
These corrections have been studied extensively for both $e p$ and $\nu N$ DIS and have been shown to have non-negligible effects on the shape of $d^2\sigma/dxdy$ and $d\sigma/dy$ for precision experiments, with a negligible impact on the total cross section \cite{Mo:1968cg, Kiskis:1973ia, BarlowWolfram, Bardin:1986bc, Seligman:1997fe, Arbuzov:2004zr, Blumlein:2012bf, Afanasev:2023gev, Plestid:2024bva}.
In the context of neutrino telescopes, FSR from the quarks is indistinguishable from the hadronic shower.
The FSR from the lepton leg transfers a fraction of the observed lepton energy to the shower.
For a recent discussion of the impacts of FSR on observables in neutrino telescopes, and the differences between the $y$ considered here and experimentally measured $y_{exp}$, we direct the reader to Ref.~\cite{Plestid:2024bva}.

In this work, we consider the following prescription of the leading logarithm approximation (LLA) for the modification of the cross section with FSR using leptonic variables from Ref.~\cite{DeRujula:1979grv}.
We denote the energy of the lepton exiting the hard scattering process as $E_{\ell} / z$ with $0 < z < 1$, where $E_{\ell}$ is the final energy of the lepton after radiating a photon.
The correction is computed by integrating over the cross section with modified kinematic variables and the QED splitting function $P_{\ell\rightarrow\ell\gamma}^{(0)}(z)$,
\begin{equation}
    \label{eq:splitting_function}
    P_{\ell\rightarrow\ell\gamma}^{(0)}(z) = \left[ \frac{1+z^2}{1-z} \right]_{+}\,.
\end{equation}
The hard scattering variables are modified to give the variables $\hat{x}$, $\hat{y}$,
\begin{equation}
    \hat{y} = \frac{1 + y - z}{z},\quad \hat{x} = \frac{xy}{1 + y - z}\,.
\end{equation}
We define the leading logarithm using the description provided by \cite{DeRujula:1979grv, Sigl:1997cq},
\begin{equation}
    L = \log\left(\frac{s(1 - y - xy)^2}{m_{\ell}^{2}}\right)\,.
\end{equation}
The complete correction term is then given by,
\begin{equation} \label{eq:fsr_corrections}
\begin{split}
    \frac{d^2\sigma^{(1)}}{dxdy} = & -\frac{\alpha}{2\pi} \log\left(\frac{s(1-y-xy)^2}{m_{\ell}^2}\right) \int_{0}^{1} dz \frac{1+z^2}{1-z} \\
    & \times \Bigg[ \frac{y\theta(z-z_{min})}{z(y+z-1)} \frac{d^2 \sigma^{(0)}}{dxdy} \Bigg\vert_{\hat{x},\hat{y}} - \frac{d^2\sigma^{(0)}}{dxdy}\Bigg]\,.
\end{split}
\end{equation}

The first term of the integral is cut off for $z < z_{min} = 1 - y + xy$, which accounts for the available kinematic phase space.
When integrating over the plus distribution in \cref{eq:splitting_function}, the modified double-differential cross section is subtracted by the unmodified cross section from \cref{eq:complete_xs}, which cancels the infrared divergence from the $z\rightarrow1$ soft photon limit.
To obtain the complete cross section, the two terms are added together,
\begin{equation}
    \label{eq:total_xs_corrected}
    \frac{d^2 \sigma}{dxdy} = \frac{d^2\sigma^{(0)}}{dxdy} + \frac{d^2\sigma^{(1)}}{dxdy}\,.
\end{equation}

\begin{figure*}[!htb]
    \centering
    \includegraphics[width=0.95\textwidth]{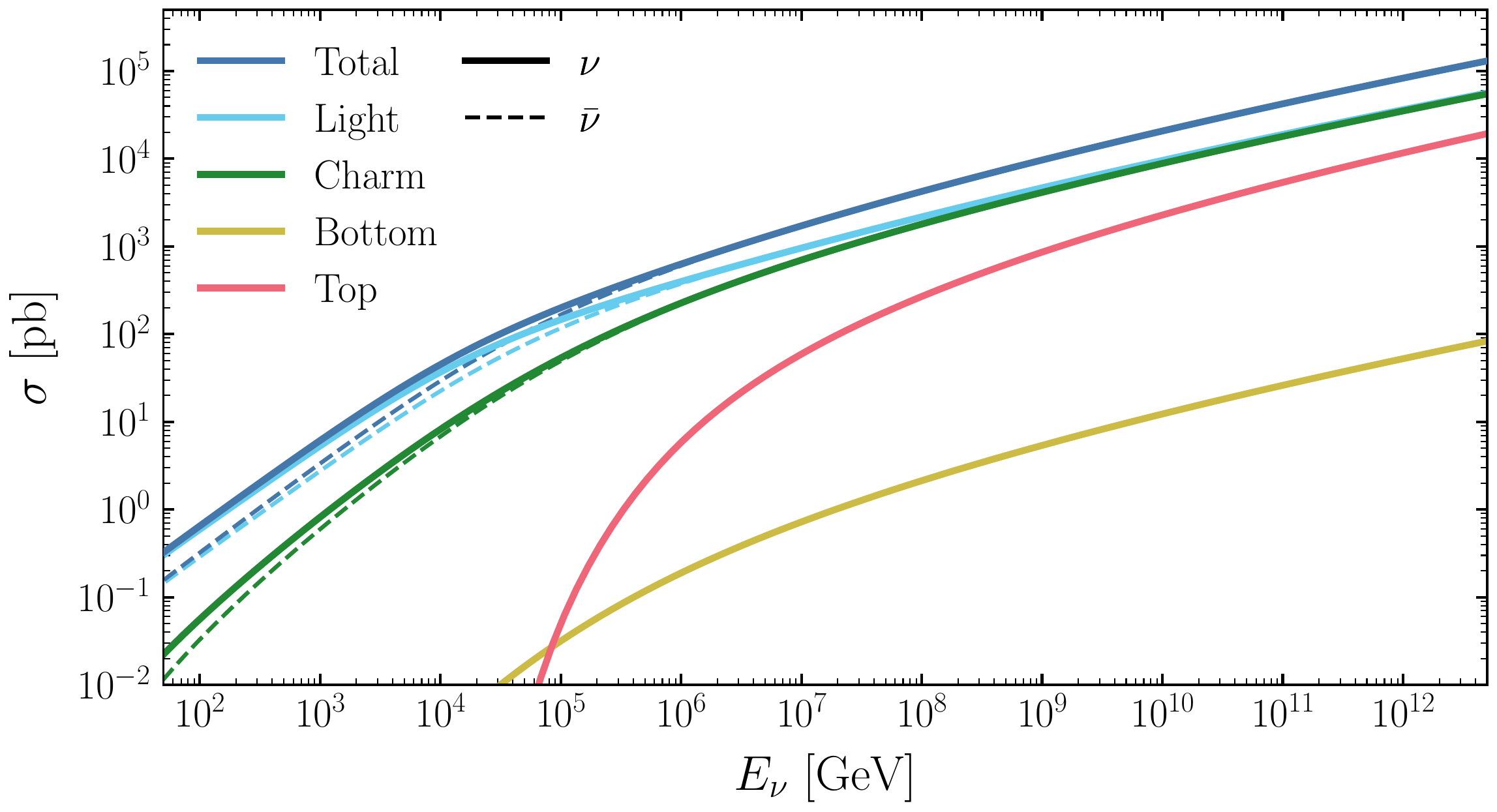}
    \caption{The $\nu$ (solid) and $\bar{\nu}$ (dashed) CC DIS isoscalar nucleon cross sections separated by the flavor components of the structure functions using the CT18ANNLO PDF set evolved with small-$x$ resummation and $n_f=6$. The cross sections are calculated with structure functions that use the CKMT+PCAC-NT parameterization described in \cref{sec:lowq2} for $Q^2 < Q_0^2 = 4~\rm{GeV}^2$ with $W_{min}^2 = m_{N}^2 + m_\pi^2$ and the small-$x$ contributions from \cref{sec:smallx}. The total cross section (dark blue) is the sum of the light, charm, bottom, and top components. The $\nu$ and $\bar{\nu}$ cross sections overlap for the bottom and top. Uncertainties are tabulated in \cref{sec:appendix_xsec}. See \cref{sec:hq_fonll} for important details about the interpretation of the flavor components.}
    \label{fig:total_xs}
\end{figure*}

We emphasize that these $\mathcal{O}(\alpha L)$ FSR corrections impact $d^2\sigma/dxdy$, not just $d\sigma/dy$.
When simulated events for experiments are created, typically the neutrino interactions are generated by sampling $x$ and $y$ values and assigned weights corresponding to the double-differential cross section \cite{IceCube:2020tcq, Schneider:2024eej}.
The impact of these changes on the production of Monte Carlo simulation should be investigated.

\subsection{Free Nucleon Cross Sections} \label{sec:nucleon_xsecs}
\begin{figure}[!h]
    \centering
    \includegraphics[width=0.48\textwidth]{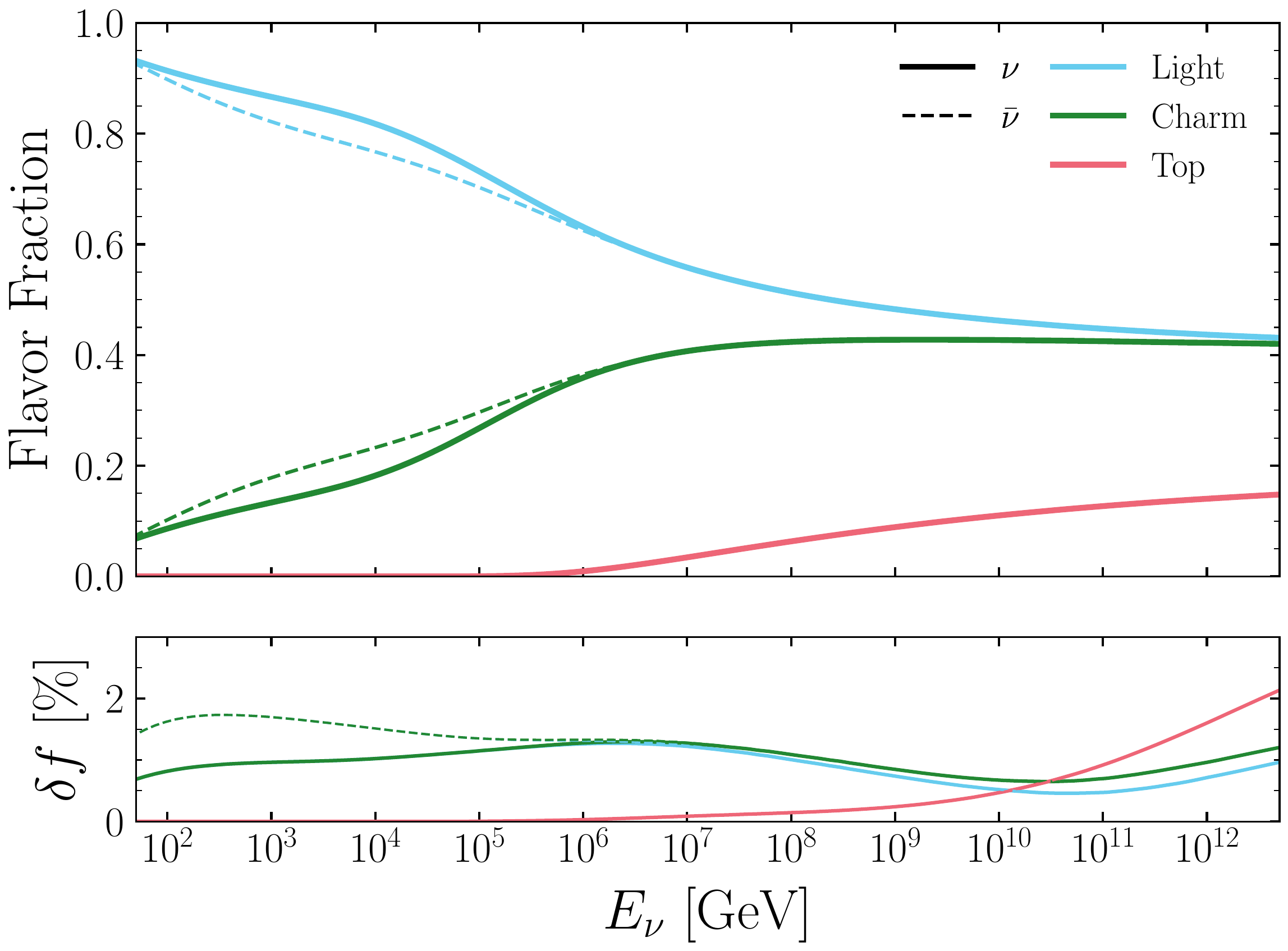}
    \caption{Top: The fractional flavor composition $f$ of the total cross section for CC DIS assuming an isoscalar nucleon target for neutrinos (solid) and antineutrinos (dashed). The bottom component is not shown because it is negligible, even at the highest energies considered here. Bottom: The PDF uncertainties for the flavor fractions.}
    \label{fig:flavor_fraction}
\end{figure}

To compute the total cross section and single-differential cross sections (also referred to as the inelasticity distribution when normalized to the total cross section), the double-differential cross section including the target mass corrections, low-$Q^2$ modifications, small-$x$ resummation, and FSR is integrated across the available phase space of $x$ and $Q^2$.
The integration of \cref{eq:complete_xs}, with the modifications from \cref{eq:fsr_corrections}, is implemented using the \texttt{VEGAS} Monte Carlo integration method \cite{Lepage:1977sw, Lepage:123074}.
DIS cross section models typically use $Q_{min}^2\sim1$ GeV$^2$ and $W_{min}^2=4$ GeV$^2$ as kinematic constraints.
In these calculations, we use $Q_{min}^2=0$ GeV and $W_{min}^2 = m_{N}^2 + m_{\pi}^2$ to include the SIS region.
In \cref{fig:total_xs}, we show the total $\nu_\mu$ and $\bar{\nu}_\mu$ CC DIS cross section for the isoscalar nucleon target and the individual contributions from the light, charm, bottom, and top structure functions.
The contribution from the bottom structure function is several orders of magnitude smaller than the other contributions and will be neglected moving forward.
\cref{fig:flavor_fraction} shows the flavor fraction, which is the cross section for each flavor component divided by the total. 

The PDF uncertainties are calculated using the symmetric Hessian method, where the uncertainty is computed using the observable $O$ evaluated with the PDF sets corresponding to the +/- variations of the $N$ eigenvectors \cite{Pumplin:2001ct, Schmidt:2018hvu},
\begin{equation} \label{eq:symmetric_hessian_unc}
    \Delta O = \frac{1}{2}\sqrt{\sum_{i=1}^{N} (O_i^+ - O_i^-)^2}\,.
\end{equation}
In the cases where the eigenvector sets are not given as $\pm68\%$ variations, such as CT18A ($\pm90\%$), a scaling factor is applied to obtain the 1$\sigma$ uncertainties.
We find that the uncertainty from the PDFs for the neutrino and antineutrino cross sections is ${\sim}2\%$ between 50 GeV and 10 PeV.
The uncertainty increases up to $54\%$ at $5\times10^{12}$ GeV which primarily comes from the poor constraints of the PDFs at low $x$.
The data for the cross sections, uncertainties, and flavor fractions for the NNLO and NLO calculations are separated into proton, neutron, and isoscalar targets in \cref{sec:appendix_xsec}.
These results will be compared to experimental data and other cross-section models in \cref{sec:discussion}.
\subsection{Inelasticity Distributions} \label{sec:inelasticity}
\begin{figure}[!h]
    \centering
    \includegraphics[width=0.48\textwidth]{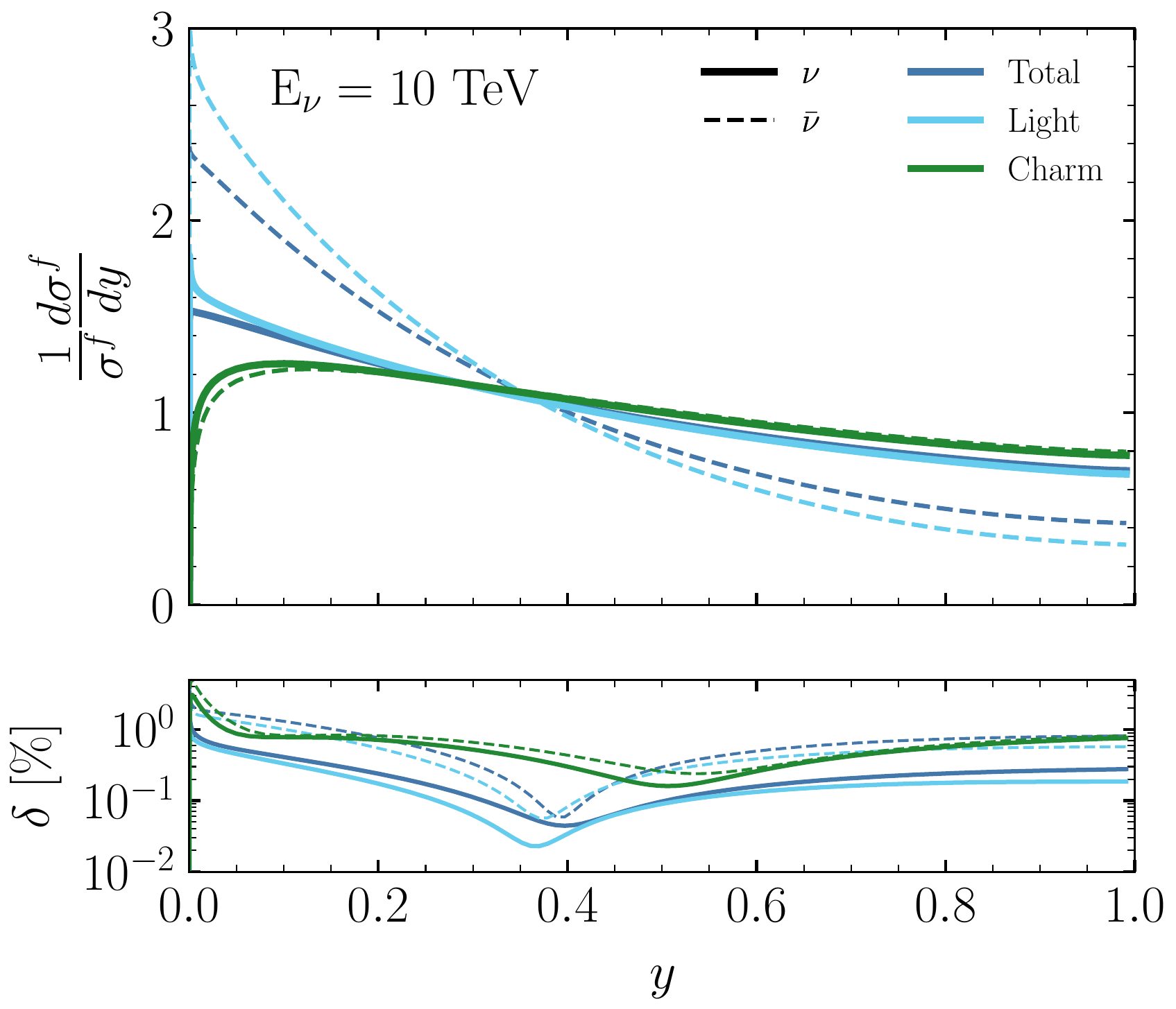}
    \caption{Top: The inelasticity distribution $(1/\sigma)d\sigma/dy$ at $E_{\nu} = 10$ TeV for neutrinos (solid) and antineutrinos (dashed), normalized to the integrated cross section for each flavor $\sigma^f$. Bottom: The corresponding PDF uncertainties. A more extensive comparison from 100 GeV to 10 PeV can be found in \cref{fig:appendix_big_inelasticity} and without FSR in \cref{fig:appendix_big_inelasticity_noFSR}.}
    \label{fig:dsdy}
\end{figure}

The shape of the differential cross section $d\sigma/dy$ differs between $\nu$ and $\bar{\nu}$ because of the sign difference of the $xF_{3}$ term in \cref{eq:complete_xs}. \cref{fig:dsdy} shows $d\sigma/dy$, including FSR, normalized to the total cross section $\sigma$ decomposed into the light and charm components for $\nu$ and $\bar{\nu}$.
The inelasticity distribution for $\nu$ is noticeably flatter over the range of $y$ compared to the $\bar{\nu}$ distribution.
The uncertainties are obtained by taking the inelasticity distributions for each PDF replica and normalizing them to the total cross section for that replica.
The $\nu$ and $\bar{\nu}$ distributions maintain their distinct shape differences until 100 TeV, where they become more narrowly peaked at low $y$.
The shape of the charm contribution is flat for both $\nu$ and $\bar{\nu}$ across most of the range in $y$.
At 1 TeV, the charm component for $\bar{\nu}$ near $y\rightarrow 1$ is $1/3$ that of the total, whereas it is closer to $1/7$ for $\bar{\nu}$.
The inelasticity distributions from 100 GeV to 10 PeV can be found in \cref{fig:appendix_big_inelasticity}.

If the inelasticity of neutrinos interacting within the instrumented volume of a neutrino telescope can be measured from the energy deposited in the hadronic shower and outgoing lepton, then neutrinos and antineutrinos could be separated statistically.
From \cref{fig:dsdy} (bottom), the uncertainty on the shape of the inelasticity is well below the reported resolution of IceCube inelasticity measurements (RMS error of 0.19) \cite{IceCube:2018pgc}.
\subsection{Mean Inelasticity} \label{sec:meany}
\begin{figure}
    \centering
    \includegraphics[width=0.48\textwidth]{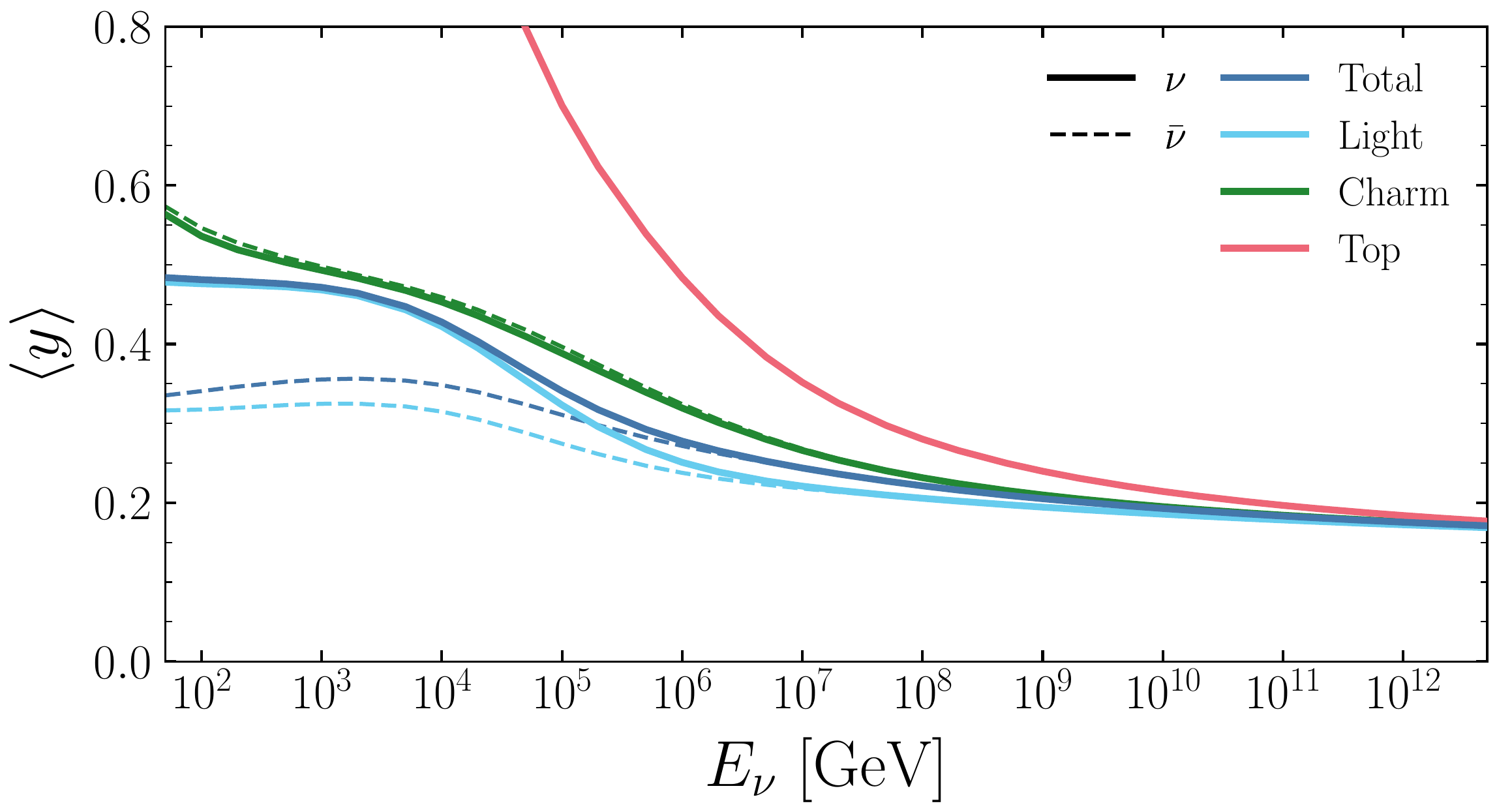}
    \caption{The mean inelasticity of the baseline prediction separated by flavor components. The curves for each flavor component correspond to averaging over $d\sigma/dy$ for the flavor. The total is computed with the differential cross section of all flavors.}
    \label{fig:meany_iso}
\end{figure}

The mean inelasticity is the average $y$ of the distributions discussed in \cref{sec:inelasticity}. It is a prediction of the differential cross sections and an observable of interest for experiments such as IceCube, and has been measured with ${\sim}5\%$ precision in a part of the TeV energy range \cite{IceCube:2018pgc}.
\cref{fig:meany_iso} shows the mean inelasticity over the full energy range.
The differences between neutrinos and antineutrinos become smaller at higher energies because of the high-energy behavior of $d\sigma/dy$.
The uncertainty of the total $\langle y \rangle$ is ${<}0.5\%$ (${<}1.5\%$) for neutrinos (antineutrinos) in the range of $50~\textrm{GeV}<E_{\nu}<100~\textrm{TeV}$, increasing to $2\%$ at $10^{9}$ GeV for both neutrinos and antineutrinos.
While the PDF-related uncertainties are small, the FSR correction to $d\sigma/dy$ leads to relative shifts of $\langle y \rangle$ from $1.4\%$ at 50 GeV to $9.4\%$ at $5\times10^{12}$ GeV.

\begin{figure}
    \centering
    \includegraphics[width=0.48\textwidth]{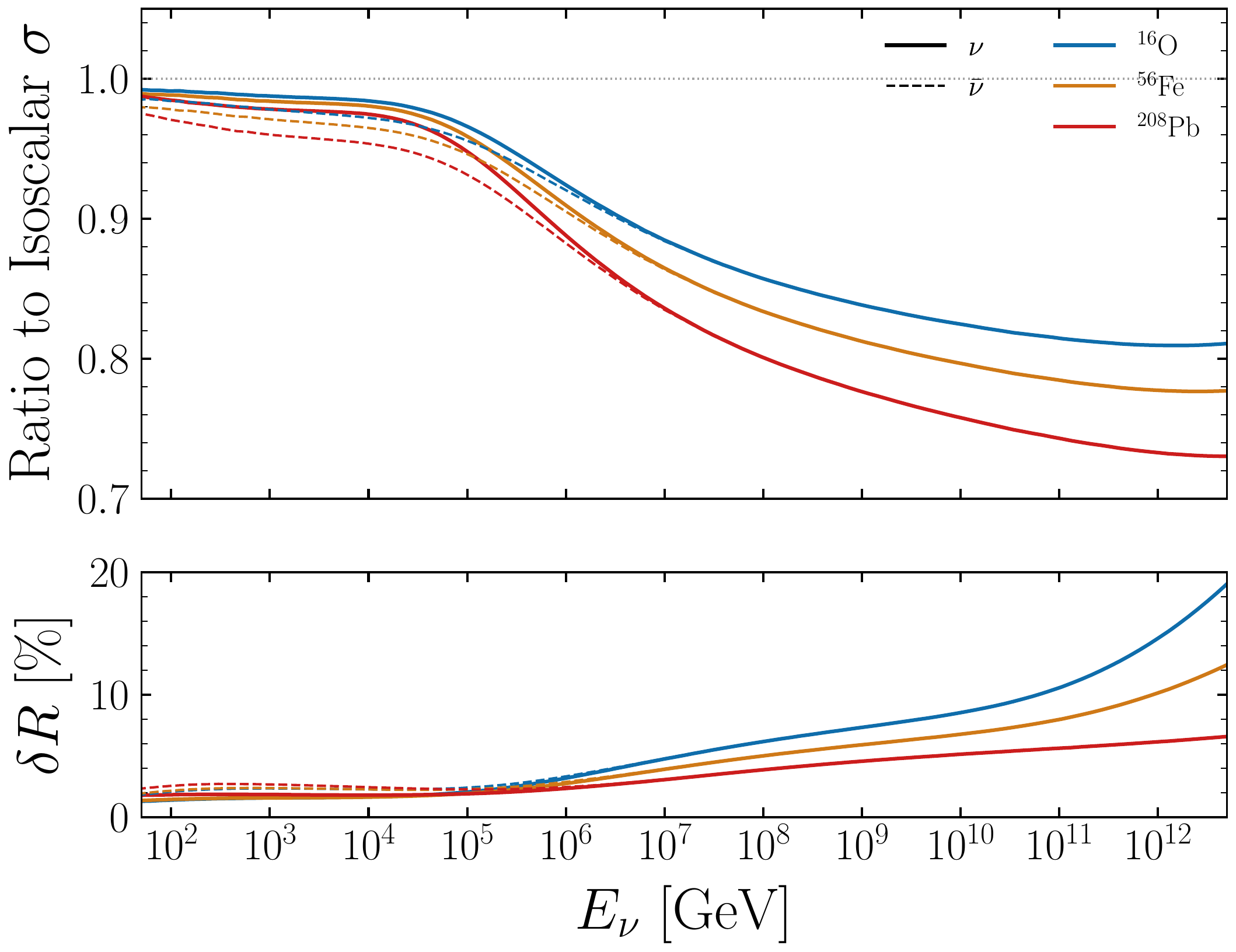}
    \caption{Top: The ratios of cross sections for different nuclear targets, divided by $A$, to the NLO isoscalar cross section (top) for neutrinos (solid) and antineutrinos (dashed). Bottom: The PDF uncertainties on the ratio of cross sections. See \cref{sec:discussion} for a discussion on the uncertainties.}
    \label{fig:nuclear_xs}
\end{figure}

\begin{figure}
    \centering
    \includegraphics[width=0.48\textwidth]{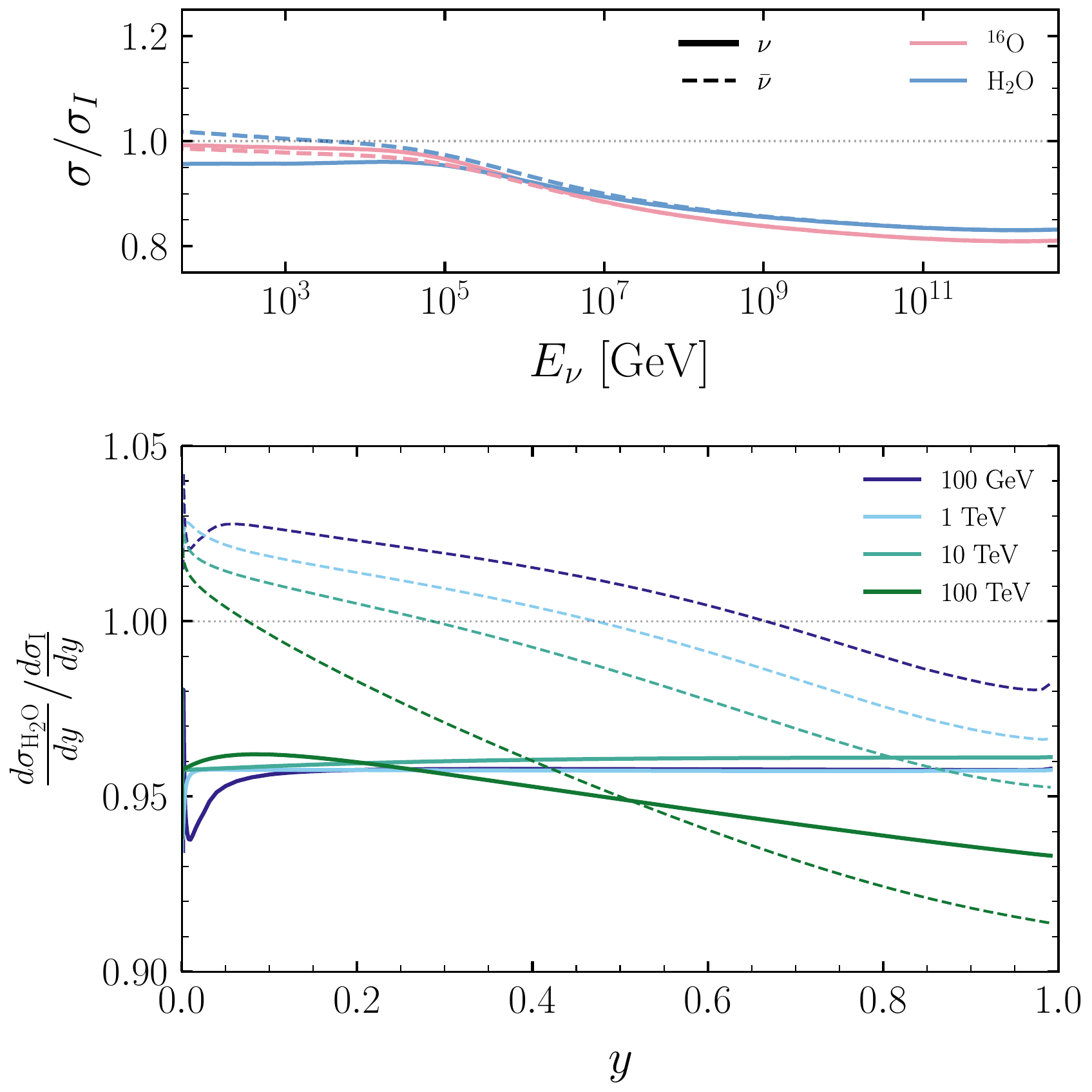}
    \caption{Top: Ratio of the oxygen (pink) and water (blue) total cross section to the NLO isoscalar cross section for neutrinos (solid) and antineutrinos (dashed). Bottom: Ratio of the single-differential cross section for H$_2$O to the isoscalar nucleon at $E_{\nu} =$ 100 GeV, 1 TeV, 10 TeV, and 100 TeV.}
    \label{fig:water_y_ratio}
\end{figure}

\subsection{Nuclear and Molecular Targets}
Above 10 TeV, the opacity of the Earth begins to significantly attenuate the atmospheric and astrophysical neutrino flux \cite{Nicolaidis:1996qu, Naumov:1998sf, Kwiecinski:1998yf, Vincent:2017svp}.
Understanding uncertainties in the absorption of neutrinos is crucial for high-energy measurements and BSM physics searches. While the uncertainty of the isoscalar cross section presented in \cref{sec:nucleon_xsecs} are $\leq2\%$ below 10 PeV, the uncertainty with nuclear targets can be much larger.
As discussed in \cref{sec:nuclearmod}, we utilize the EPPS21 nPDF sets for our studies of neutrino cross sections with nuclear targets. In \cref{fig:nuclear_xs}, we show the ratio of the cross sections on nuclear targets to that of the isoscalar target for a selection of nuclei: $^{16}\rm{O}$, $^{56}\rm{Fe}$, $^{208}\rm{Pb}$.
While $^{208}\rm{Pb}$ is not a target material for neutrino telescopes, it provides a useful reference for the nuclear modifications of the cross sections in the extreme case.
At energies below 10 TeV, the nuclear cross sections are only a few percent below that of the isoscalar cross section. This difference grows up to a $-19\%$ ($-22\%$) reduction of the total cross section at $10^{12}$ GeV for oxygen (iron).

In the case of molecular targets, we simply sum the contributions for each individual target and divide by the sum of mass numbers to compare to the isoscalar cross section.
H$_2$O is an interesting target since there are two additional proton targets with no nuclear effects, which will lead to shape differences in the inelasticity distribution. 
In \cref{fig:water_y_ratio}, we show the ratios of the oxygen and water total cross section to the isoscalar cross section, and the ratios of the single-differential cross sections for the two targets. 
In this comparison, the H$_{2}$O and oxygen cross sections are normalized to $A$ (taken to be 18 for H$_{2}$O).
At low energies, the $\nu{-}$H$_{2}$O cross section is reduced relative to the isoscalar and oxygen targets since $\sigma^{\nu p} < \sigma^{\nu I}$ in this energy range. Similarly, the antineutrino cross section is larger since $\sigma^{\bar{\nu}p} > \sigma^{\bar{\nu}I}$.
There are non-trivial differences between the single-differential cross-section shapes of $\nu$ and $\bar{\nu}$, which demonstrates the need for properly modeling these effects for inelasticity-based measurements.
\begin{figure}[h]
    \centering
    \includegraphics[width=0.48\textwidth]{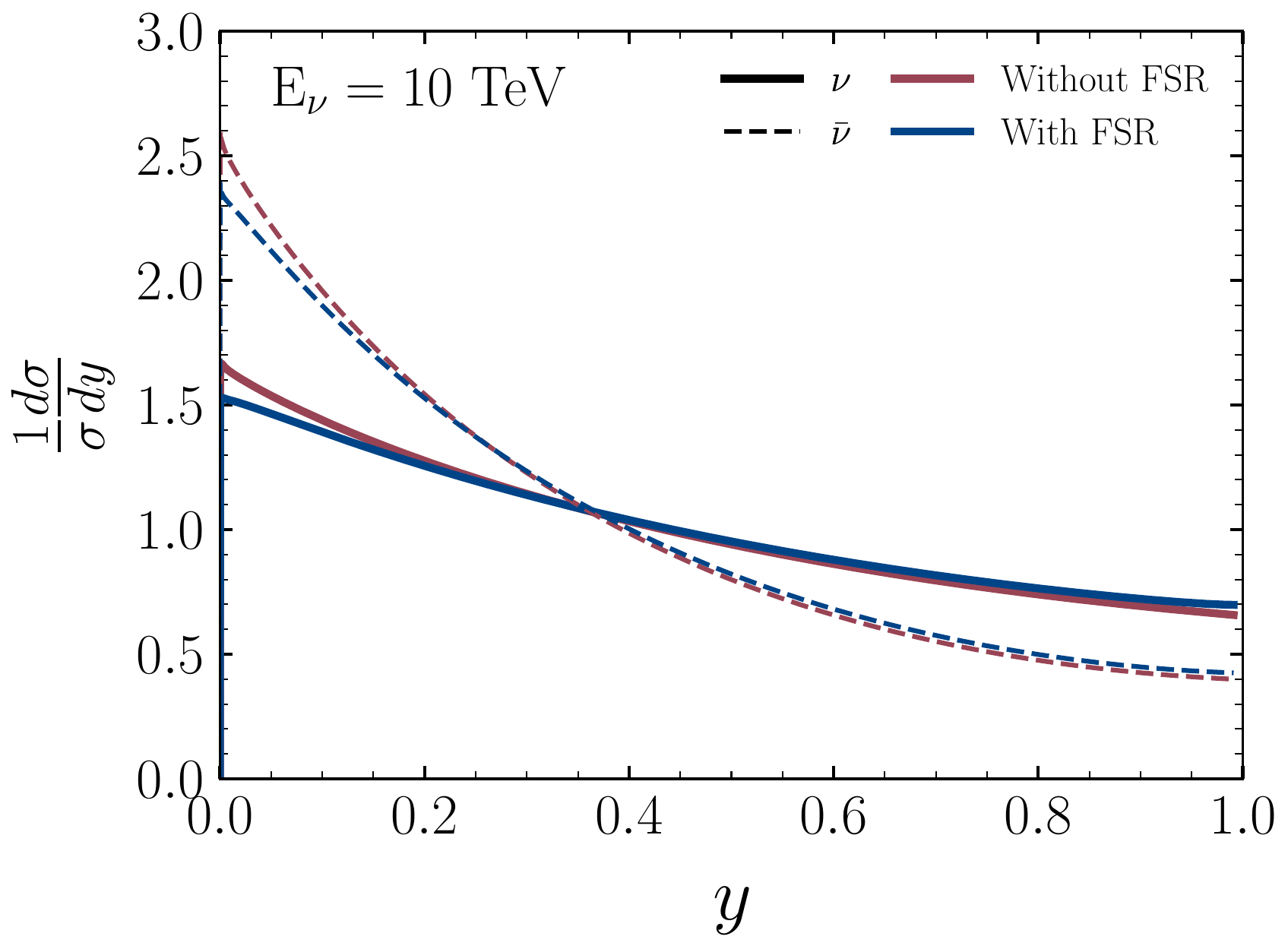}
    \caption{Inelasticity distribution evaluated at $E_\nu = 10$ TeV for neutrinos (solid) and antineutrinos (dashed), including the effects of final state radiation (blue) and without (red).}
    \label{fig:dsdy_rad_corr}
\end{figure}
\subsection{Effects of Final State Radiation} \label{sec:fsr_xsecs}
The cross sections presented in the preceding sections have included corrections from final state radiation.
In \cref{fig:dsdy_rad_corr}, we show the differences between the inelasticity distributions with and without final state radiation.
At low $y$, the inclusion of FSR decreases the cross section and increases it at higher $y$.
This is expected from the inclusive nature of FSR, which effectively transfers energy away from the outgoing charged lepton.
Experimentally, the energy of the $\gamma$ is indistinguishable from the experimental signature of the outgoing hadronic shower. 
Neglecting this effect can lead to a bias in the reconstructed neutrino energy since the muon energy will be lower than expected \cite{Plestid:2024bva}. 
While this would not have a significant impact on starting tracks, $\nu_{\mu}$ CC events that interact inside the detector volume, since it is an inclusive measurement of track and shower energies, throughgoing tracks can be systematically mis-reconstructed. 
Lastly, we emphasize the importance of these shifts in Monte Carlo simulation.
With improperly modeled DIS kinematics, simulated events at higher energies may differ from observed data for the same reasons. 
In cases where inelasticity is being reconstructed, even larger disagreements may be observed.

\begin{figure}
    \centering
    \includegraphics[width=0.48\textwidth]{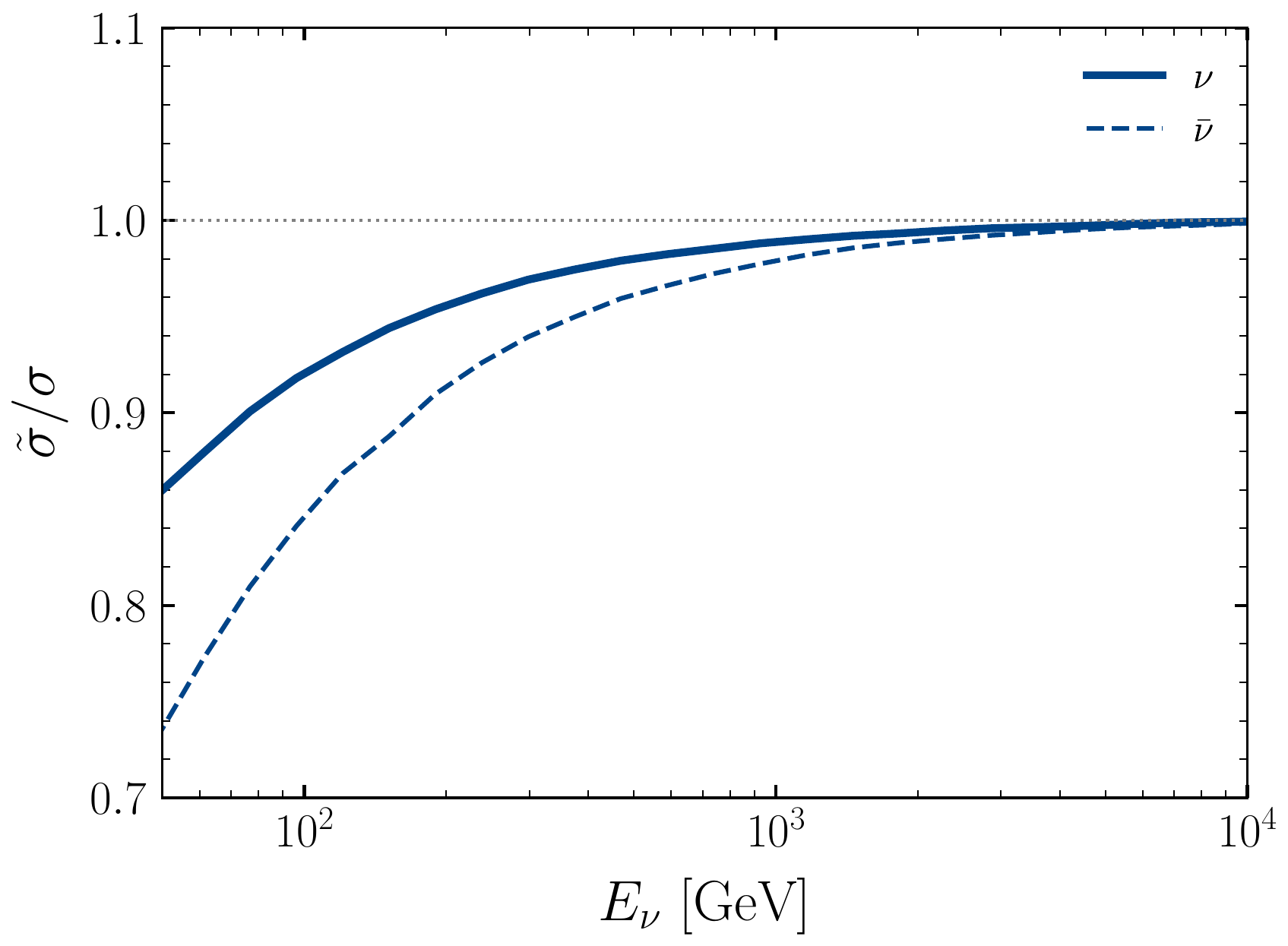}
    \caption{Ratio of the total isoscalar cross section without low-$Q^2$ modifications ($\tilde{\sigma}$) to the full calculation ($\sigma$) for neutrinos (solid) and antineutrinos (dashed).}
    \label{fig:lowq_xs_comparison}
\end{figure}

\subsection{Effects of Low-$Q^2$ Modifications} \label{sec:lowq_xsecs}
In \cref{fig:lowq_xs_comparison}, we show the ratio between the total CC DIS cross section with and without the low-$Q^2$ modifications to the structure functions from \cref{sec:lowq2}.
When we exclude these modifications, we use $Q_{min}^2 = 1.69~\rm{GeV}^2$ and $W_{min}^2=4~\rm{GeV}^2$ for a more practical comparison.
Neglecting the modifications leads to a reduction in the total cross section by $-14\%$ and $-26\%$ for $\nu$ and $\bar{\nu}$ respectively at 50 GeV.
At 1 TeV, these corrections are ${<}2\%$ effects.
The enhancement of the cross section due to these modifications to the structure functions can also be observed when compared to other pQCD-based models, which fall off at low $E_{\nu}$.
The notable exceptions are models that use alternative descriptions of the low-$Q^2$ region, such as the JR \cite{Jeong:2023hwe} and NNSF$\nu$ \cite{Candido:2023utz} models, which will be discussed in \cref{sec:discussion}.
We conclude that including the shallow inelastic scattering region is important for properly modeling the cross section, even at energies as high as 1 TeV.

\section{Discussion} \label{sec:discussion}

\begin{figure}
    \centering
    \includegraphics[width=0.48\textwidth]{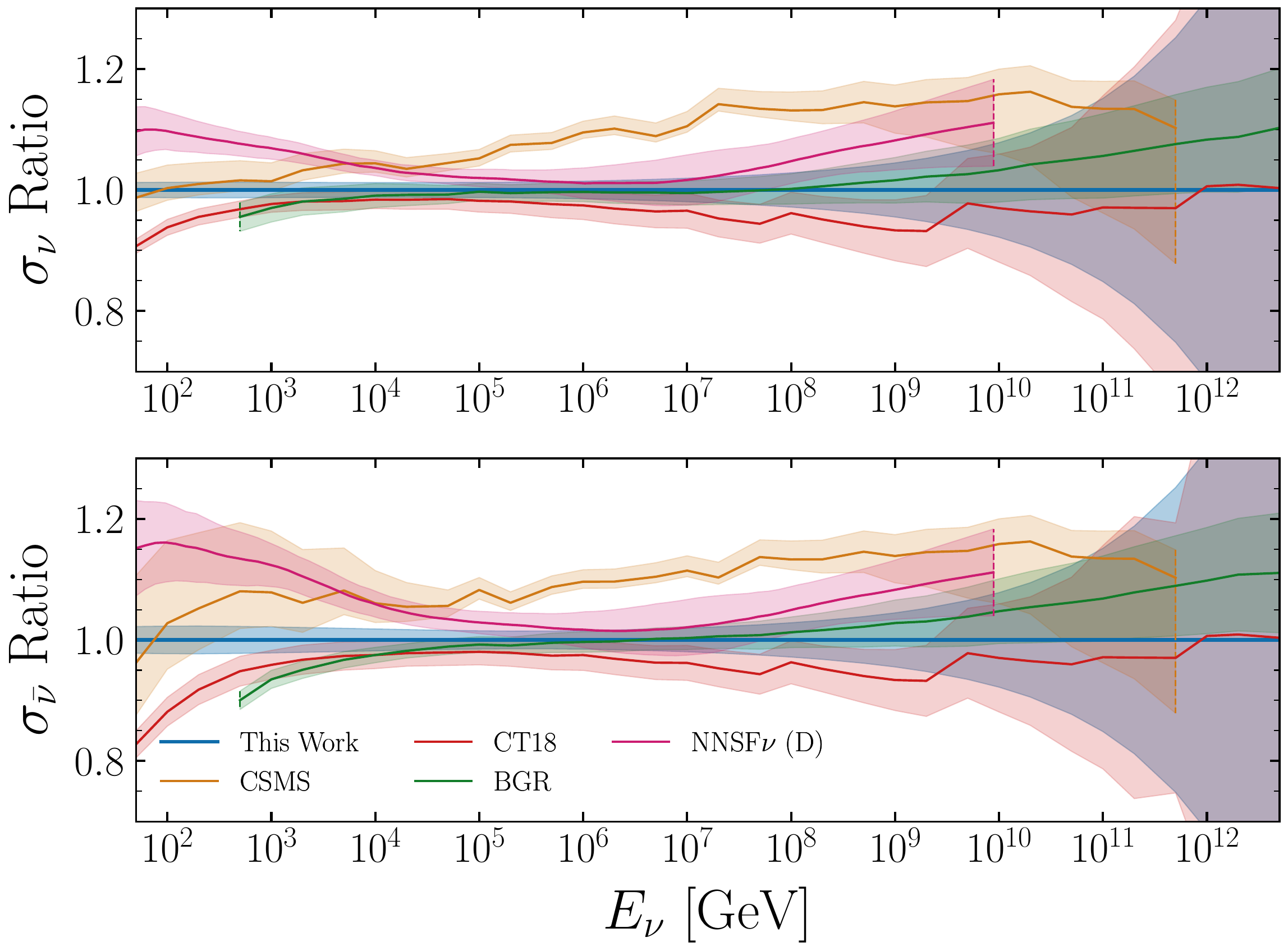}
    \caption{Ratio of the CSMS, CT18, BGR, and NNSF$\nu$ cross section models to the NNLO isoscalar model presented in this work.}
    \label{fig:he_xsec_comparison}
\end{figure}

We now turn to comparisons of the cross-section model from this work to other models. 
In \cref{fig:he_xsec_comparison}, we take the ratio of the other cross-section models to the total cross section shown in \cref{fig:total_xs}. 
The CT18 model \cite{Xie:2023suk} is a NNLO calculation using the CT18NNLO PDF set and the S-ACOT-$\chi$ GM-VFNS \cite{Guzzi:2011ew} and also includes small-$x$ resummation.
We note that the CT18 model is most similar to the one presented here in terms of PDF sets and methods for computing the cross sections at high energies, which served as a starting point for this work.
The BGR model \cite{Bertone:2017gds} is based on a modified NNPDF3.1sx PDF set \cite{NNPDF:2017mvq, Ball:2017otu} and uses the FONLL GM-VFNS with small-$x$ resummation.
The CSMS model \cite{Cooper-Sarkar:2011jtt} is based on the HERAPDF1.5 PDF set \cite{Radescu:2013mka} and uses the ZM-VFNS at NLO, which is frequently used as the baseline cross-section model in IceCube analyses \cite{IceCube:2017roe,IceCube:2020rnc}.
Lastly, the NNSF$\nu$ model \cite{Candido:2023utz} is a mixture of fits to data using the NNPDF methodology \cite{Forte:2002fg, DelDebbio:2004xtd, DelDebbio:2007ee, NNPDF:2021njg} and pQCD calculations.
A comparison to the isoscalar, oxygen, iron, and lead cross sections from NNSF$\nu$ can be found in \cref{sec:appendix_nnsfnu_comparison}.

\begin{figure}
    \centering
    \includegraphics[width=0.48\textwidth]{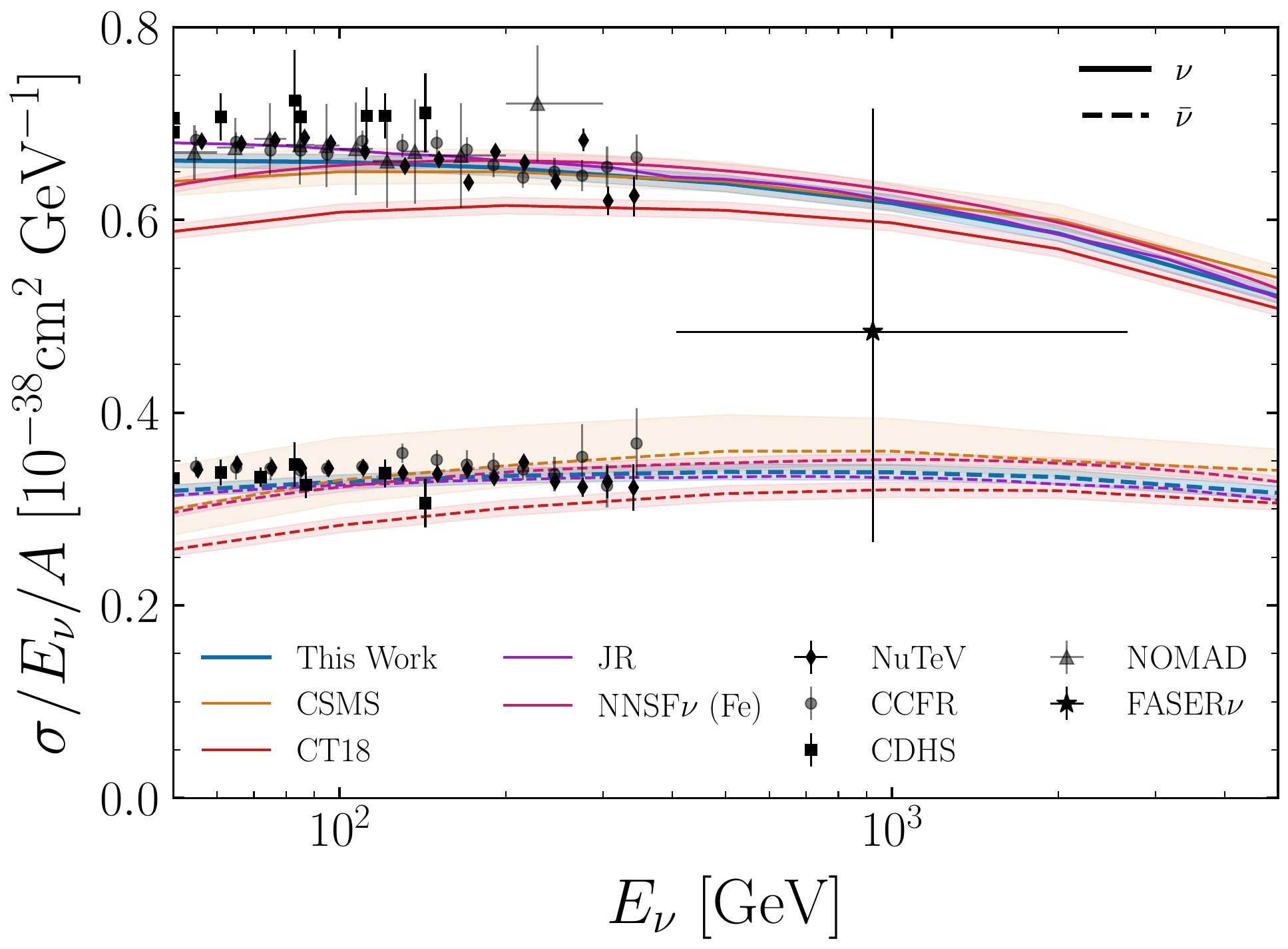}
    \caption{The low-energy cross sections per nucleon divided by neutrino energy, $\sigma/E_\nu/A$ for this work, the CSMS, CT18, JR, NNSF$\nu$ models and reported uncertainties. Neutrino cross sections are shown as solid lines, and antineutrino as dashed. The experimental data from NuTeV, CCFR, CDHS, NOMAD, and FASER$\nu$ are shown as black points with their uncertainties. Note: the FASER$\nu$ measurement is flux averaged.}
    \label{fig:le_xsec_comparison}
\end{figure}

The low-energy cross sections are shown in \cref{fig:le_xsec_comparison}, which features several of the same models as \cref{fig:he_xsec_comparison}.
Notably, we also compare against the JR model \cite{Jeong:2023hwe}, which uses the CKMT-PCAC-NT method for constructing the structure functions for $Q^2 < 4$ GeV$^2$.
Also shown are measurements from NuTeV \cite{NuTeV:2005wsg}, CCFR \cite{Seligman:1997fe}, CDHS \cite{Berge:1987zw}, NOMAD \cite{NOMAD:2007krq}, and FASER$\nu$ \cite{FASER:2024hoe}. Note that the FASER$\nu$ measurement is a flux-averaged measurement.

\begin{figure}
    \centering
    \includegraphics[width=0.48\textwidth]{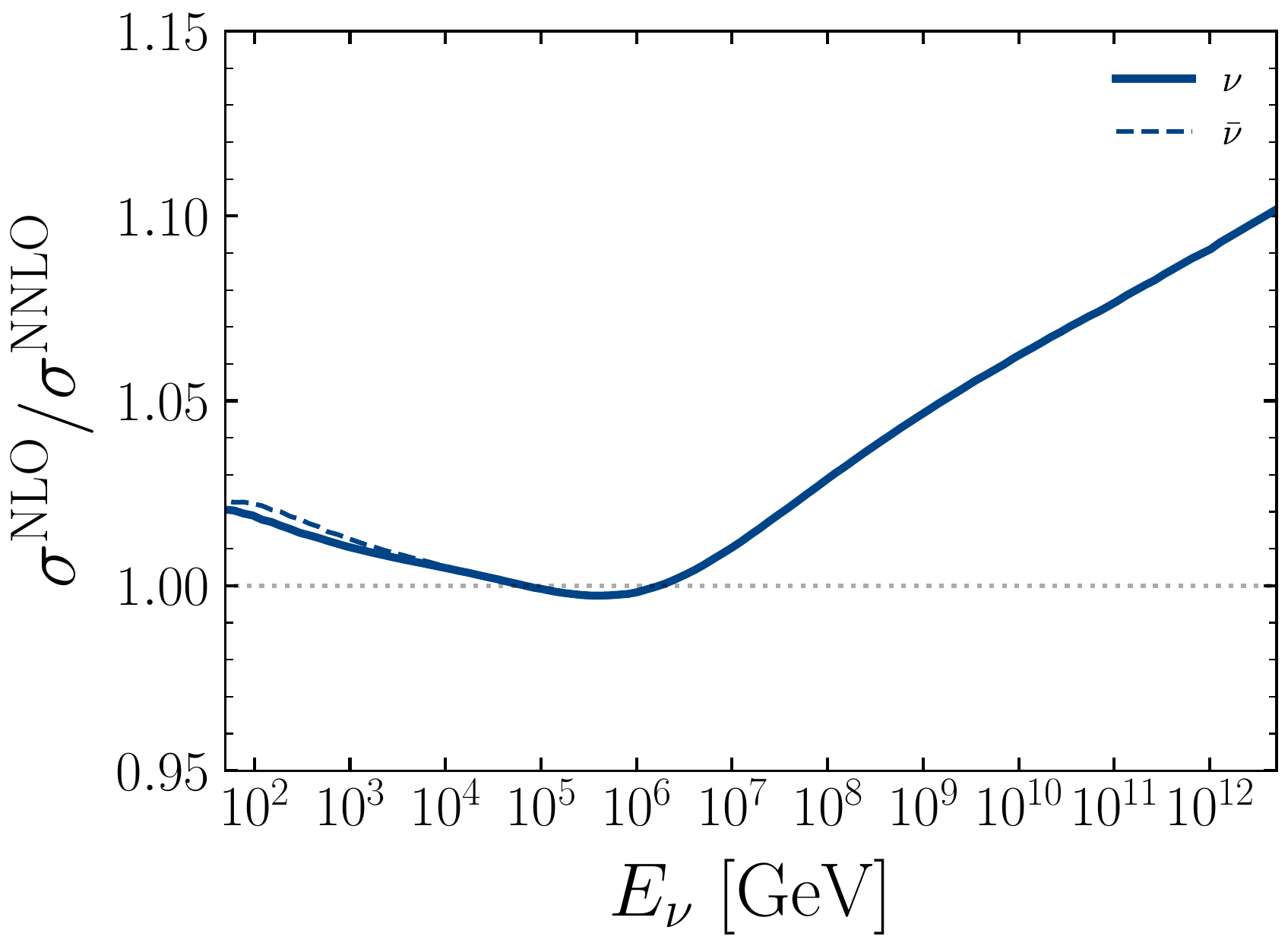}
    \caption{Ratio of the NLO cross section to the NNLO cross section for neutrinos (solid) and antineutrinos (dashed). The NLO cross section was computed using the FONLL-B mass scheme with the CT18ANLO PDF set, and the NNLO cross section uses FONLL-C scheme with the CT18ANNLO PDF set.}
    \label{fig:nlo_nnlo_comparison}
\end{figure}

In \cref{fig:nlo_nnlo_comparison}, we compare the calculation of the total cross section using the NNLO calculation (CT18ANNLO PDF set and FONLL-C) with the NLO calculation (CT18ANLO PDF set and FONLL-B).
Alternative choices of PDF sets may also be of interest due to the differences in fitting methodologies and datasets.
We include calculations of the cross sections using the baseline model and an alternative model using the NNPDF4.0 MHOU PDF set \cite{NNPDF:2024dpb} in \cref{sec:appendix_xsec}.
The missing higher-order uncertainties (MHOU) variant is an improvement over previous releases of the NNPDF4 PDF set \cite{NNPDF:2021njg}, which includes estimated theoretical uncertainties from higher-order terms beyond NNLO \cite{NNPDF:2019vjt, NNPDF:2019ubu}.
We note that the cross section calculated with this PDF set is similar to that of the CT18A PDF set but has much smaller PDF uncertainties. 

Significant work has been done recently on improving the calculations of the DIS structure functions and the frameworks for computing them; we highlight a few of them here.
\texttt{yadism} \cite{Candido:2024rkr} is a code developed for the computation of DIS structure functions written in Python and is used by the NNPDF collaboration.
Additionally, \texttt{APFEL++} \cite{Bertone:2017gds} is a C++ rewrite of the original \texttt{APFEL} code that provides additional flexibility and speed improvements to the calculations, with additional applications beyond DIS structure functions.
Improvements in the methods for determinations of the PDFs and the DIS structure functions will improve our understanding of the high-energy neutrino cross sections.
Already, there has been a significant effort towards producing N$^3$LO PDF sets from the MSHT and NNPDF groups \cite{McGowan:2022nag, McGowan:2023xyu, NNPDF:2024nan}, and implementations of some N$^3$LO structure functions in the aforementioned codes.

The choice of the FONLL-B and FONLL-C mass schemes was driven by the availability of the structure function calculations in \texttt{APFEL} to perform these calculations.
In principle, similar predictions can be obtained using other mass schemes such as S-ACOT-$\chi$ \cite{Guzzi:2011ew,Gao:2021fle} or TR' \cite{Thorne:2006qt}.
We note that the inclusive nature of the charm cross sections in these calculations arises from the usage of the FONLL mass scheme, as discussed in \cref{sec:hq_fonll}, and is not ideal for making predictions of charm production at high energies. 
A different methodology is required to obtain more exclusive flavor separation of the cross sections (i.e. only outgoing charm). 

The effects of nuclear shadowing result in an overall decrease in the total cross section, a few percent at low energies and as large as $-20\%$ at the highest energies for oxygen.
In these calculations, we found that the uncertainties at high energies for the nPDF-based calculations are substantially lower than those of the free nucleons.
This was traced to member 57 of the CT18A PDF sets, which is included in the EPPS21 nPDF sets with nuclear modifications as member 106.
The gluon and singlet PDFs at $x < 10^{-4}$ are substantially larger than the other replicas in the set, but these differences become smaller with increasing $A$ in the EPPS21 sets.
While the nuclear cross sections presented here have a larger uncertainty over most of the energy range ($E_{\nu} < 2\times10^{10}~\rm{GeV}$), this indicates that the EPPS21 fit is reducing the uncertainty from this PDF replica.
We speculate that this is due to the inclusion of an LHCb prompt $D$ meson production dataset with $p$Pb collisions \cite{LHCb:2017yua}, which probes down to $x=10^{-5}$ \cite{Eskola:2021nhw}.
This is consistent with the findings of Refs.~\cite{Bertone:2018dse, Garcia:2020jwr}, where the impact of including the LHCb $D$ meson data at low $x$ was quantified on the NNPDF3.1sx PDF sets.

\section{Conclusions} \label{sec:conclusions}
In this work, we have presented calculations of the high-energy charged-current neutrino deep inelastic scattering cross sections and inelasticity distributions, improving various parts of the calculation over previous models.
This work primarily focused on quantifying the contributions of the flavor components, specifically charm and top, and PDF uncertainties of the cross sections and inelasticity distributions.
The NNLO structure functions were constructed to account for heavy quark masses through the FONLL GM-VFNS for each mass threshold and to include target mass corrections, effects at low $Q^2$ using the CKMT-PCAC-NT method, and small-$x$ resummation.
It was determined that the uncertainty of the total CC DIS cross section is less than $2\%$ in the energy range of neutrinos measured by neutrino telescopes ($E_\nu \lesssim 10$ PeV) for free nucleons and less than $5\%$ for oxygen.
The uncertainty on the shape of the inelasticity distribution and mean inelasticity $\langle y \rangle$ was found to be smaller than the current resolution of inelasticity measurements from IceCube \cite{IceCube:2018pgc}, though the shifts from FSR were found to be non-negligible.

In our evaluation of the modifications to the cross section from nuclear effects, we find that there is a small reduction in the total cross section at low energies but larger shadowing effects are seen at $E_{\nu} \gtrsim 100$ TeV.
The oxygen and free proton cross sections were used to construct the H$_2$O inelasticity distribution, which showed non-negligible shape effects when compared to the isoscalar target case.
These differences and the FSR effects demonstrate the need for properly modeling the single-differential cross section for inelasticity-related observables since they can lead to substantial systematic shifts when compared to the PDF uncertainties.

\section*{Acknowledgements}
The authors gratefully acknowledge M. H. Reno for discussions and guidance related to the low-$Q^2$ structure functions and target mass corrections; F. Olness for assistance in checks against the ACOT charm cross sections; R. Plestid for discussions on final state radiation; and F. Kling and M. Bustamante for providing the data files for the experimental cross sections. P.L.R.W. and J.M.C. are supported by NSF under NSF-PHY-2310050, and A.G. is supported by the CDEIGENT grant No. CIDEIG/2023/20.

\bibliographystyle{apsrev4-1}
\bibliography{bibliography}

\appendix
\counterwithin{figure}{section}
\counterwithin{table}{section}

\section{Comparisons to NNSF$\nu$} \label{sec:appendix_nnsfnu_comparison}
\begin{figure*}
    \centering
    \includegraphics[width=0.8\textwidth]{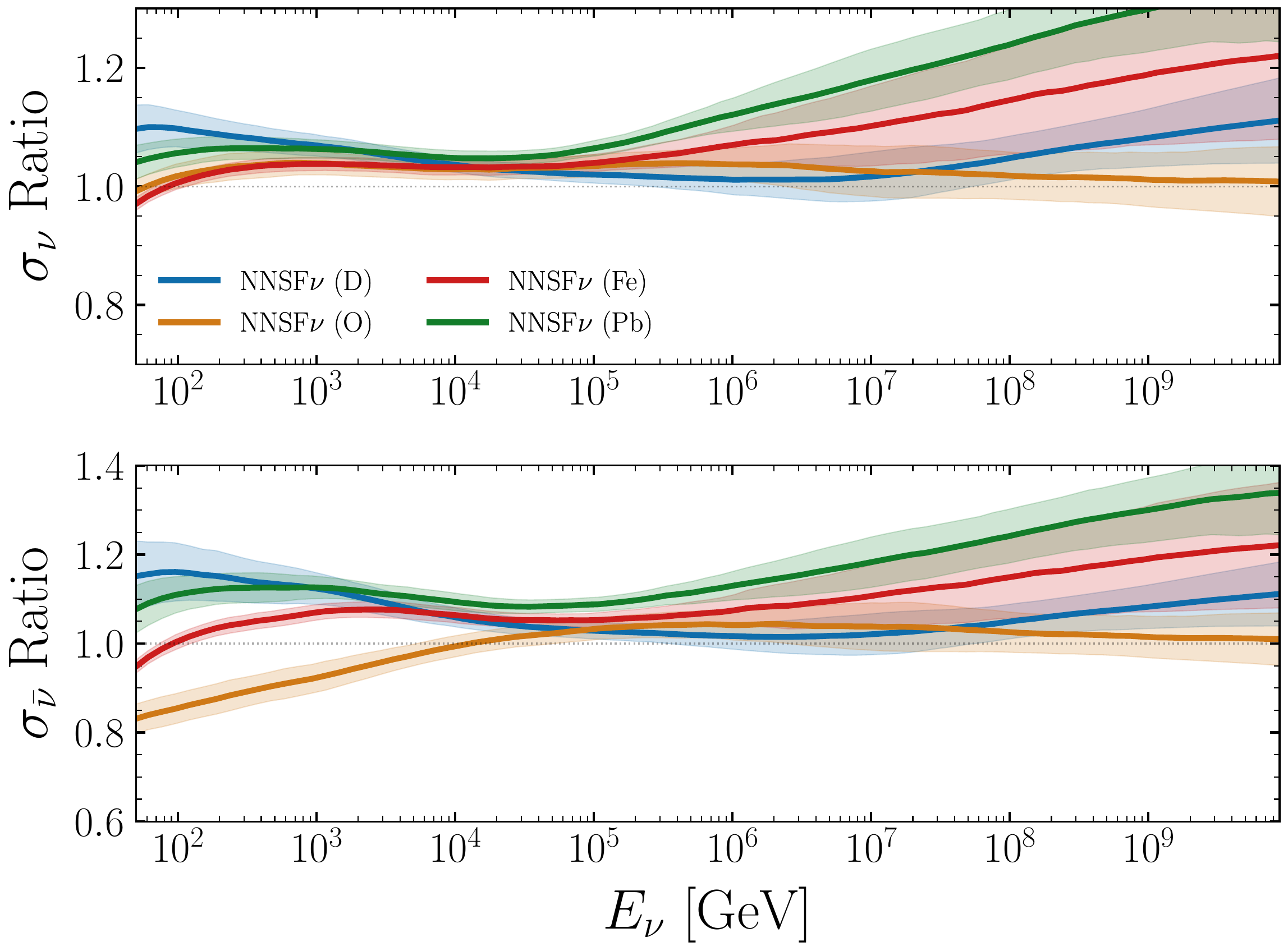}
    \caption{Comparison between the isoscalar nucleon cross section (\cref{fig:total_xs}) and nuclear cross sections (\cref{fig:nuclear_xs}) with those of the NNSF$\nu$ model \cite{Candido:2023utz}.}
    \label{fig:appendix_nnsfnu_comparison}
\end{figure*}

The NNSF$\nu$ model provides neutrino CC DIS cross sections for the same targets as shown in \cref{fig:nuclear_xs}. In \cref{fig:appendix_nnsfnu_comparison}, we take the ratio of the NNSF$\nu$ nuclear cross sections to the ones from \cref{fig:nuclear_xs}.
We also show the ratio of the isoscalar cross section in \cref{fig:total_xs} with the (scaled) deuteron cross section from NNSF$\nu$.

\section{Inelasticity Distributions} \label{sec:appendix_inelasticity_distributions}
\begin{figure*}
    \centering
    \includegraphics[width=\textwidth]{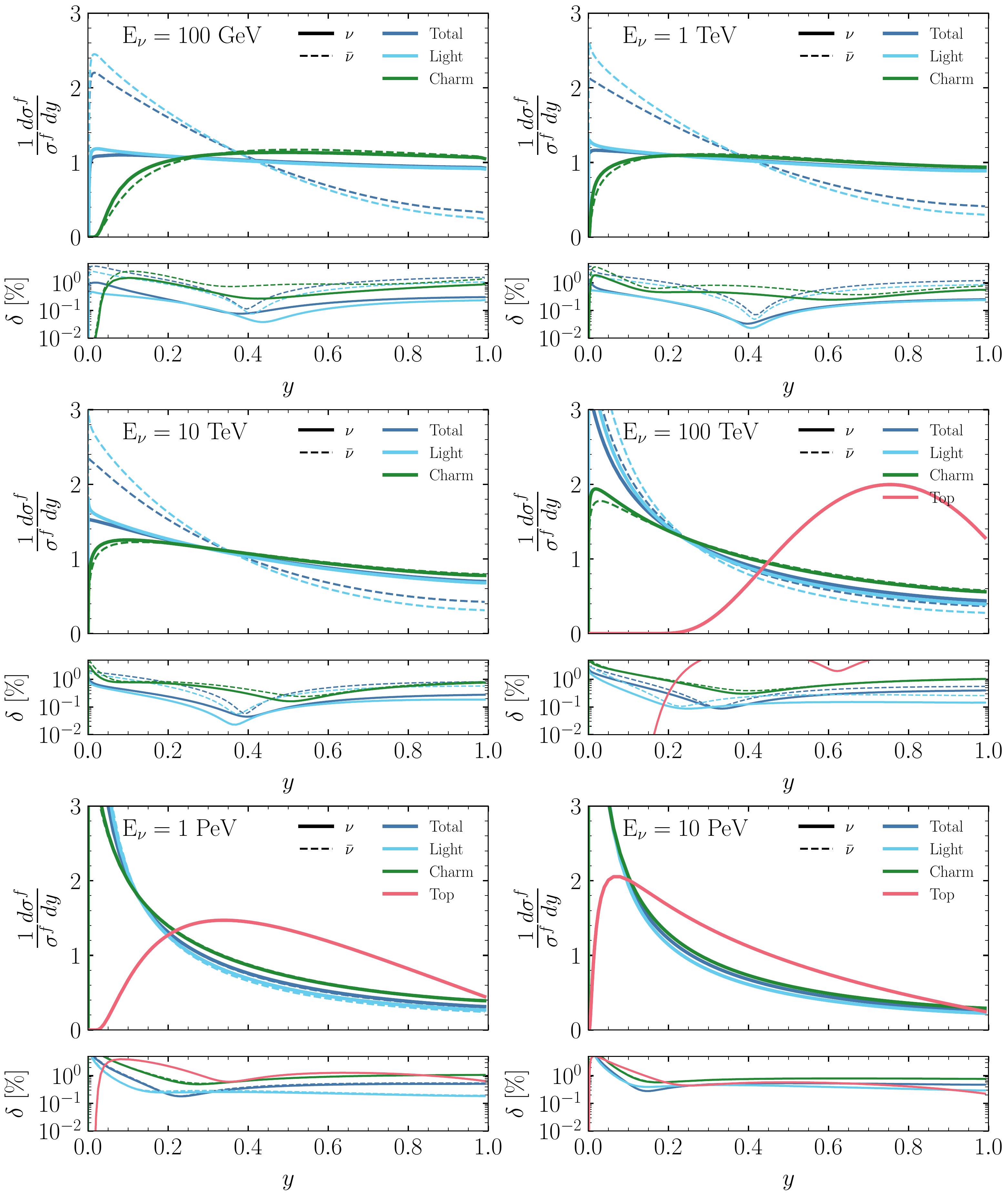}
    \caption{Inelasticity distributions for each flavor component of the cross section (top panel) and PDF uncertainties (bottom panel) as shown in \cref{fig:dsdy} for $E_{\nu} =$ 100 GeV, 1 TeV, 10 TeV, 100 TeV, 1 PeV, and 10 PeV.}
    \label{fig:appendix_big_inelasticity}
\end{figure*}

\begin{figure*}
    \centering
    \includegraphics[width=\textwidth]{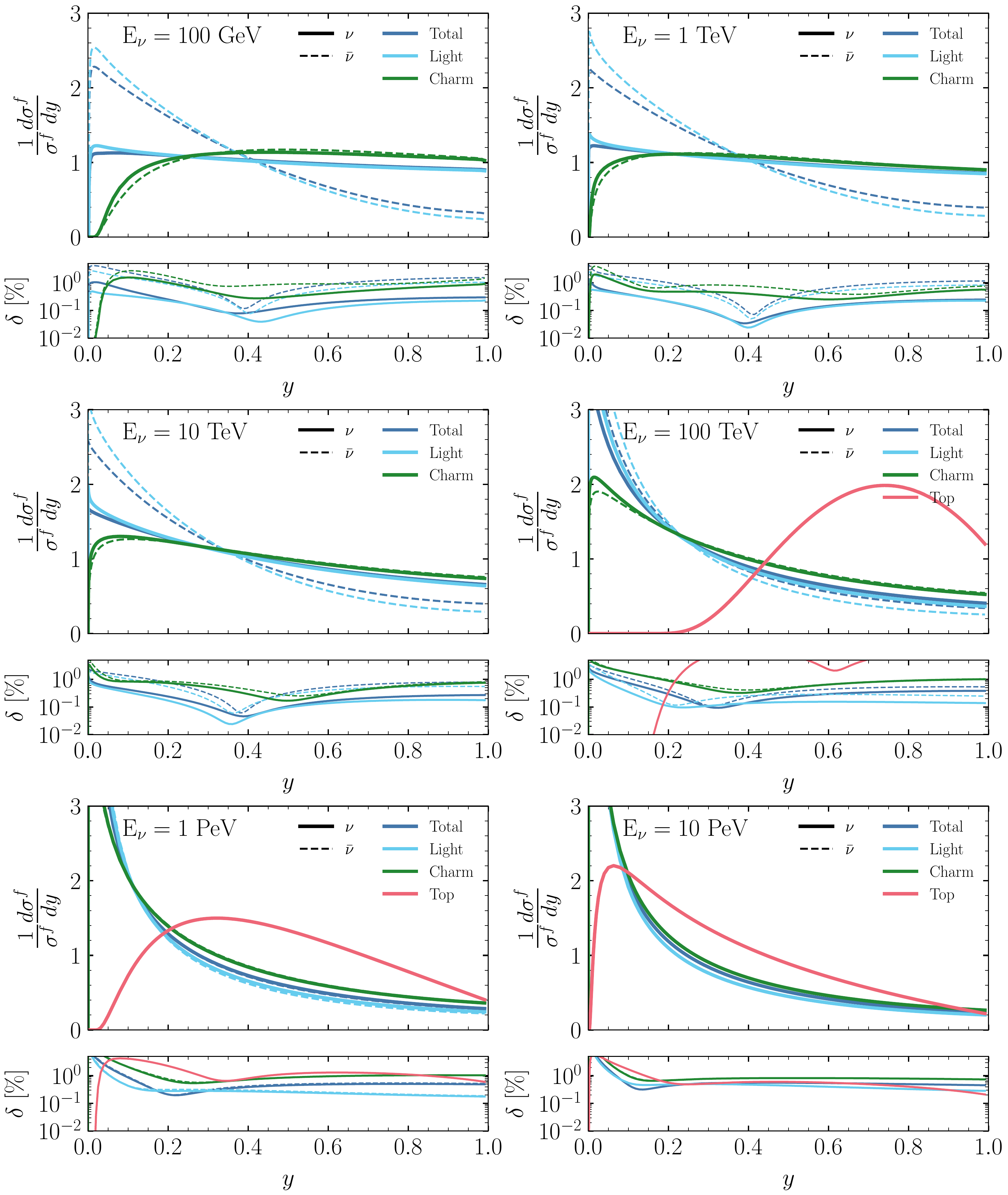}
    \caption{Same as \cref{fig:appendix_big_inelasticity} without final state radiation.}
    \label{fig:appendix_big_inelasticity_noFSR}
\end{figure*}

\counterwithin{figure}{section}
In \cref{fig:appendix_big_inelasticity}, we show the inelasticity distributions and the PDF uncertainties shown in \cref{fig:dsdy} over a range of energies. Additionally, we show the same quantities but without final state radiation in \cref{fig:appendix_big_inelasticity_noFSR} as it may be useful.

\section{Neutrino Cross Sections} \label{sec:appendix_xsec}
\counterwithin{figure}{section}
In this appendix, we include tables of the cross sections shown in this work. For all nuclear targets, the cross sections are reported as $\sigma/A$. We provide the following tables:
\begin{itemize}
    \item NNLO free nucleon cross sections: proton (\cref{tab:appendix_ct18annlo_proton_xs}), neutron (\cref{tab:appendix_ct18annlo_neutron_xs}), isoscalar (\cref{tab:appendix_ct18annlo_isoscalar_xs})
    \item NLO free nucleon cross sections: proton (\cref{tab:appendix_ct18anlo_proton_xs}), neutron (\cref{tab:appendix_ct18anlo_neutron_xs}), isoscalar (\cref{tab:appendix_ct18anlo_isoscalar_xs})
    \item NLO $^{16}\rm{O}$ cross sections (\cref{tab:appendix_epps_oxygen_xs})
    \item NLO $^{56}\rm{Fe}$ cross sections (\cref{tab:appendix_epps_iron_xs})
    \item NLO $^{208}\rm{Pb}$ cross sections (\cref{tab:appendix_epps_lead_xs})
    \item NNLO free nucleon cross sections using the NNPDF40\_nnlo\_as\_01180\_mhou PDF set: proton (\cref{tab:appendix_nnpdf4_proton_xs}), neutron (\cref{tab:appendix_nnpdf4_neutron_xs}), isoscalar (\cref{tab:appendix_nnpdf4_isoscalar_xs})
\end{itemize}
This data will be available at:
\begin{center}
    \href{https://github.com/pweigel/WCG_Cross_Sections}{https://github.com/pweigel/WCG\_Cross\_Sections}
\end{center}

A neutral-current cross section compatible with the charged-current cross section presented here and the differential cross sections are also planned for a future release.

\setlength\extrarowheight{0.5pt}
\setlength{\tabcolsep}{7pt} 
\renewcommand{\arraystretch}{1.25}

\clearpage
\begin{table*}[h!]
\begin{tabular}{|c||c|c|c|c|c||c|c|c|c|c|} 
 \hline
 $E_{\nu}~\textrm{[GeV]}$ & $\sigma^{\nu p}~\textrm{[pb]}$ & $\delta \sigma^{\nu p}~[\%]$ & $f_{c}^{\nu p}$ & $f_{b}^{\nu p}$ & $f_{t}^{\nu p}$ & $\sigma^{\bar{\nu} p}~\textrm{[pb]}$ & $\delta \sigma^{\bar{\nu} p}~[\%]$ & $f_{c}^{\bar{\nu} p}$ & $f_{b}^{\bar{\nu} p}$ & $f_{t}^{\bar{\nu} p}$ \\
 \hline
5e1 & 2.17e-1 & $2.3$ & 8.2e-2 & 6.1e-9 & 0 & 2.00e-1 & $1.8$ & 5.9e-2 & 5.3e-7 & 0 \\ 
1e2 & 4.41e-1 & $2.3$ & 1.1e-1 & 1.9e-7 & 0 & 4.06e-1 & $1.9$ & 8.2e-2 & 2.2e-6 & 0 \\ 
2e2 & 8.88e-1 & $2.3$ & 1.3e-1 & 1.4e-6 & 0 & 8.19e-1 & $1.9$ & 1.0e-1 & 5.2e-6 & 0 \\ 
5e2 & 2.22e0 & $2.3$ & 1.5e-1 & 6.7e-6 & 0 & 2.05e0 & $1.8$ & 1.3e-1 & 1.3e-5 & 0 \\ 
1e3 & 4.36e0 & $2.2$ & 1.7e-1 & 1.5e-5 & 0 & 4.06e0 & $1.8$ & 1.5e-1 & 2.2e-5 & 0 \\ 
2e3 & 8.44e0 & $2.1$ & 1.8e-1 & 2.8e-5 & 0 & 7.94e0 & $1.7$ & 1.6e-1 & 3.6e-5 & 0 \\ 
5e3 & 1.93e1 & $2.0$ & 2.0e-1 & 5.2e-5 & 0 & 1.87e1 & $1.6$ & 1.8e-1 & 6.0e-5 & 0 \\ 
1e4 & 3.46e1 & $1.9$ & 2.2e-1 & 7.4e-5 & 0 & 3.43e1 & $1.5$ & 2.0e-1 & 8.1e-5 & 0 \\ 
2e4 & 5.90e1 & $1.8$ & 2.4e-1 & 1.0e-4 & 3.0e-9 & 6.04e1 & $1.5$ & 2.2e-1 & 1.0e-4 & 2.7e-9 \\ 
5e4 & 1.11e2 & $1.7$ & 2.8e-1 & 1.5e-4 & 2.4e-5 & 1.18e2 & $1.4$ & 2.4e-1 & 1.4e-4 & 2.2e-5 \\ 
1e5 & 1.71e2 & $1.7$ & 3.0e-1 & 1.8e-4 & 2.9e-4 & 1.85e2 & $1.3$ & 2.7e-1 & 1.7e-4 & 2.7e-4 \\ 
2e5 & 2.55e2 & $1.6$ & 3.3e-1 & 2.3e-4 & 1.3e-3 & 2.79e2 & $1.3$ & 2.9e-1 & 2.1e-4 & 1.2e-3 \\ 
5e5 & 4.15e2 & $1.6$ & 3.6e-1 & 2.8e-4 & 5.0e-3 & 4.54e2 & $1.4$ & 3.2e-1 & 2.6e-4 & 4.6e-3 \\ 
1e6 & 5.86e2 & $1.7$ & 3.8e-1 & 3.2e-4 & 9.9e-3 & 6.38e2 & $1.5$ & 3.4e-1 & 3.0e-4 & 9.1e-3 \\ 
2e6 & 8.16e2 & $1.8$ & 4.0e-1 & 3.6e-4 & 1.6e-2 & 8.79e2 & $1.6$ & 3.6e-1 & 3.4e-4 & 1.5e-2 \\ 
5e6 & 1.24e3 & $1.9$ & 4.1e-1 & 4.0e-4 & 2.7e-2 & 1.31e3 & $1.8$ & 3.8e-1 & 3.8e-4 & 2.5e-2 \\ 
1e7 & 1.67e3 & $2.0$ & 4.2e-1 & 4.3e-4 & 3.5e-2 & 1.75e3 & $1.9$ & 4.0e-1 & 4.1e-4 & 3.4e-2 \\ 
2e7 & 2.22e3 & $2.2$ & 4.2e-1 & 4.6e-4 & 4.4e-2 & 2.31e3 & $2.1$ & 4.0e-1 & 4.4e-4 & 4.2e-2 \\ 
5e7 & 3.21e3 & $2.6$ & 4.3e-1 & 4.9e-4 & 5.6e-2 & 3.30e3 & $2.5$ & 4.1e-1 & 4.8e-4 & 5.4e-2 \\ 
1e8 & 4.19e3 & $2.9$ & 4.3e-1 & 5.1e-4 & 6.4e-2 & 4.29e3 & $2.8$ & 4.2e-1 & 5.0e-4 & 6.3e-2 \\ 
2e8 & 5.42e3 & $3.2$ & 4.3e-1 & 5.3e-4 & 7.2e-2 & 5.52e3 & $3.2$ & 4.2e-1 & 5.2e-4 & 7.1e-2 \\ 
5e8 & 7.54e3 & $3.9$ & 4.3e-1 & 5.5e-4 & 8.2e-2 & 7.64e3 & $3.8$ & 4.2e-1 & 5.4e-4 & 8.1e-2 \\ 
1e9 & 9.61e3 & $4.5$ & 4.3e-1 & 5.6e-4 & 9.0e-2 & 9.71e3 & $4.4$ & 4.2e-1 & 5.5e-4 & 8.8e-2 \\ 
2e9 & 1.22e4 & $5.2$ & 4.3e-1 & 5.7e-4 & 9.6e-2 & 1.23e4 & $5.2$ & 4.3e-1 & 5.7e-4 & 9.5e-2 \\ 
5e9 & 1.65e4 & $6.5$ & 4.3e-1 & 5.9e-4 & 1.0e-1 & 1.66e4 & $6.5$ & 4.3e-1 & 5.8e-4 & 1.0e-1 \\ 
1e10 & 2.07e4 & $7.8$ & 4.3e-1 & 5.9e-4 & 1.1e-1 & 2.08e4 & $7.8$ & 4.3e-1 & 5.9e-4 & 1.1e-1 \\ 
2e10 & 2.57e4 & $9.5$ & 4.3e-1 & 6.0e-4 & 1.2e-1 & 2.59e4 & $9.4$ & 4.3e-1 & 6.0e-4 & 1.2e-1 \\ 
5e10 & 3.42e4 & $12$ & 4.3e-1 & 6.1e-4 & 1.2e-1 & 3.43e4 & $12$ & 4.2e-1 & 6.1e-4 & 1.2e-1 \\ 
1e11 & 4.23e4 & $15$ & 4.3e-1 & 6.2e-4 & 1.3e-1 & 4.24e4 & $15$ & 4.2e-1 & 6.2e-4 & 1.3e-1 \\ 
2e11 & 5.20e4 & $19$ & 4.2e-1 & 6.2e-4 & 1.3e-1 & 5.21e4 & $19$ & 4.2e-1 & 6.2e-4 & 1.3e-1 \\ 
5e11 & 6.80e4 & $25$ & 4.2e-1 & 6.3e-4 & 1.4e-1 & 6.81e4 & $25$ & 4.2e-1 & 6.3e-4 & 1.4e-1 \\ 
1e12 & 8.30e4 & $32$ & 4.2e-1 & 6.3e-4 & 1.4e-1 & 8.31e4 & $31$ & 4.2e-1 & 6.3e-4 & 1.4e-1 \\ 
2e12 & 1.01e5 & $40$ & 4.2e-1 & 6.4e-4 & 1.4e-1 & 1.01e5 & $40$ & 4.2e-1 & 6.3e-4 & 1.4e-1 \\ 
5e12 & 1.31e5 & $54$ & 4.2e-1 & 6.4e-4 & 1.5e-1 & 1.31e5 & $53$ & 4.2e-1 & 6.4e-4 & 1.5e-1 \\ 
 \hline
\end{tabular}
\caption{NNLO neutrino-proton and antineutrino-proton CC DIS cross section, cross section uncertainty, and flavor ratios shown in \cref{fig:total_xs} and \cref{fig:flavor_fraction}. These cross sections were computed with the CT18ANNLO PDF set evolved with $n_f = 6$ and small-$x$ resummation. See \cref{sec:structure_functions} and \cref{sec:xsecs} for details on the structure function and cross-section calculations.}
\label{tab:appendix_ct18annlo_proton_xs}
\end{table*}

\begin{table*}[h!]
\begin{tabular}{|c||c|c|c|c|c||c|c|c|c|c|} 
 \hline
 $E_{\nu}~\textrm{[GeV]}$ & $\sigma^{\nu n}~\textrm{[pb]}$ & $\delta \sigma^{\nu n}~[\%]$ & $f_{c}^{\nu n}$ & $f_{b}^{\nu n}$ & $f_{t}^{\nu n}$ & $\sigma^{\bar{\nu} n}~\textrm{[pb]}$ & $\delta \sigma^{\bar{\nu} n}~[\%]$ & $f_{c}^{\bar{\nu} n}$ & $f_{b}^{\bar{\nu} n}$ & $f_{t}^{\bar{\nu} n}$ \\
 \hline
5e1 & 4.31e-1 & $1.2$ & 6.2e-2 & 3.1e-9 & 0 & 1.12e-1 & $3.4$ & 1.0e-1 & 2.0e-7 & 0 \\ 
1e2 & 8.55e-1 & $1.2$ & 7.6e-2 & 1.0e-7 & 0 & 2.36e-1 & $3.4$ & 1.4e-1 & 1.4e-6 & 0 \\ 
2e2 & 1.69e0 & $1.2$ & 8.9e-2 & 7.3e-7 & 0 & 4.93e-1 & $3.3$ & 1.7e-1 & 4.7e-6 & 0 \\ 
5e2 & 4.08e0 & $1.2$ & 1.0e-1 & 3.7e-6 & 0 & 1.28e0 & $3.2$ & 2.0e-1 & 1.5e-5 & 0 \\ 
1e3 & 7.86e0 & $1.2$ & 1.1e-1 & 8.6e-6 & 0 & 2.62e0 & $3.0$ & 2.3e-1 & 3.0e-5 & 0 \\ 
2e3 & 1.48e1 & $1.2$ & 1.3e-1 & 1.6e-5 & 0 & 5.25e0 & $2.8$ & 2.4e-1 & 5.0e-5 & 0 \\ 
5e3 & 3.24e1 & $1.1$ & 1.4e-1 & 3.1e-5 & 0 & 1.28e1 & $2.6$ & 2.7e-1 & 8.3e-5 & 0 \\ 
1e4 & 5.55e1 & $1.1$ & 1.6e-1 & 4.7e-5 & 0 & 2.42e1 & $2.4$ & 2.8e-1 & 1.1e-4 & 0 \\ 
2e4 & 8.98e1 & $1.1$ & 1.8e-1 & 6.7e-5 & 2.9e-9 & 4.39e1 & $2.2$ & 3.0e-1 & 1.4e-4 & 3.8e-9 \\ 
5e4 & 1.57e2 & $1.1$ & 2.1e-1 & 1.0e-4 & 1.8e-5 & 9.02e1 & $1.9$ & 3.2e-1 & 1.8e-4 & 2.9e-5 \\ 
1e5 & 2.28e2 & $1.1$ & 2.4e-1 & 1.4e-4 & 2.2e-4 & 1.47e2 & $1.8$ & 3.3e-1 & 2.2e-4 & 3.4e-4 \\ 
2e5 & 3.22e2 & $1.2$ & 2.7e-1 & 1.8e-4 & 1.1e-3 & 2.30e2 & $1.6$ & 3.5e-1 & 2.5e-4 & 1.5e-3 \\ 
5e5 & 4.95e2 & $1.2$ & 3.1e-1 & 2.4e-4 & 4.2e-3 & 3.92e2 & $1.6$ & 3.7e-1 & 3.0e-4 & 5.3e-3 \\ 
1e6 & 6.74e2 & $1.3$ & 3.4e-1 & 2.8e-4 & 8.6e-3 & 5.66e2 & $1.6$ & 3.9e-1 & 3.4e-4 & 1.0e-2 \\ 
2e6 & 9.10e2 & $1.5$ & 3.6e-1 & 3.2e-4 & 1.5e-2 & 8.00e2 & $1.7$ & 4.0e-1 & 3.7e-4 & 1.7e-2 \\ 
5e6 & 1.34e3 & $1.7$ & 3.8e-1 & 3.7e-4 & 2.5e-2 & 1.22e3 & $1.8$ & 4.1e-1 & 4.1e-4 & 2.7e-2 \\ 
1e7 & 1.77e3 & $1.9$ & 4.0e-1 & 4.1e-4 & 3.3e-2 & 1.66e3 & $2.0$ & 4.2e-1 & 4.4e-4 & 3.6e-2 \\ 
2e7 & 2.33e3 & $2.1$ & 4.1e-1 & 4.4e-4 & 4.2e-2 & 2.22e3 & $2.2$ & 4.2e-1 & 4.6e-4 & 4.4e-2 \\ 
5e7 & 3.32e3 & $2.5$ & 4.1e-1 & 4.7e-4 & 5.4e-2 & 3.21e3 & $2.5$ & 4.3e-1 & 4.9e-4 & 5.6e-2 \\ 
1e8 & 4.30e3 & $2.8$ & 4.2e-1 & 5.0e-4 & 6.3e-2 & 4.19e3 & $2.9$ & 4.3e-1 & 5.1e-4 & 6.4e-2 \\ 
2e8 & 5.53e3 & $3.2$ & 4.2e-1 & 5.2e-4 & 7.1e-2 & 5.42e3 & $3.2$ & 4.3e-1 & 5.3e-4 & 7.2e-2 \\ 
5e8 & 7.66e3 & $3.8$ & 4.2e-1 & 5.4e-4 & 8.1e-2 & 7.54e3 & $3.9$ & 4.3e-1 & 5.5e-4 & 8.2e-2 \\ 
1e9 & 9.72e3 & $4.4$ & 4.3e-1 & 5.5e-4 & 8.9e-2 & 9.61e3 & $4.5$ & 4.3e-1 & 5.6e-4 & 8.9e-2 \\ 
2e9 & 1.23e4 & $5.2$ & 4.3e-1 & 5.7e-4 & 9.5e-2 & 1.22e4 & $5.2$ & 4.3e-1 & 5.7e-4 & 9.6e-2 \\ 
5e9 & 1.66e4 & $6.5$ & 4.3e-1 & 5.8e-4 & 1.0e-1 & 1.65e4 & $6.5$ & 4.3e-1 & 5.9e-4 & 1.0e-1 \\ 
1e10 & 2.08e4 & $7.8$ & 4.3e-1 & 5.9e-4 & 1.1e-1 & 2.07e4 & $7.8$ & 4.3e-1 & 5.9e-4 & 1.1e-1 \\ 
2e10 & 2.59e4 & $9.4$ & 4.3e-1 & 6.0e-4 & 1.2e-1 & 2.58e4 & $9.5$ & 4.3e-1 & 6.0e-4 & 1.2e-1 \\ 
5e10 & 3.44e4 & $12$ & 4.3e-1 & 6.1e-4 & 1.2e-1 & 3.42e4 & $12$ & 4.3e-1 & 6.1e-4 & 1.2e-1 \\ 
1e11 & 4.24e4 & $15$ & 4.2e-1 & 6.2e-4 & 1.3e-1 & 4.23e4 & $15$ & 4.3e-1 & 6.2e-4 & 1.3e-1 \\ 
2e11 & 5.21e4 & $19$ & 4.2e-1 & 6.2e-4 & 1.3e-1 & 5.20e4 & $19$ & 4.2e-1 & 6.2e-4 & 1.3e-1 \\ 
5e11 & 6.81e4 & $25$ & 4.2e-1 & 6.3e-4 & 1.4e-1 & 6.80e4 & $25$ & 4.2e-1 & 6.3e-4 & 1.4e-1 \\ 
1e12 & 8.32e4 & $32$ & 4.2e-1 & 6.3e-4 & 1.4e-1 & 8.30e4 & $32$ & 4.2e-1 & 6.3e-4 & 1.4e-1 \\ 
2e12 & 1.01e5 & $40$ & 4.2e-1 & 6.3e-4 & 1.4e-1 & 1.01e5 & $40$ & 4.2e-1 & 6.4e-4 & 1.4e-1 \\ 
5e12 & 1.31e5 & $53$ & 4.2e-1 & 6.4e-4 & 1.5e-1 & 1.31e5 & $54$ & 4.2e-1 & 6.4e-4 & 1.5e-1 \\ 
 \hline
\end{tabular}
\caption{Same as \cref{tab:appendix_ct18annlo_proton_xs} for a neutron target.}
\label{tab:appendix_ct18annlo_neutron_xs}
\end{table*}

\begin{table*}[h!]
\begin{tabular}{|c||c|c|c|c|c||c|c|c|c|c|} 
 \hline
 $E_{\nu}~\textrm{[GeV]}$ & $\sigma^{\nu I}~\textrm{[pb]}$ & $\delta \sigma^{\nu I}~[\%]$ & $f_{c}^{\nu I}$ & $f_{b}^{\nu I}$ & $f_{t}^{\nu I}$ & $\sigma^{\bar{\nu} I}~\textrm{[pb]}$ & $\delta \sigma^{\bar{\nu} I}~[\%]$ & $f_{c}^{\bar{\nu} I}$ & $f_{b}^{\bar{\nu} I}$ & $f_{t}^{\bar{\nu} I}$ \\
 \hline
5e1 & 3.24e-1 & $1.2$ & 6.8e-2 & 4.1e-9 & 0 & 1.56e-1 & $2.2$ & 7.4e-2 & 4.1e-7 & 0 \\ 
1e2 & 6.48e-1 & $1.3$ & 8.6e-2 & 1.3e-7 & 0 & 3.21e-1 & $2.3$ & 1.0e-1 & 1.9e-6 & 0 \\ 
2e2 & 1.29e0 & $1.3$ & 1.0e-1 & 9.5e-7 & 0 & 6.56e-1 & $2.3$ & 1.3e-1 & 5.0e-6 & 0 \\ 
5e2 & 3.15e0 & $1.3$ & 1.2e-1 & 4.7e-6 & 0 & 1.67e0 & $2.2$ & 1.6e-1 & 1.4e-5 & 0 \\ 
1e3 & 6.11e0 & $1.3$ & 1.3e-1 & 1.1e-5 & 0 & 3.34e0 & $2.1$ & 1.8e-1 & 2.5e-5 & 0 \\ 
2e3 & 1.16e1 & $1.2$ & 1.5e-1 & 2.1e-5 & 0 & 6.60e0 & $2.1$ & 2.0e-1 & 4.2e-5 & 0 \\ 
5e3 & 2.59e1 & $1.2$ & 1.6e-1 & 3.9e-5 & 0 & 1.57e1 & $1.9$ & 2.2e-1 & 6.9e-5 & 0 \\ 
1e4 & 4.50e1 & $1.2$ & 1.8e-1 & 5.7e-5 & 0 & 2.92e1 & $1.8$ & 2.3e-1 & 9.3e-5 & 0 \\ 
2e4 & 7.44e1 & $1.2$ & 2.0e-1 & 8.1e-5 & 3.0e-9 & 5.21e1 & $1.7$ & 2.5e-1 & 1.2e-4 & 3.2e-9 \\ 
5e4 & 1.34e2 & $1.2$ & 2.4e-1 & 1.2e-4 & 2.0e-5 & 1.04e2 & $1.6$ & 2.8e-1 & 1.6e-4 & 2.5e-5 \\ 
1e5 & 2.00e2 & $1.3$ & 2.7e-1 & 1.6e-4 & 2.5e-4 & 1.66e2 & $1.5$ & 3.0e-1 & 1.9e-4 & 3.0e-4 \\ 
2e5 & 2.88e2 & $1.3$ & 3.0e-1 & 2.0e-4 & 1.2e-3 & 2.54e2 & $1.5$ & 3.2e-1 & 2.3e-4 & 1.3e-3 \\ 
5e5 & 4.55e2 & $1.4$ & 3.4e-1 & 2.6e-4 & 4.6e-3 & 4.23e2 & $1.5$ & 3.5e-1 & 2.8e-4 & 4.9e-3 \\ 
1e6 & 6.30e2 & $1.5$ & 3.6e-1 & 3.0e-4 & 9.2e-3 & 6.02e2 & $1.5$ & 3.6e-1 & 3.2e-4 & 9.7e-3 \\ 
2e6 & 8.63e2 & $1.6$ & 3.8e-1 & 3.4e-4 & 1.6e-2 & 8.39e2 & $1.6$ & 3.8e-1 & 3.5e-4 & 1.6e-2 \\ 
5e6 & 1.29e3 & $1.8$ & 4.0e-1 & 3.9e-4 & 2.6e-2 & 1.27e3 & $1.8$ & 4.0e-1 & 3.9e-4 & 2.6e-2 \\ 
1e7 & 1.72e3 & $1.9$ & 4.1e-1 & 4.2e-4 & 3.4e-2 & 1.70e3 & $2.0$ & 4.1e-1 & 4.2e-4 & 3.5e-2 \\ 
2e7 & 2.28e3 & $2.2$ & 4.1e-1 & 4.5e-4 & 4.3e-2 & 2.27e3 & $2.2$ & 4.1e-1 & 4.5e-4 & 4.3e-2 \\ 
5e7 & 3.26e3 & $2.5$ & 4.2e-1 & 4.8e-4 & 5.5e-2 & 3.25e3 & $2.5$ & 4.2e-1 & 4.8e-4 & 5.5e-2 \\ 
1e8 & 4.24e3 & $2.8$ & 4.2e-1 & 5.0e-4 & 6.3e-2 & 4.24e3 & $2.8$ & 4.2e-1 & 5.0e-4 & 6.3e-2 \\ 
2e8 & 5.48e3 & $3.2$ & 4.3e-1 & 5.2e-4 & 7.2e-2 & 5.47e3 & $3.2$ & 4.2e-1 & 5.2e-4 & 7.1e-2 \\ 
5e8 & 7.60e3 & $3.9$ & 4.3e-1 & 5.4e-4 & 8.2e-2 & 7.59e3 & $3.9$ & 4.3e-1 & 5.4e-4 & 8.2e-2 \\ 
1e9 & 9.66e3 & $4.5$ & 4.3e-1 & 5.6e-4 & 8.9e-2 & 9.66e3 & $4.5$ & 4.3e-1 & 5.6e-4 & 8.9e-2 \\ 
2e9 & 1.22e4 & $5.2$ & 4.3e-1 & 5.7e-4 & 9.6e-2 & 1.22e4 & $5.2$ & 4.3e-1 & 5.7e-4 & 9.6e-2 \\ 
5e9 & 1.66e4 & $6.5$ & 4.3e-1 & 5.8e-4 & 1.0e-1 & 1.66e4 & $6.5$ & 4.3e-1 & 5.8e-4 & 1.0e-1 \\ 
1e10 & 2.07e4 & $7.8$ & 4.3e-1 & 5.9e-4 & 1.1e-1 & 2.07e4 & $7.8$ & 4.3e-1 & 5.9e-4 & 1.1e-1 \\ 
2e10 & 2.58e4 & $9.4$ & 4.3e-1 & 6.0e-4 & 1.2e-1 & 2.58e4 & $9.4$ & 4.3e-1 & 6.0e-4 & 1.2e-1 \\ 
5e10 & 3.43e4 & $12$ & 4.3e-1 & 6.1e-4 & 1.2e-1 & 3.43e4 & $12$ & 4.3e-1 & 6.1e-4 & 1.2e-1 \\ 
1e11 & 4.23e4 & $15$ & 4.2e-1 & 6.2e-4 & 1.3e-1 & 4.23e4 & $15$ & 4.2e-1 & 6.2e-4 & 1.3e-1 \\ 
2e11 & 5.20e4 & $19$ & 4.2e-1 & 6.2e-4 & 1.3e-1 & 5.20e4 & $19$ & 4.2e-1 & 6.2e-4 & 1.3e-1 \\ 
5e11 & 6.80e4 & $25$ & 4.2e-1 & 6.3e-4 & 1.4e-1 & 6.80e4 & $25$ & 4.2e-1 & 6.3e-4 & 1.4e-1 \\ 
1e12 & 8.31e4 & $32$ & 4.2e-1 & 6.3e-4 & 1.4e-1 & 8.31e4 & $32$ & 4.2e-1 & 6.3e-4 & 1.4e-1 \\ 
2e12 & 1.01e5 & $40$ & 4.2e-1 & 6.3e-4 & 1.4e-1 & 1.01e5 & $40$ & 4.2e-1 & 6.3e-4 & 1.4e-1 \\ 
5e12 & 1.31e5 & $54$ & 4.2e-1 & 6.4e-4 & 1.5e-1 & 1.31e5 & $54$ & 4.2e-1 & 6.4e-4 & 1.5e-1 \\ 
 \hline
\end{tabular}
\caption{Same as \cref{tab:appendix_ct18annlo_proton_xs} for an isoscalar target.}
\label{tab:appendix_ct18annlo_isoscalar_xs}
\end{table*}

\begin{table*}[h!]
\begin{tabular}{|c||c|c|c|c|c||c|c|c|c|c|} 
 \hline
 $E_{\nu}~\textrm{[GeV]}$ & $\sigma^{\nu p}~\textrm{[pb]}$ & $\delta \sigma^{\nu p}~[\%]$ & $f_{c}^{\nu p}$ & $f_{b}^{\nu p}$ & $f_{t}^{\nu p}$ & $\sigma^{\bar{\nu} p}~\textrm{[pb]}$ & $\delta \sigma^{\bar{\nu} p}~[\%]$ & $f_{c}^{\bar{\nu} p}$ & $f_{b}^{\bar{\nu} p}$ & $f_{t}^{\bar{\nu} p}$ \\
 \hline
5e1 & 2.22e-1 & $2.0$ & 8.0e-2 & 3.4e-9 & 0 & 2.04e-1 & $1.5$ & 5.7e-2 & 4.4e-7 & 0 \\ 
1e2 & 4.50e-1 & $2.2$ & 1.0e-1 & 1.2e-8 & 0 & 4.14e-1 & $1.7$ & 7.9e-2 & 1.6e-6 & 0 \\ 
2e2 & 9.04e-1 & $2.2$ & 1.2e-1 & 2.6e-7 & 0 & 8.33e-1 & $1.8$ & 9.9e-2 & 3.8e-6 & 0 \\ 
5e2 & 2.24e0 & $2.3$ & 1.4e-1 & 3.9e-6 & 0 & 2.08e0 & $1.9$ & 1.2e-1 & 9.6e-6 & 0 \\ 
1e3 & 4.41e0 & $2.3$ & 1.6e-1 & 1.1e-5 & 0 & 4.11e0 & $1.9$ & 1.4e-1 & 1.8e-5 & 0 \\ 
2e3 & 8.51e0 & $2.2$ & 1.7e-1 & 2.3e-5 & 0 & 8.03e0 & $1.8$ & 1.5e-1 & 3.1e-5 & 0 \\ 
5e3 & 1.94e1 & $2.2$ & 1.9e-1 & 4.6e-5 & 0 & 1.88e1 & $1.8$ & 1.7e-1 & 5.4e-5 & 0 \\ 
1e4 & 3.47e1 & $2.1$ & 2.1e-1 & 6.9e-5 & 0 & 3.45e1 & $1.7$ & 1.9e-1 & 7.5e-5 & 0 \\ 
2e4 & 5.91e1 & $2.1$ & 2.3e-1 & 9.7e-5 & 2.1e-10 & 6.07e1 & $1.7$ & 2.1e-1 & 1.0e-4 & 8.3e-11 \\ 
5e4 & 1.11e2 & $2.0$ & 2.7e-1 & 1.4e-4 & 1.3e-5 & 1.18e2 & $1.6$ & 2.3e-1 & 1.4e-4 & 1.2e-5 \\ 
1e5 & 1.71e2 & $2.0$ & 2.9e-1 & 1.8e-4 & 2.3e-4 & 1.86e2 & $1.6$ & 2.6e-1 & 1.7e-4 & 2.1e-4 \\ 
2e5 & 2.54e2 & $2.0$ & 3.2e-1 & 2.3e-4 & 1.2e-3 & 2.79e2 & $1.7$ & 2.8e-1 & 2.1e-4 & 1.1e-3 \\ 
5e5 & 4.13e2 & $2.1$ & 3.5e-1 & 2.9e-4 & 4.6e-3 & 4.54e2 & $1.8$ & 3.1e-1 & 2.7e-4 & 4.2e-3 \\ 
1e6 & 5.85e2 & $2.1$ & 3.7e-1 & 3.3e-4 & 9.3e-3 & 6.37e2 & $1.9$ & 3.4e-1 & 3.1e-4 & 8.5e-3 \\ 
2e6 & 8.16e2 & $2.2$ & 3.9e-1 & 3.7e-4 & 1.6e-2 & 8.79e2 & $2.0$ & 3.6e-1 & 3.5e-4 & 1.5e-2 \\ 
5e6 & 1.24e3 & $2.4$ & 4.1e-1 & 4.2e-4 & 2.6e-2 & 1.32e3 & $2.3$ & 3.8e-1 & 4.0e-4 & 2.4e-2 \\ 
1e7 & 1.68e3 & $2.6$ & 4.2e-1 & 4.5e-4 & 3.5e-2 & 1.77e3 & $2.5$ & 3.9e-1 & 4.3e-4 & 3.3e-2 \\ 
2e7 & 2.26e3 & $2.9$ & 4.2e-1 & 4.8e-4 & 4.3e-2 & 2.35e3 & $2.8$ & 4.0e-1 & 4.6e-4 & 4.2e-2 \\ 
5e7 & 3.28e3 & $3.3$ & 4.3e-1 & 5.1e-4 & 5.4e-2 & 3.38e3 & $3.2$ & 4.1e-1 & 4.9e-4 & 5.3e-2 \\ 
1e8 & 4.31e3 & $3.6$ & 4.3e-1 & 5.3e-4 & 6.3e-2 & 4.41e3 & $3.6$ & 4.2e-1 & 5.1e-4 & 6.1e-2 \\ 
2e8 & 5.61e3 & $4.1$ & 4.3e-1 & 5.4e-4 & 7.0e-2 & 5.71e3 & $4.0$ & 4.2e-1 & 5.3e-4 & 6.9e-2 \\ 
5e8 & 7.85e3 & $5.0$ & 4.3e-1 & 5.6e-4 & 8.0e-2 & 7.96e3 & $4.9$ & 4.3e-1 & 5.5e-4 & 7.9e-2 \\ 
1e9 & 1.01e4 & $5.9$ & 4.3e-1 & 5.7e-4 & 8.6e-2 & 1.02e4 & $5.8$ & 4.3e-1 & 5.7e-4 & 8.5e-2 \\ 
2e9 & 1.28e4 & $7.0$ & 4.3e-1 & 5.8e-4 & 9.3e-2 & 1.29e4 & $7.0$ & 4.3e-1 & 5.8e-4 & 9.2e-2 \\ 
5e9 & 1.75e4 & $9.2$ & 4.3e-1 & 5.9e-4 & 1.0e-1 & 1.76e4 & $9.1$ & 4.3e-1 & 5.9e-4 & 9.9e-2 \\ 
1e10 & 2.20e4 & $11$ & 4.3e-1 & 6.0e-4 & 1.1e-1 & 2.21e4 & $11$ & 4.3e-1 & 6.0e-4 & 1.0e-1 \\ 
2e10 & 2.75e4 & $15$ & 4.3e-1 & 6.1e-4 & 1.1e-1 & 2.76e4 & $15$ & 4.3e-1 & 6.0e-4 & 1.1e-1 \\ 
5e10 & 3.67e4 & $20$ & 4.3e-1 & 6.1e-4 & 1.2e-1 & 3.68e4 & $20$ & 4.3e-1 & 6.1e-4 & 1.1e-1 \\ 
1e11 & 4.55e4 & $26$ & 4.3e-1 & 6.2e-4 & 1.2e-1 & 4.56e4 & $26$ & 4.3e-1 & 6.2e-4 & 1.2e-1 \\ 
2e11 & 5.62e4 & $35$ & 4.3e-1 & 6.2e-4 & 1.2e-1 & 5.63e4 & $35$ & 4.3e-1 & 6.2e-4 & 1.2e-1 \\ 
5e11 & 7.39e4 & $50$ & 4.3e-1 & 6.3e-4 & 1.3e-1 & 7.40e4 & $50$ & 4.3e-1 & 6.3e-4 & 1.3e-1 \\ 
1e12 & 9.06e4 & $66$ & 4.3e-1 & 6.3e-4 & 1.3e-1 & 9.06e4 & $66$ & 4.3e-1 & 6.3e-4 & 1.3e-1 \\ 
2e12 & 1.11e5 & $88$ & 4.3e-1 & 6.4e-4 & 1.3e-1 & 1.11e5 & $88$ & 4.3e-1 & 6.4e-4 & 1.3e-1 \\ 
5e12 & 1.44e5 & $128$ & 4.2e-1 & 6.4e-4 & 1.4e-1 & 1.44e5 & $128$ & 4.2e-1 & 6.4e-4 & 1.4e-1 \\ 
 \hline
\end{tabular}
\caption{Same as \cref{tab:appendix_ct18annlo_proton_xs} for the NLO calculation using the CT18ANLO PDF set.}
\label{tab:appendix_ct18anlo_proton_xs}
\end{table*}

\begin{table*} [h!]
\begin{tabular}{|c||c|c|c|c|c||c|c|c|c|c|} 
 \hline
 $E_{\nu}~\textrm{[GeV]}$ & $\sigma^{\nu n}~\textrm{[pb]}$ & $\delta \sigma^{\nu n}~[\%]$ & $f_{c}^{\nu n}$ & $f_{b}^{\nu n}$ & $f_{t}^{\nu n}$ & $\sigma^{\bar{\nu} n}~\textrm{[pb]}$ & $\delta \sigma^{\bar{\nu} n}~[\%]$ & $f_{c}^{\bar{\nu} n}$ & $f_{b}^{\bar{\nu} n}$ & $f_{t}^{\bar{\nu} n}$ \\
 \hline
5e1 & 4.40e-1 & $1.1$ & 6.1e-2 & 1.7e-9 & 0 & 1.15e-1 & $3.1$ & 9.9e-2 & 5.9e-8 & 0 \\ 
1e2 & 8.71e-1 & $1.1$ & 7.5e-2 & 6.2e-9 & 0 & 2.43e-1 & $3.4$ & 1.3e-1 & 5.3e-7 & 0 \\ 
2e2 & 1.71e0 & $1.2$ & 8.7e-2 & 1.5e-7 & 0 & 5.04e-1 & $3.5$ & 1.6e-1 & 2.5e-6 & 0 \\ 
5e2 & 4.13e0 & $1.2$ & 1.0e-1 & 2.2e-6 & 0 & 1.31e0 & $3.4$ & 1.9e-1 & 1.0e-5 & 0 \\ 
1e3 & 7.94e0 & $1.2$ & 1.1e-1 & 6.4e-6 & 0 & 2.65e0 & $3.3$ & 2.1e-1 & 2.3e-5 & 0 \\ 
2e3 & 1.49e1 & $1.2$ & 1.2e-1 & 1.3e-5 & 0 & 5.30e0 & $3.2$ & 2.3e-1 & 4.2e-5 & 0 \\ 
5e3 & 3.26e1 & $1.2$ & 1.4e-1 & 2.8e-5 & 0 & 1.28e1 & $3.0$ & 2.5e-1 & 7.5e-5 & 0 \\ 
1e4 & 5.58e1 & $1.3$ & 1.5e-1 & 4.3e-5 & 0 & 2.42e1 & $2.8$ & 2.7e-1 & 1.0e-4 & 0 \\ 
2e4 & 9.02e1 & $1.3$ & 1.7e-1 & 6.4e-5 & 1.1e-9 & 4.40e1 & $2.6$ & 2.8e-1 & 1.3e-4 & 1.2e-10 \\ 
5e4 & 1.57e2 & $1.3$ & 2.0e-1 & 1.0e-4 & 1.0e-5 & 9.00e1 & $2.3$ & 3.0e-1 & 1.8e-4 & 1.6e-5 \\ 
1e5 & 2.28e2 & $1.4$ & 2.3e-1 & 1.4e-4 & 1.7e-4 & 1.47e2 & $2.2$ & 3.2e-1 & 2.2e-4 & 2.6e-4 \\ 
2e5 & 3.22e2 & $1.4$ & 2.6e-1 & 1.8e-4 & 9.2e-4 & 2.29e2 & $2.0$ & 3.4e-1 & 2.6e-4 & 1.3e-3 \\ 
5e5 & 4.94e2 & $1.6$ & 3.0e-1 & 2.4e-4 & 3.8e-3 & 3.90e2 & $2.0$ & 3.6e-1 & 3.1e-4 & 4.8e-3 \\ 
1e6 & 6.73e2 & $1.7$ & 3.3e-1 & 2.9e-4 & 8.1e-3 & 5.65e2 & $2.0$ & 3.8e-1 & 3.5e-4 & 9.6e-3 \\ 
2e6 & 9.10e2 & $1.9$ & 3.6e-1 & 3.4e-4 & 1.4e-2 & 7.99e2 & $2.2$ & 3.9e-1 & 3.8e-4 & 1.6e-2 \\ 
5e6 & 1.34e3 & $2.2$ & 3.8e-1 & 3.9e-4 & 2.4e-2 & 1.23e3 & $2.4$ & 4.1e-1 & 4.3e-4 & 2.6e-2 \\ 
1e7 & 1.79e3 & $2.4$ & 3.9e-1 & 4.3e-4 & 3.3e-2 & 1.68e3 & $2.6$ & 4.1e-1 & 4.5e-4 & 3.5e-2 \\ 
2e7 & 2.37e3 & $2.7$ & 4.1e-1 & 4.6e-4 & 4.1e-2 & 2.26e3 & $2.8$ & 4.2e-1 & 4.8e-4 & 4.3e-2 \\ 
5e7 & 3.39e3 & $3.1$ & 4.2e-1 & 4.9e-4 & 5.3e-2 & 3.28e3 & $3.2$ & 4.3e-1 & 5.1e-4 & 5.4e-2 \\ 
1e8 & 4.42e3 & $3.5$ & 4.2e-1 & 5.1e-4 & 6.1e-2 & 4.31e3 & $3.6$ & 4.3e-1 & 5.3e-4 & 6.3e-2 \\ 
2e8 & 5.72e3 & $4.0$ & 4.2e-1 & 5.3e-4 & 6.9e-2 & 5.61e3 & $4.1$ & 4.3e-1 & 5.4e-4 & 7.0e-2 \\ 
5e8 & 7.97e3 & $4.9$ & 4.3e-1 & 5.5e-4 & 7.9e-2 & 7.86e3 & $5.0$ & 4.3e-1 & 5.6e-4 & 8.0e-2 \\ 
1e9 & 1.02e4 & $5.8$ & 4.3e-1 & 5.6e-4 & 8.5e-2 & 1.01e4 & $5.9$ & 4.3e-1 & 5.7e-4 & 8.6e-2 \\ 
2e9 & 1.29e4 & $7.0$ & 4.3e-1 & 5.8e-4 & 9.2e-2 & 1.28e4 & $7.0$ & 4.3e-1 & 5.8e-4 & 9.2e-2 \\ 
5e9 & 1.76e4 & $9.1$ & 4.3e-1 & 5.9e-4 & 9.9e-2 & 1.75e4 & $9.2$ & 4.3e-1 & 5.9e-4 & 1.0e-1 \\ 
1e10 & 2.21e4 & $11$ & 4.3e-1 & 6.0e-4 & 1.0e-1 & 2.20e4 & $11$ & 4.3e-1 & 6.0e-4 & 1.0e-1 \\ 
2e10 & 2.76e4 & $15$ & 4.3e-1 & 6.0e-4 & 1.1e-1 & 2.75e4 & $15$ & 4.3e-1 & 6.1e-4 & 1.1e-1 \\ 
5e10 & 3.68e4 & $20$ & 4.3e-1 & 6.1e-4 & 1.2e-1 & 3.67e4 & $20$ & 4.3e-1 & 6.1e-4 & 1.2e-1 \\ 
1e11 & 4.56e4 & $26$ & 4.3e-1 & 6.2e-4 & 1.2e-1 & 4.55e4 & $26$ & 4.3e-1 & 6.2e-4 & 1.2e-1 \\ 
2e11 & 5.63e4 & $35$ & 4.3e-1 & 6.2e-4 & 1.2e-1 & 5.62e4 & $35$ & 4.3e-1 & 6.2e-4 & 1.2e-1 \\ 
5e11 & 7.40e4 & $50$ & 4.3e-1 & 6.3e-4 & 1.3e-1 & 7.39e4 & $50$ & 4.3e-1 & 6.3e-4 & 1.3e-1 \\ 
1e12 & 9.07e4 & $66$ & 4.3e-1 & 6.3e-4 & 1.3e-1 & 9.06e4 & $66$ & 4.3e-1 & 6.3e-4 & 1.3e-1 \\ 
2e12 & 1.11e5 & $88$ & 4.3e-1 & 6.4e-4 & 1.3e-1 & 1.11e5 & $88$ & 4.3e-1 & 6.4e-4 & 1.3e-1 \\ 
5e12 & 1.44e5 & $128$ & 4.2e-1 & 6.4e-4 & 1.4e-1 & 1.44e5 & $128$ & 4.3e-1 & 6.4e-4 & 1.4e-1 \\ 
 \hline
\end{tabular}
\caption{Same as \cref{tab:appendix_ct18anlo_proton_xs} for a neutron target.}
\label{tab:appendix_ct18anlo_neutron_xs}
\end{table*}

\begin{table*}[h!]
\begin{tabular}{|c||c|c|c|c|c||c|c|c|c|c|} 
 \hline
 $E_{\nu}~\textrm{[GeV]}$ & $\sigma^{\nu I}~\textrm{[pb]}$ & $\delta \sigma^{\nu I}~[\%]$ & $f_{c}^{\nu I}$ & $f_{b}^{\nu I}$ & $f_{t}^{\nu I}$ & $\sigma^{\bar{\nu} I}~\textrm{[pb]}$ & $\delta \sigma^{\bar{\nu} I}~[\%]$ & $f_{c}^{\bar{\nu} I}$ & $f_{b}^{\bar{\nu} I}$ & $f_{t}^{\bar{\nu} I}$ \\
 \hline
5e1 & 3.31e-1 & $1.1$ & 6.7e-2 & 2.3e-9 & 0 & 1.59e-1 & $1.9$ & 7.2e-2 & 3.0e-7 & 0 \\ 
1e2 & 6.60e-1 & $1.2$ & 8.4e-2 & 8.1e-9 & 0 & 3.28e-1 & $2.2$ & 9.8e-2 & 1.2e-6 & 0 \\ 
2e2 & 1.31e0 & $1.3$ & 9.9e-2 & 1.9e-7 & 0 & 6.69e-1 & $2.3$ & 1.2e-1 & 3.3e-6 & 0 \\ 
5e2 & 3.19e0 & $1.4$ & 1.2e-1 & 2.8e-6 & 0 & 1.69e0 & $2.4$ & 1.5e-1 & 9.9e-6 & 0 \\ 
1e3 & 6.18e0 & $1.4$ & 1.3e-1 & 8.1e-6 & 0 & 3.38e0 & $2.4$ & 1.7e-1 & 2.0e-5 & 0 \\ 
2e3 & 1.17e1 & $1.4$ & 1.4e-1 & 1.7e-5 & 0 & 6.66e0 & $2.3$ & 1.8e-1 & 3.5e-5 & 0 \\ 
5e3 & 2.60e1 & $1.4$ & 1.6e-1 & 3.5e-5 & 0 & 1.58e1 & $2.2$ & 2.1e-1 & 6.2e-5 & 0 \\ 
1e4 & 4.52e1 & $1.4$ & 1.7e-1 & 5.3e-5 & 0 & 2.94e1 & $2.1$ & 2.2e-1 & 8.7e-5 & 0 \\ 
2e4 & 7.46e1 & $1.4$ & 2.0e-1 & 7.7e-5 & 7.4e-10 & 5.23e1 & $2.0$ & 2.4e-1 & 1.1e-4 & 9.8e-11 \\ 
5e4 & 1.34e2 & $1.5$ & 2.3e-1 & 1.2e-4 & 1.1e-5 & 1.04e2 & $1.9$ & 2.6e-1 & 1.6e-4 & 1.3e-5 \\ 
1e5 & 1.99e2 & $1.6$ & 2.6e-1 & 1.6e-4 & 2.0e-4 & 1.66e2 & $1.8$ & 2.8e-1 & 1.9e-4 & 2.3e-4 \\ 
2e5 & 2.88e2 & $1.6$ & 2.9e-1 & 2.0e-4 & 1.0e-3 & 2.54e2 & $1.8$ & 3.1e-1 & 2.3e-4 & 1.2e-3 \\ 
5e5 & 4.53e2 & $1.8$ & 3.3e-1 & 2.6e-4 & 4.2e-3 & 4.22e2 & $1.9$ & 3.4e-1 & 2.9e-4 & 4.5e-3 \\ 
1e6 & 6.29e2 & $1.9$ & 3.5e-1 & 3.1e-4 & 8.6e-3 & 6.01e2 & $2.0$ & 3.6e-1 & 3.3e-4 & 9.0e-3 \\ 
2e6 & 8.63e2 & $2.0$ & 3.7e-1 & 3.5e-4 & 1.5e-2 & 8.39e2 & $2.1$ & 3.7e-1 & 3.7e-4 & 1.5e-2 \\ 
5e6 & 1.29e3 & $2.3$ & 3.9e-1 & 4.0e-4 & 2.5e-2 & 1.27e3 & $2.3$ & 3.9e-1 & 4.1e-4 & 2.5e-2 \\ 
1e7 & 1.74e3 & $2.5$ & 4.0e-1 & 4.4e-4 & 3.4e-2 & 1.72e3 & $2.5$ & 4.0e-1 & 4.4e-4 & 3.4e-2 \\ 
2e7 & 2.31e3 & $2.8$ & 4.1e-1 & 4.7e-4 & 4.2e-2 & 2.30e3 & $2.8$ & 4.1e-1 & 4.7e-4 & 4.2e-2 \\ 
5e7 & 3.34e3 & $3.2$ & 4.2e-1 & 5.0e-4 & 5.4e-2 & 3.33e3 & $3.2$ & 4.2e-1 & 5.0e-4 & 5.4e-2 \\ 
1e8 & 4.37e3 & $3.6$ & 4.2e-1 & 5.2e-4 & 6.2e-2 & 4.36e3 & $3.6$ & 4.2e-1 & 5.2e-4 & 6.2e-2 \\ 
2e8 & 5.67e3 & $4.1$ & 4.3e-1 & 5.4e-4 & 7.0e-2 & 5.66e3 & $4.1$ & 4.3e-1 & 5.4e-4 & 7.0e-2 \\ 
5e8 & 7.91e3 & $4.9$ & 4.3e-1 & 5.6e-4 & 7.9e-2 & 7.91e3 & $4.9$ & 4.3e-1 & 5.6e-4 & 7.9e-2 \\ 
1e9 & 1.01e4 & $5.8$ & 4.3e-1 & 5.7e-4 & 8.6e-2 & 1.01e4 & $5.8$ & 4.3e-1 & 5.7e-4 & 8.6e-2 \\ 
2e9 & 1.29e4 & $7.0$ & 4.3e-1 & 5.8e-4 & 9.2e-2 & 1.29e4 & $7.0$ & 4.3e-1 & 5.8e-4 & 9.2e-2 \\ 
5e9 & 1.75e4 & $9.2$ & 4.3e-1 & 5.9e-4 & 1.0e-1 & 1.75e4 & $9.2$ & 4.3e-1 & 5.9e-4 & 9.9e-2 \\ 
1e10 & 2.20e4 & $11$ & 4.3e-1 & 6.0e-4 & 1.0e-1 & 2.20e4 & $11$ & 4.3e-1 & 6.0e-4 & 1.0e-1 \\ 
2e10 & 2.75e4 & $15$ & 4.3e-1 & 6.1e-4 & 1.1e-1 & 2.75e4 & $15$ & 4.3e-1 & 6.1e-4 & 1.1e-1 \\ 
5e10 & 3.68e4 & $20$ & 4.3e-1 & 6.1e-4 & 1.2e-1 & 3.68e4 & $20$ & 4.3e-1 & 6.1e-4 & 1.2e-1 \\ 
1e11 & 4.56e4 & $26$ & 4.3e-1 & 6.2e-4 & 1.2e-1 & 4.55e4 & $26$ & 4.3e-1 & 6.2e-4 & 1.2e-1 \\ 
2e11 & 5.63e4 & $35$ & 4.3e-1 & 6.2e-4 & 1.2e-1 & 5.62e4 & $35$ & 4.3e-1 & 6.2e-4 & 1.2e-1 \\ 
5e11 & 7.39e4 & $50$ & 4.3e-1 & 6.3e-4 & 1.3e-1 & 7.39e4 & $50$ & 4.3e-1 & 6.3e-4 & 1.3e-1 \\ 
1e12 & 9.06e4 & $66$ & 4.3e-1 & 6.3e-4 & 1.3e-1 & 9.06e4 & $66$ & 4.3e-1 & 6.3e-4 & 1.3e-1 \\ 
2e12 & 1.11e5 & $88$ & 4.3e-1 & 6.4e-4 & 1.3e-1 & 1.11e5 & $88$ & 4.3e-1 & 6.4e-4 & 1.3e-1 \\ 
5e12 & 1.44e5 & $128$ & 4.2e-1 & 6.4e-4 & 1.4e-1 & 1.44e5 & $128$ & 4.2e-1 & 6.4e-4 & 1.4e-1 \\ 
 \hline
\end{tabular}
\caption{Same as \cref{tab:appendix_ct18anlo_proton_xs} for an isoscalar target.}
\label{tab:appendix_ct18anlo_isoscalar_xs}
\end{table*}


\begin{table*}[h!]
\begin{tabular}{|c||c|c|c|c|c||c|c|c|c|c|} 
 \hline
 $E_{\nu}~\textrm{[GeV]}$ & $\sigma^{\nu A}~\textrm{[pb]}$ & $\delta \sigma^{\nu A}~[\%]$ & $f_{c}^{\nu A}$ & $f_{b}^{\nu A}$ & $f_{t}^{\nu A}$ & $\sigma^{\bar{\nu} A}~\textrm{[pb]}$ & $\delta \sigma^{\bar{\nu} A}~[\%]$ & $f_{c}^{\bar{\nu} A}$ & $f_{b}^{\bar{\nu} A}$ & $f_{t}^{\bar{\nu} A}$ \\
 \hline
5e1 & 3.28e-1 & $1.3$ & 6.7e-2 & 1.8e-9 & 0 & 1.57e-1 & $1.9$ & 7.3e-2 & 2.9e-7 & 0 \\ 
1e2 & 6.54e-1 & $1.4$ & 8.4e-2 & 6.3e-9 & 0 & 3.23e-1 & $2.1$ & 9.7e-2 & 1.2e-6 & 0 \\ 
2e2 & 1.30e0 & $1.5$ & 9.7e-2 & 1.5e-7 & 0 & 6.57e-1 & $2.3$ & 1.2e-1 & 3.2e-6 & 0 \\ 
5e2 & 3.15e0 & $1.6$ & 1.1e-1 & 2.7e-6 & 0 & 1.66e0 & $2.4$ & 1.5e-1 & 9.9e-6 & 0 \\ 
1e3 & 6.10e0 & $1.6$ & 1.2e-1 & 8.2e-6 & 0 & 3.31e0 & $2.4$ & 1.6e-1 & 2.0e-5 & 0 \\ 
2e3 & 1.16e1 & $1.6$ & 1.4e-1 & 1.8e-5 & 0 & 6.51e0 & $2.4$ & 1.8e-1 & 3.7e-5 & 0 \\ 
5e3 & 2.57e1 & $1.6$ & 1.5e-1 & 3.6e-5 & 0 & 1.54e1 & $2.4$ & 2.0e-1 & 6.5e-5 & 0 \\ 
1e4 & 4.45e1 & $1.7$ & 1.7e-1 & 5.5e-5 & 0 & 2.86e1 & $2.3$ & 2.1e-1 & 9.1e-5 & 0 \\ 
2e4 & 7.33e1 & $1.8$ & 1.9e-1 & 7.9e-5 & 2.3e-9 & 5.07e1 & $2.3$ & 2.3e-1 & 1.2e-4 & 1.4e-10 \\ 
5e4 & 1.31e2 & $1.9$ & 2.2e-1 & 1.2e-4 & 1.1e-5 & 1.00e2 & $2.4$ & 2.5e-1 & 1.6e-4 & 1.3e-5 \\ 
1e5 & 1.93e2 & $2.1$ & 2.4e-1 & 1.6e-4 & 1.9e-4 & 1.59e2 & $2.5$ & 2.7e-1 & 2.0e-4 & 2.3e-4 \\ 
2e5 & 2.75e2 & $2.4$ & 2.7e-1 & 2.1e-4 & 1.1e-3 & 2.40e2 & $2.7$ & 2.9e-1 & 2.4e-4 & 1.2e-3 \\ 
5e5 & 4.25e2 & $3.0$ & 3.1e-1 & 2.7e-4 & 4.5e-3 & 3.93e2 & $3.2$ & 3.2e-1 & 3.0e-4 & 4.8e-3 \\ 
1e6 & 5.81e2 & $3.5$ & 3.3e-1 & 3.2e-4 & 9.5e-3 & 5.53e2 & $3.6$ & 3.4e-1 & 3.4e-4 & 1.0e-2 \\ 
2e6 & 7.86e2 & $4.0$ & 3.5e-1 & 3.6e-4 & 1.7e-2 & 7.63e2 & $4.1$ & 3.5e-1 & 3.8e-4 & 1.7e-2 \\ 
5e6 & 1.16e3 & $4.8$ & 3.7e-1 & 4.2e-4 & 2.8e-2 & 1.14e3 & $4.9$ & 3.7e-1 & 4.2e-4 & 2.8e-2 \\ 
1e7 & 1.54e3 & $5.4$ & 3.8e-1 & 4.5e-4 & 3.7e-2 & 1.52e3 & $5.4$ & 3.8e-1 & 4.6e-4 & 3.8e-2 \\ 
2e7 & 2.02e3 & $6.0$ & 3.9e-1 & 4.8e-4 & 4.7e-2 & 2.01e3 & $6.0$ & 3.9e-1 & 4.8e-4 & 4.7e-2 \\ 
5e7 & 2.89e3 & $6.7$ & 4.0e-1 & 5.1e-4 & 5.9e-2 & 2.88e3 & $6.7$ & 4.0e-1 & 5.1e-4 & 5.9e-2 \\ 
1e8 & 3.74e3 & $7.2$ & 4.1e-1 & 5.3e-4 & 6.7e-2 & 3.74e3 & $7.2$ & 4.1e-1 & 5.3e-4 & 6.7e-2 \\ 
2e8 & 4.82e3 & $7.7$ & 4.1e-1 & 5.5e-4 & 7.5e-2 & 4.82e3 & $7.7$ & 4.1e-1 & 5.5e-4 & 7.5e-2 \\ 
5e8 & 6.68e3 & $8.3$ & 4.1e-1 & 5.6e-4 & 8.4e-2 & 6.67e3 & $8.3$ & 4.1e-1 & 5.7e-4 & 8.4e-2 \\ 
1e9 & 8.48e3 & $8.7$ & 4.2e-1 & 5.8e-4 & 9.1e-2 & 8.48e3 & $8.7$ & 4.2e-1 & 5.8e-4 & 9.1e-2 \\ 
2e9 & 1.07e4 & $9.2$ & 4.2e-1 & 5.9e-4 & 9.7e-2 & 1.07e4 & $9.2$ & 4.2e-1 & 5.9e-4 & 9.7e-2 \\ 
5e9 & 1.45e4 & $9.8$ & 4.2e-1 & 6.0e-4 & 1.0e-1 & 1.45e4 & $9.8$ & 4.2e-1 & 6.0e-4 & 1.0e-1 \\ 
1e10 & 1.82e4 & $10$ & 4.2e-1 & 6.0e-4 & 1.1e-1 & 1.82e4 & $10$ & 4.2e-1 & 6.0e-4 & 1.1e-1 \\ 
2e10 & 2.26e4 & $11$ & 4.2e-1 & 6.1e-4 & 1.1e-1 & 2.26e4 & $11$ & 4.2e-1 & 6.1e-4 & 1.1e-1 \\ 
5e10 & 3.01e4 & $12$ & 4.2e-1 & 6.2e-4 & 1.2e-1 & 3.00e4 & $12$ & 4.2e-1 & 6.2e-4 & 1.2e-1 \\ 
1e11 & 3.71e4 & $13$ & 4.2e-1 & 6.2e-4 & 1.2e-1 & 3.71e4 & $13$ & 4.2e-1 & 6.2e-4 & 1.2e-1 \\ 
2e11 & 4.57e4 & $14$ & 4.2e-1 & 6.3e-4 & 1.3e-1 & 4.57e4 & $14$ & 4.2e-1 & 6.3e-4 & 1.3e-1 \\ 
5e11 & 5.99e4 & $16$ & 4.2e-1 & 6.3e-4 & 1.3e-1 & 5.99e4 & $16$ & 4.2e-1 & 6.3e-4 & 1.3e-1 \\ 
1e12 & 7.34e4 & $18$ & 4.2e-1 & 6.4e-4 & 1.3e-1 & 7.34e4 & $18$ & 4.2e-1 & 6.4e-4 & 1.3e-1 \\ 
2e12 & 8.97e4 & $20$ & 4.2e-1 & 6.4e-4 & 1.3e-1 & 8.97e4 & $20$ & 4.2e-1 & 6.4e-4 & 1.3e-1 \\ 
5e12 & 1.17e5 & $23$ & 4.2e-1 & 6.5e-4 & 1.4e-1 & 1.17e5 & $23$ & 4.2e-1 & 6.5e-4 & 1.4e-1 \\ 
 \hline
\end{tabular}
\caption{\cref{tab:appendix_ct18annlo_proton_xs} for the NLO calculation of the $^{16}\rm{O}$ cross section using the EPPS21nlo\_CT18Anlo\_O16 PDF set.}
\label{tab:appendix_epps_oxygen_xs}
\end{table*}

\begin{table*}[h!]
\begin{tabular}{|c||c|c|c|c|c||c|c|c|c|c|} 
 \hline
 $E_{\nu}~\textrm{[GeV]}$ & $\sigma^{\nu A}~\textrm{[pb]}$ & $\delta \sigma^{\nu A}~[\%]$ & $f_{c}^{\nu A}$ & $f_{b}^{\nu A}$ & $f_{t}^{\nu A}$ & $\sigma^{\bar{\nu} A}~\textrm{[pb]}$ & $\delta \sigma^{\bar{\nu} A}~[\%]$ & $f_{c}^{\bar{\nu} A}$ & $f_{b}^{\bar{\nu} A}$ & $f_{t}^{\bar{\nu} A}$ \\
 \hline
5e1 & 3.27e-1 & $1.4$ & 6.7e-2 & 2.0e-9 & 0 & 1.56e-1 & $2.0$ & 7.3e-2 & 3.0e-7 & 0 \\ 
1e2 & 6.52e-1 & $1.5$ & 8.4e-2 & 6.4e-9 & 0 & 3.21e-1 & $2.2$ & 9.8e-2 & 1.2e-6 & 0 \\ 
2e2 & 1.29e0 & $1.5$ & 9.8e-2 & 1.5e-7 & 0 & 6.53e-1 & $2.4$ & 1.2e-1 & 3.2e-6 & 0 \\ 
5e2 & 3.14e0 & $1.6$ & 1.1e-1 & 2.8e-6 & 0 & 1.65e0 & $2.5$ & 1.5e-1 & 1.0e-5 & 0 \\ 
1e3 & 6.08e0 & $1.6$ & 1.2e-1 & 8.4e-6 & 0 & 3.28e0 & $2.5$ & 1.6e-1 & 2.1e-5 & 0 \\ 
2e3 & 1.15e1 & $1.6$ & 1.4e-1 & 1.8e-5 & 0 & 6.46e0 & $2.4$ & 1.8e-1 & 3.7e-5 & 0 \\ 
5e3 & 2.56e1 & $1.6$ & 1.5e-1 & 3.6e-5 & 0 & 1.53e1 & $2.4$ & 2.0e-1 & 6.6e-5 & 0 \\ 
1e4 & 4.44e1 & $1.7$ & 1.7e-1 & 5.6e-5 & 0 & 2.84e1 & $2.4$ & 2.1e-1 & 9.2e-5 & 0 \\ 
2e4 & 7.30e1 & $1.7$ & 1.9e-1 & 8.0e-5 & 2.0e-9 & 5.03e1 & $2.3$ & 2.3e-1 & 1.2e-4 & 6.9e-11 \\ 
5e4 & 1.30e2 & $1.9$ & 2.2e-1 & 1.2e-4 & 1.0e-5 & 9.95e1 & $2.3$ & 2.5e-1 & 1.7e-4 & 1.3e-5 \\ 
1e5 & 1.91e2 & $2.0$ & 2.4e-1 & 1.6e-4 & 1.9e-4 & 1.57e2 & $2.4$ & 2.7e-1 & 2.0e-4 & 2.3e-4 \\ 
2e5 & 2.72e2 & $2.2$ & 2.7e-1 & 2.1e-4 & 1.1e-3 & 2.37e2 & $2.6$ & 2.9e-1 & 2.4e-4 & 1.2e-3 \\ 
5e5 & 4.20e2 & $2.7$ & 3.0e-1 & 2.8e-4 & 4.6e-3 & 3.87e2 & $2.9$ & 3.1e-1 & 3.0e-4 & 4.9e-3 \\ 
1e6 & 5.72e2 & $3.0$ & 3.3e-1 & 3.2e-4 & 9.8e-3 & 5.44e2 & $3.2$ & 3.3e-1 & 3.4e-4 & 1.0e-2 \\ 
2e6 & 7.72e2 & $3.5$ & 3.5e-1 & 3.7e-4 & 1.7e-2 & 7.49e2 & $3.6$ & 3.5e-1 & 3.8e-4 & 1.8e-2 \\ 
5e6 & 1.13e3 & $4.1$ & 3.7e-1 & 4.2e-4 & 2.9e-2 & 1.12e3 & $4.1$ & 3.7e-1 & 4.3e-4 & 2.9e-2 \\ 
1e7 & 1.50e3 & $4.5$ & 3.8e-1 & 4.6e-4 & 3.8e-2 & 1.49e3 & $4.6$ & 3.8e-1 & 4.6e-4 & 3.9e-2 \\ 
2e7 & 1.98e3 & $5.0$ & 3.9e-1 & 4.9e-4 & 4.8e-2 & 1.97e3 & $5.0$ & 3.9e-1 & 4.9e-4 & 4.8e-2 \\ 
5e7 & 2.81e3 & $5.6$ & 4.0e-1 & 5.2e-4 & 6.0e-2 & 2.80e3 & $5.6$ & 4.0e-1 & 5.2e-4 & 6.0e-2 \\ 
1e8 & 3.64e3 & $6.0$ & 4.1e-1 & 5.4e-4 & 6.9e-2 & 3.63e3 & $6.0$ & 4.0e-1 & 5.4e-4 & 6.9e-2 \\ 
2e8 & 4.68e3 & $6.4$ & 4.1e-1 & 5.5e-4 & 7.7e-2 & 4.68e3 & $6.4$ & 4.1e-1 & 5.5e-4 & 7.6e-2 \\ 
5e8 & 6.48e3 & $6.9$ & 4.1e-1 & 5.7e-4 & 8.6e-2 & 6.47e3 & $6.9$ & 4.1e-1 & 5.7e-4 & 8.6e-2 \\ 
1e9 & 8.22e3 & $7.3$ & 4.2e-1 & 5.8e-4 & 9.3e-2 & 8.22e3 & $7.3$ & 4.1e-1 & 5.8e-4 & 9.2e-2 \\ 
2e9 & 1.04e4 & $7.6$ & 4.2e-1 & 5.9e-4 & 9.9e-2 & 1.04e4 & $7.6$ & 4.2e-1 & 5.9e-4 & 9.8e-2 \\ 
5e9 & 1.40e4 & $8.1$ & 4.2e-1 & 6.0e-4 & 1.1e-1 & 1.40e4 & $8.1$ & 4.2e-1 & 6.0e-4 & 1.1e-1 \\ 
1e10 & 1.75e4 & $8.5$ & 4.2e-1 & 6.1e-4 & 1.1e-1 & 1.75e4 & $8.5$ & 4.2e-1 & 6.1e-4 & 1.1e-1 \\ 
2e10 & 2.18e4 & $8.9$ & 4.2e-1 & 6.1e-4 & 1.2e-1 & 2.18e4 & $8.9$ & 4.2e-1 & 6.1e-4 & 1.1e-1 \\ 
5e10 & 2.90e4 & $9.6$ & 4.2e-1 & 6.2e-4 & 1.2e-1 & 2.90e4 & $9.6$ & 4.2e-1 & 6.2e-4 & 1.2e-1 \\ 
1e11 & 3.58e4 & $10$ & 4.2e-1 & 6.3e-4 & 1.2e-1 & 3.57e4 & $10$ & 4.2e-1 & 6.3e-4 & 1.2e-1 \\ 
2e11 & 4.40e4 & $11$ & 4.2e-1 & 6.3e-4 & 1.3e-1 & 4.40e4 & $11$ & 4.2e-1 & 6.3e-4 & 1.3e-1 \\ 
5e11 & 5.76e4 & $12$ & 4.2e-1 & 6.4e-4 & 1.3e-1 & 5.76e4 & $12$ & 4.2e-1 & 6.4e-4 & 1.3e-1 \\ 
1e12 & 7.05e4 & $13$ & 4.2e-1 & 6.4e-4 & 1.3e-1 & 7.05e4 & $13$ & 4.2e-1 & 6.4e-4 & 1.3e-1 \\ 
2e12 & 8.61e4 & $14$ & 4.2e-1 & 6.4e-4 & 1.4e-1 & 8.60e4 & $14$ & 4.2e-1 & 6.4e-4 & 1.4e-1 \\ 
5e12 & 1.12e5 & $16$ & 4.2e-1 & 6.5e-4 & 1.4e-1 & 1.12e5 & $16$ & 4.2e-1 & 6.5e-4 & 1.4e-1 \\ 
 \hline
\end{tabular}
\caption{\cref{tab:appendix_ct18annlo_proton_xs} for the NLO calculation of the $^{56}\rm{Fe}$ cross section using the EPPS21nlo\_CT18Anlo\_Fe56 PDF set.}
\label{tab:appendix_epps_iron_xs}
\end{table*}

\begin{table*}[h!]
\begin{tabular}{|c||c|c|c|c|c||c|c|c|c|c|} 
 \hline
 $E_{\nu}~\textrm{[GeV]}$ & $\sigma^{\nu A}~\textrm{[pb]}$ & $\delta \sigma^{\nu A}~[\%]$ & $f_{c}^{\nu A}$ & $f_{b}^{\nu A}$ & $f_{t}^{\nu A}$ & $\sigma^{\bar{\nu} A}~\textrm{[pb]}$ & $\delta \sigma^{\bar{\nu} A}~[\%]$ & $f_{c}^{\bar{\nu} A}$ & $f_{b}^{\bar{\nu} A}$ & $f_{t}^{\bar{\nu} A}$ \\
 \hline
5e1 & 3.27e-1 & $1.8$ & 6.8e-2 & 1.6e-9 & 0 & 1.55e-1 & $2.4$ & 7.4e-2 & 2.8e-7 & 0 \\ 
1e2 & 6.50e-1 & $1.9$ & 8.4e-2 & 5.9e-9 & 0 & 3.19e-1 & $2.6$ & 9.9e-2 & 1.1e-6 & 0 \\ 
2e2 & 1.28e0 & $1.9$ & 9.8e-2 & 1.3e-7 & 0 & 6.47e-1 & $2.8$ & 1.2e-1 & 3.1e-6 & 0 \\ 
5e2 & 3.12e0 & $1.9$ & 1.1e-1 & 2.7e-6 & 0 & 1.63e0 & $2.8$ & 1.5e-1 & 9.9e-6 & 0 \\ 
1e3 & 6.04e0 & $1.9$ & 1.2e-1 & 8.3e-6 & 0 & 3.25e0 & $2.8$ & 1.6e-1 & 2.1e-5 & 0 \\ 
2e3 & 1.15e1 & $1.9$ & 1.3e-1 & 1.8e-5 & 0 & 6.38e0 & $2.7$ & 1.8e-1 & 3.8e-5 & 0 \\ 
5e3 & 2.54e1 & $1.9$ & 1.5e-1 & 3.7e-5 & 0 & 1.51e1 & $2.6$ & 2.0e-1 & 6.8e-5 & 0 \\ 
1e4 & 4.41e1 & $1.9$ & 1.7e-1 & 5.7e-5 & 0 & 2.80e1 & $2.6$ & 2.1e-1 & 9.4e-5 & 0 \\ 
2e4 & 7.25e1 & $1.9$ & 1.9e-1 & 8.2e-5 & 2.4e-9 & 4.97e1 & $2.5$ & 2.3e-1 & 1.2e-4 & 1.4e-10 \\ 
5e4 & 1.29e2 & $1.9$ & 2.2e-1 & 1.3e-4 & 1.0e-5 & 9.81e1 & $2.4$ & 2.5e-1 & 1.7e-4 & 1.2e-5 \\ 
1e5 & 1.89e2 & $2.0$ & 2.4e-1 & 1.7e-4 & 1.9e-4 & 1.55e2 & $2.4$ & 2.7e-1 & 2.1e-4 & 2.2e-4 \\ 
2e5 & 2.68e2 & $2.2$ & 2.7e-1 & 2.2e-4 & 1.1e-3 & 2.33e2 & $2.5$ & 2.9e-1 & 2.5e-4 & 1.2e-3 \\ 
5e5 & 4.11e2 & $2.4$ & 3.0e-1 & 2.8e-4 & 4.7e-3 & 3.79e2 & $2.6$ & 3.1e-1 & 3.1e-4 & 5.1e-3 \\ 
1e6 & 5.59e2 & $2.6$ & 3.3e-1 & 3.3e-4 & 1.0e-2 & 5.30e2 & $2.8$ & 3.3e-1 & 3.5e-4 & 1.1e-2 \\ 
2e6 & 7.51e2 & $2.9$ & 3.5e-1 & 3.8e-4 & 1.8e-2 & 7.28e2 & $3.0$ & 3.5e-1 & 3.9e-4 & 1.8e-2 \\ 
5e6 & 1.10e3 & $3.3$ & 3.7e-1 & 4.3e-4 & 3.0e-2 & 1.08e3 & $3.4$ & 3.7e-1 & 4.4e-4 & 3.1e-2 \\ 
1e7 & 1.45e3 & $3.7$ & 3.8e-1 & 4.7e-4 & 4.0e-2 & 1.44e3 & $3.7$ & 3.8e-1 & 4.7e-4 & 4.0e-2 \\ 
2e7 & 1.91e3 & $4.0$ & 3.9e-1 & 5.0e-4 & 5.0e-2 & 1.90e3 & $4.0$ & 3.9e-1 & 5.0e-4 & 5.0e-2 \\ 
5e7 & 2.70e3 & $4.5$ & 4.0e-1 & 5.3e-4 & 6.2e-2 & 2.70e3 & $4.5$ & 4.0e-1 & 5.3e-4 & 6.2e-2 \\ 
1e8 & 3.50e3 & $4.8$ & 4.1e-1 & 5.5e-4 & 7.1e-2 & 3.49e3 & $4.8$ & 4.0e-1 & 5.5e-4 & 7.1e-2 \\ 
2e8 & 4.49e3 & $5.2$ & 4.1e-1 & 5.6e-4 & 7.9e-2 & 4.49e3 & $5.2$ & 4.1e-1 & 5.6e-4 & 7.9e-2 \\ 
5e8 & 6.20e3 & $5.6$ & 4.1e-1 & 5.8e-4 & 8.9e-2 & 6.19e3 & $5.6$ & 4.1e-1 & 5.8e-4 & 8.9e-2 \\ 
1e9 & 7.85e3 & $5.9$ & 4.1e-1 & 5.9e-4 & 9.5e-2 & 7.85e3 & $5.9$ & 4.1e-1 & 5.9e-4 & 9.5e-2 \\ 
2e9 & 9.91e3 & $6.2$ & 4.2e-1 & 6.0e-4 & 1.0e-1 & 9.90e3 & $6.2$ & 4.1e-1 & 6.0e-4 & 1.0e-1 \\ 
5e9 & 1.34e4 & $6.5$ & 4.2e-1 & 6.1e-4 & 1.1e-1 & 1.34e4 & $6.5$ & 4.2e-1 & 6.1e-4 & 1.1e-1 \\ 
1e10 & 1.67e4 & $6.8$ & 4.2e-1 & 6.2e-4 & 1.1e-1 & 1.67e4 & $6.8$ & 4.2e-1 & 6.2e-4 & 1.1e-1 \\ 
2e10 & 2.07e4 & $7.0$ & 4.2e-1 & 6.2e-4 & 1.2e-1 & 2.07e4 & $7.0$ & 4.2e-1 & 6.2e-4 & 1.2e-1 \\ 
5e10 & 2.75e4 & $7.3$ & 4.2e-1 & 6.3e-4 & 1.2e-1 & 2.75e4 & $7.3$ & 4.2e-1 & 6.3e-4 & 1.2e-1 \\ 
1e11 & 3.39e4 & $7.6$ & 4.2e-1 & 6.3e-4 & 1.3e-1 & 3.38e4 & $7.6$ & 4.2e-1 & 6.3e-4 & 1.3e-1 \\ 
2e11 & 4.16e4 & $7.8$ & 4.2e-1 & 6.4e-4 & 1.3e-1 & 4.16e4 & $7.8$ & 4.2e-1 & 6.4e-4 & 1.3e-1 \\ 
5e11 & 5.44e4 & $8.1$ & 4.2e-1 & 6.4e-4 & 1.3e-1 & 5.44e4 & $8.1$ & 4.2e-1 & 6.4e-4 & 1.3e-1 \\ 
1e12 & 6.64e4 & $8.4$ & 4.2e-1 & 6.4e-4 & 1.4e-1 & 6.64e4 & $8.4$ & 4.2e-1 & 6.4e-4 & 1.4e-1 \\ 
2e12 & 8.10e4 & $8.7$ & 4.2e-1 & 6.5e-4 & 1.4e-1 & 8.10e4 & $8.7$ & 4.2e-1 & 6.5e-4 & 1.4e-1 \\ 
5e12 & 1.05e5 & $9.0$ & 4.2e-1 & 6.5e-4 & 1.4e-1 & 1.05e5 & $9.0$ & 4.2e-1 & 6.5e-4 & 1.4e-1 \\ 
 \hline
\end{tabular}
\caption{\cref{tab:appendix_ct18annlo_proton_xs} for the NLO calculation of the $^{208}\rm{Pb}$ cross section using the EPPS21nlo\_CT18Anlo\_Pb208 PDF set.}
\label{tab:appendix_epps_lead_xs}
\end{table*}

\begin{table*}[h!]
\begin{tabular}{|c||c|c|c|c|c||c|c|c|c|c|} 
 \hline
 $E_{\nu}~\textrm{[GeV]}$ & $\sigma^{\nu p}~\textrm{[pb]}$ & $\delta \sigma^{\nu p}~[\%]$ & $f_{c}^{\nu p}$ & $f_{b}^{\nu p}$ & $f_{t}^{\nu p}$ & $\sigma^{\bar{\nu} p}~\textrm{[pb]}$ & $\delta \sigma^{\bar{\nu} p}~[\%]$ & $f_{c}^{\bar{\nu} p}$ & $f_{b}^{\bar{\nu} p}$ & $f_{t}^{\bar{\nu} p}$ \\
 \hline
5e1 & 2.17e-1 & $1.1$ & 1.1e-1 & 9.9e-7 & 0 & 2.04e-1 & $0.85$ & 5.4e-2 & 1.6e-6 & 0 \\ 
1e2 & 4.43e-1 & $1.0$ & 1.4e-1 & 3.8e-6 & 0 & 4.13e-1 & $0.81$ & 7.7e-2 & 6.1e-6 & 0 \\ 
2e2 & 8.96e-1 & $0.94$ & 1.6e-1 & 7.2e-6 & 0 & 8.33e-1 & $0.78$ & 1.0e-1 & 1.1e-5 & 0 \\ 
5e2 & 2.24e0 & $0.88$ & 1.9e-1 & 1.3e-5 & 0 & 2.08e0 & $0.74$ & 1.3e-1 & 1.9e-5 & 0 \\ 
1e3 & 4.42e0 & $0.84$ & 2.0e-1 & 1.9e-5 & 0 & 4.13e0 & $0.71$ & 1.4e-1 & 2.6e-5 & 0 \\ 
2e3 & 8.55e0 & $0.80$ & 2.2e-1 & 3.0e-5 & 0 & 8.07e0 & $0.67$ & 1.6e-1 & 3.8e-5 & 0 \\ 
5e3 & 1.96e1 & $0.75$ & 2.3e-1 & 5.1e-5 & 0 & 1.89e1 & $0.62$ & 1.8e-1 & 5.9e-5 & 0 \\ 
1e4 & 3.50e1 & $0.71$ & 2.5e-1 & 7.2e-5 & 0 & 3.48e1 & $0.57$ & 1.9e-1 & 7.8e-5 & 0 \\ 
2e4 & 5.97e1 & $0.67$ & 2.7e-1 & 9.8e-5 & 1.4e-9 & 6.12e1 & $0.53$ & 2.1e-1 & 1.0e-4 & 1.1e-9 \\ 
5e4 & 1.12e2 & $0.62$ & 2.9e-1 & 1.4e-4 & 1.6e-5 & 1.20e2 & $0.47$ & 2.4e-1 & 1.4e-4 & 1.4e-5 \\ 
1e5 & 1.73e2 & $0.59$ & 3.2e-1 & 1.8e-4 & 2.3e-4 & 1.88e2 & $0.44$ & 2.6e-1 & 1.7e-4 & 2.1e-4 \\ 
2e5 & 2.57e2 & $0.55$ & 3.4e-1 & 2.2e-4 & 1.2e-3 & 2.82e2 & $0.41$ & 2.9e-1 & 2.0e-4 & 1.0e-3 \\ 
5e5 & 4.19e2 & $0.51$ & 3.7e-1 & 2.8e-4 & 4.5e-3 & 4.59e2 & $0.40$ & 3.2e-1 & 2.5e-4 & 4.1e-3 \\ 
1e6 & 5.93e2 & $0.48$ & 3.8e-1 & 3.2e-4 & 9.3e-3 & 6.45e2 & $0.40$ & 3.4e-1 & 2.9e-4 & 8.5e-3 \\ 
2e6 & 8.25e2 & $0.47$ & 4.0e-1 & 3.6e-4 & 1.6e-2 & 8.88e2 & $0.41$ & 3.6e-1 & 3.3e-4 & 1.5e-2 \\ 
5e6 & 1.25e3 & $0.5$ & 4.1e-1 & 4.0e-4 & 2.6e-2 & 1.32e3 & $0.46$ & 3.8e-1 & 3.8e-4 & 2.5e-2 \\ 
1e7 & 1.68e3 & $0.56$ & 4.2e-1 & 4.3e-4 & 3.5e-2 & 1.77e3 & $0.53$ & 4.0e-1 & 4.1e-4 & 3.3e-2 \\ 
2e7 & 2.24e3 & $0.67$ & 4.2e-1 & 4.6e-4 & 4.4e-2 & 2.33e3 & $0.65$ & 4.0e-1 & 4.4e-4 & 4.2e-2 \\ 
5e7 & 3.23e3 & $0.87$ & 4.3e-1 & 4.9e-4 & 5.6e-2 & 3.32e3 & $0.86$ & 4.1e-1 & 4.8e-4 & 5.4e-2 \\ 
1e8 & 4.20e3 & $1.1$ & 4.3e-1 & 5.1e-4 & 6.4e-2 & 4.30e3 & $1.0$ & 4.2e-1 & 5.0e-4 & 6.3e-2 \\ 
2e8 & 5.43e3 & $1.3$ & 4.3e-1 & 5.3e-4 & 7.3e-2 & 5.53e3 & $1.2$ & 4.2e-1 & 5.2e-4 & 7.1e-2 \\ 
5e8 & 7.52e3 & $1.5$ & 4.3e-1 & 5.5e-4 & 8.3e-2 & 7.62e3 & $1.5$ & 4.3e-1 & 5.4e-4 & 8.2e-2 \\ 
1e9 & 9.56e3 & $1.7$ & 4.3e-1 & 5.6e-4 & 9.0e-2 & 9.66e3 & $1.7$ & 4.3e-1 & 5.6e-4 & 8.9e-2 \\ 
2e9 & 1.21e4 & $1.9$ & 4.3e-1 & 5.7e-4 & 9.7e-2 & 1.22e4 & $1.9$ & 4.3e-1 & 5.7e-4 & 9.6e-2 \\ 
5e9 & 1.63e4 & $2.2$ & 4.3e-1 & 5.9e-4 & 1.1e-1 & 1.64e4 & $2.2$ & 4.3e-1 & 5.8e-4 & 1.0e-1 \\ 
1e10 & 2.04e4 & $2.4$ & 4.3e-1 & 5.9e-4 & 1.1e-1 & 2.05e4 & $2.4$ & 4.3e-1 & 5.9e-4 & 1.1e-1 \\ 
2e10 & 2.53e4 & $2.6$ & 4.3e-1 & 6.0e-4 & 1.2e-1 & 2.54e4 & $2.5$ & 4.3e-1 & 6.0e-4 & 1.2e-1 \\ 
5e10 & 3.35e4 & $2.8$ & 4.3e-1 & 6.1e-4 & 1.2e-1 & 3.36e4 & $2.8$ & 4.3e-1 & 6.1e-4 & 1.2e-1 \\ 
1e11 & 4.13e4 & $2.9$ & 4.3e-1 & 6.2e-4 & 1.3e-1 & 4.14e4 & $2.9$ & 4.2e-1 & 6.1e-4 & 1.3e-1 \\ 
2e11 & 5.07e4 & $3.1$ & 4.3e-1 & 6.2e-4 & 1.3e-1 & 5.08e4 & $3.1$ & 4.2e-1 & 6.2e-4 & 1.3e-1 \\ 
5e11 & 6.62e4 & $3.3$ & 4.2e-1 & 6.3e-4 & 1.4e-1 & 6.63e4 & $3.2$ & 4.2e-1 & 6.3e-4 & 1.4e-1 \\ 
1e12 & 8.08e4 & $3.4$ & 4.2e-1 & 6.3e-4 & 1.4e-1 & 8.09e4 & $3.4$ & 4.2e-1 & 6.3e-4 & 1.4e-1 \\ 
2e12 & 9.82e4 & $3.5$ & 4.2e-1 & 6.3e-4 & 1.4e-1 & 9.83e4 & $3.5$ & 4.2e-1 & 6.3e-4 & 1.4e-1 \\ 
5e12 & 1.27e5 & $3.6$ & 4.2e-1 & 6.4e-4 & 1.5e-1 & 1.27e5 & $3.6$ & 4.2e-1 & 6.4e-4 & 1.5e-1 \\ 
 \hline
\end{tabular}
\caption{Same as \cref{tab:appendix_ct18annlo_proton_xs} using the NNPDF4 MHOU PDF set.}
\label{tab:appendix_nnpdf4_proton_xs}
\end{table*}

\begin{table*}[h!]
\begin{tabular}{|c||c|c|c|c|c||c|c|c|c|c|} 
 \hline
 $E_{\nu}~\textrm{[GeV]}$ & $\sigma^{\nu n}~\textrm{[pb]}$ & $\delta \sigma^{\nu n}~[\%]$ & $f_{c}^{\nu n}$ & $f_{b}^{\nu n}$ & $f_{t}^{\nu n}$ & $\sigma^{\bar{\nu} n}~\textrm{[pb]}$ & $\delta \sigma^{\bar{\nu} n}~[\%]$ & $f_{c}^{\bar{\nu} n}$ & $f_{b}^{\bar{\nu} n}$ & $f_{t}^{\bar{\nu} n}$ \\
 \hline
5e1 & 4.35e-1 & $0.78$ & 7.6e-2 & 5.0e-7 & 0 & 1.10e-1 & $1.8$ & 9.6e-2 & 2.2e-6 & 0 \\ 
1e2 & 8.66e-1 & $0.81$ & 9.4e-2 & 2.0e-6 & 0 & 2.34e-1 & $1.7$ & 1.3e-1 & 8.4e-6 & 0 \\ 
2e2 & 1.71e0 & $0.82$ & 1.1e-1 & 3.8e-6 & 0 & 4.88e-1 & $1.6$ & 1.7e-1 & 1.5e-5 & 0 \\ 
5e2 & 4.15e0 & $0.82$ & 1.2e-1 & 6.9e-6 & 0 & 1.27e0 & $1.5$ & 2.0e-1 & 2.5e-5 & 0 \\ 
1e3 & 7.99e0 & $0.81$ & 1.3e-1 & 1.1e-5 & 0 & 2.60e0 & $1.4$ & 2.3e-1 & 3.7e-5 & 0 \\ 
2e3 & 1.51e1 & $0.80$ & 1.4e-1 & 1.7e-5 & 0 & 5.21e0 & $1.3$ & 2.4e-1 & 5.3e-5 & 0 \\ 
5e3 & 3.30e1 & $0.78$ & 1.6e-1 & 3.0e-5 & 0 & 1.27e1 & $1.1$ & 2.6e-1 & 8.3e-5 & 0 \\ 
1e4 & 5.65e1 & $0.76$ & 1.7e-1 & 4.5e-5 & 0 & 2.40e1 & $1.0$ & 2.8e-1 & 1.1e-4 & 0 \\ 
2e4 & 9.15e1 & $0.74$ & 1.9e-1 & 6.4e-5 & 1.6e-9 & 4.37e1 & $0.92$ & 2.9e-1 & 1.4e-4 & 1.6e-9 \\ 
5e4 & 1.60e2 & $0.70$ & 2.2e-1 & 9.9e-5 & 1.2e-5 & 8.99e1 & $0.78$ & 3.1e-1 & 1.8e-4 & 1.8e-5 \\ 
1e5 & 2.32e2 & $0.66$ & 2.5e-1 & 1.3e-4 & 1.8e-4 & 1.47e2 & $0.69$ & 3.3e-1 & 2.1e-4 & 2.7e-4 \\ 
2e5 & 3.28e2 & $0.62$ & 2.8e-1 & 1.7e-4 & 9.1e-4 & 2.30e2 & $0.61$ & 3.5e-1 & 2.5e-4 & 1.3e-3 \\ 
5e5 & 5.02e2 & $0.55$ & 3.1e-1 & 2.3e-4 & 3.8e-3 & 3.93e2 & $0.54$ & 3.7e-1 & 3.0e-4 & 4.9e-3 \\ 
1e6 & 6.83e2 & $0.49$ & 3.4e-1 & 2.8e-4 & 8.0e-3 & 5.69e2 & $0.50$ & 3.9e-1 & 3.3e-4 & 9.7e-3 \\ 
2e6 & 9.21e2 & $0.46$ & 3.6e-1 & 3.2e-4 & 1.4e-2 & 8.04e2 & $0.49$ & 4.0e-1 & 3.7e-4 & 1.6e-2 \\ 
5e6 & 1.35e3 & $0.46$ & 3.8e-1 & 3.7e-4 & 2.4e-2 & 1.23e3 & $0.51$ & 4.1e-1 & 4.1e-4 & 2.6e-2 \\ 
1e7 & 1.79e3 & $0.52$ & 4.0e-1 & 4.1e-4 & 3.3e-2 & 1.67e3 & $0.58$ & 4.2e-1 & 4.4e-4 & 3.5e-2 \\ 
2e7 & 2.35e3 & $0.63$ & 4.1e-1 & 4.4e-4 & 4.2e-2 & 2.23e3 & $0.69$ & 4.2e-1 & 4.6e-4 & 4.4e-2 \\ 
5e7 & 3.34e3 & $0.84$ & 4.2e-1 & 4.7e-4 & 5.4e-2 & 3.22e3 & $0.89$ & 4.3e-1 & 4.9e-4 & 5.6e-2 \\ 
1e8 & 4.31e3 & $1.0$ & 4.2e-1 & 5.0e-4 & 6.3e-2 & 4.20e3 & $1.1$ & 4.3e-1 & 5.1e-4 & 6.4e-2 \\ 
2e8 & 5.54e3 & $1.2$ & 4.2e-1 & 5.2e-4 & 7.1e-2 & 5.42e3 & $1.3$ & 4.3e-1 & 5.3e-4 & 7.3e-2 \\ 
5e8 & 7.64e3 & $1.5$ & 4.3e-1 & 5.4e-4 & 8.2e-2 & 7.52e3 & $1.5$ & 4.3e-1 & 5.5e-4 & 8.3e-2 \\ 
1e9 & 9.67e3 & $1.7$ & 4.3e-1 & 5.5e-4 & 8.9e-2 & 9.55e3 & $1.7$ & 4.3e-1 & 5.6e-4 & 9.0e-2 \\ 
2e9 & 1.22e4 & $1.9$ & 4.3e-1 & 5.7e-4 & 9.6e-2 & 1.21e4 & $1.9$ & 4.3e-1 & 5.7e-4 & 9.7e-2 \\ 
5e9 & 1.64e4 & $2.2$ & 4.3e-1 & 5.8e-4 & 1.0e-1 & 1.63e4 & $2.2$ & 4.3e-1 & 5.9e-4 & 1.1e-1 \\ 
1e10 & 2.05e4 & $2.4$ & 4.3e-1 & 5.9e-4 & 1.1e-1 & 2.04e4 & $2.4$ & 4.3e-1 & 5.9e-4 & 1.1e-1 \\ 
2e10 & 2.54e4 & $2.5$ & 4.3e-1 & 6.0e-4 & 1.2e-1 & 2.53e4 & $2.6$ & 4.3e-1 & 6.0e-4 & 1.2e-1 \\ 
5e10 & 3.37e4 & $2.8$ & 4.3e-1 & 6.1e-4 & 1.2e-1 & 3.35e4 & $2.8$ & 4.3e-1 & 6.1e-4 & 1.2e-1 \\ 
1e11 & 4.15e4 & $2.9$ & 4.3e-1 & 6.1e-4 & 1.3e-1 & 4.13e4 & $2.9$ & 4.3e-1 & 6.2e-4 & 1.3e-1 \\ 
2e11 & 5.09e4 & $3.1$ & 4.2e-1 & 6.2e-4 & 1.3e-1 & 5.07e4 & $3.1$ & 4.3e-1 & 6.2e-4 & 1.3e-1 \\ 
5e11 & 6.64e4 & $3.2$ & 4.2e-1 & 6.3e-4 & 1.4e-1 & 6.62e4 & $3.3$ & 4.2e-1 & 6.3e-4 & 1.4e-1 \\ 
1e12 & 8.09e4 & $3.4$ & 4.2e-1 & 6.3e-4 & 1.4e-1 & 8.08e4 & $3.4$ & 4.2e-1 & 6.3e-4 & 1.4e-1 \\ 
2e12 & 9.84e4 & $3.5$ & 4.2e-1 & 6.3e-4 & 1.4e-1 & 9.82e4 & $3.5$ & 4.2e-1 & 6.3e-4 & 1.4e-1 \\ 
5e12 & 1.27e5 & $3.6$ & 4.2e-1 & 6.4e-4 & 1.5e-1 & 1.27e5 & $3.6$ & 4.2e-1 & 6.4e-4 & 1.5e-1 \\ 
 \hline
\end{tabular}
\caption{Same as \cref{tab:appendix_ct18annlo_neutron_xs} using the NNPDF4 MHOU PDF set.}
\label{tab:appendix_nnpdf4_neutron_xs}
\end{table*}

\begin{table*}[h!]
\begin{tabular}{|c||c|c|c|c|c||c|c|c|c|c|} 
 \hline
 $E_{\nu}~\textrm{[GeV]}$ & $\sigma^{\nu I}~\textrm{[pb]}$ & $\delta \sigma^{\nu I}~[\%]$ & $f_{c}^{\nu I}$ & $f_{b}^{\nu I}$ & $f_{t}^{\nu I}$ & $\sigma^{\bar{\nu} I}~\textrm{[pb]}$ & $\delta \sigma^{\bar{\nu} I}~[\%]$ & $f_{c}^{\bar{\nu} I}$ & $f_{b}^{\bar{\nu} I}$ & $f_{t}^{\bar{\nu} I}$ \\
 \hline
5e1 & 3.26e-1 & $0.64$ & 8.8e-2 & 6.6e-7 & 0 & 1.57e-1 & $1.0$ & 6.9e-2 & 1.8e-6 & 0 \\ 
1e2 & 6.55e-1 & $0.66$ & 1.1e-1 & 2.6e-6 & 0 & 3.23e-1 & $1.0$ & 9.7e-2 & 6.9e-6 & 0 \\ 
2e2 & 1.30e0 & $0.66$ & 1.3e-1 & 5.0e-6 & 0 & 6.61e-1 & $0.97$ & 1.2e-1 & 1.3e-5 & 0 \\ 
5e2 & 3.20e0 & $0.66$ & 1.5e-1 & 8.8e-6 & 0 & 1.68e0 & $0.92$ & 1.6e-1 & 2.1e-5 & 0 \\ 
1e3 & 6.20e0 & $0.64$ & 1.6e-1 & 1.4e-5 & 0 & 3.36e0 & $0.87$ & 1.8e-1 & 3.0e-5 & 0 \\ 
2e3 & 1.18e1 & $0.63$ & 1.7e-1 & 2.2e-5 & 0 & 6.64e0 & $0.82$ & 1.9e-1 & 4.4e-5 & 0 \\ 
5e3 & 2.63e1 & $0.60$ & 1.9e-1 & 3.8e-5 & 0 & 1.58e1 & $0.74$ & 2.1e-1 & 6.8e-5 & 0 \\ 
1e4 & 4.57e1 & $0.57$ & 2.0e-1 & 5.5e-5 & 0 & 2.94e1 & $0.68$ & 2.3e-1 & 9.1e-5 & 0 \\ 
2e4 & 7.56e1 & $0.54$ & 2.2e-1 & 7.7e-5 & 1.5e-9 & 5.24e1 & $0.62$ & 2.5e-1 & 1.2e-4 & 1.3e-9 \\ 
5e4 & 1.36e2 & $0.51$ & 2.5e-1 & 1.2e-4 & 1.3e-5 & 1.05e2 & $0.55$ & 2.7e-1 & 1.5e-4 & 1.6e-5 \\ 
1e5 & 2.03e2 & $0.49$ & 2.8e-1 & 1.5e-4 & 2.0e-4 & 1.67e2 & $0.51$ & 2.9e-1 & 1.9e-4 & 2.4e-4 \\ 
2e5 & 2.93e2 & $0.47$ & 3.0e-1 & 1.9e-4 & 1.0e-3 & 2.56e2 & $0.48$ & 3.1e-1 & 2.2e-4 & 1.2e-3 \\ 
5e5 & 4.61e2 & $0.44$ & 3.4e-1 & 2.5e-4 & 4.1e-3 & 4.26e2 & $0.45$ & 3.4e-1 & 2.7e-4 & 4.5e-3 \\ 
1e6 & 6.38e2 & $0.43$ & 3.6e-1 & 2.9e-4 & 8.6e-3 & 6.07e2 & $0.44$ & 3.6e-1 & 3.1e-4 & 9.0e-3 \\ 
2e6 & 8.73e2 & $0.42$ & 3.8e-1 & 3.4e-4 & 1.5e-2 & 8.46e2 & $0.44$ & 3.8e-1 & 3.5e-4 & 1.5e-2 \\ 
5e6 & 1.30e3 & $0.46$ & 4.0e-1 & 3.9e-4 & 2.5e-2 & 1.28e3 & $0.48$ & 4.0e-1 & 3.9e-4 & 2.5e-2 \\ 
1e7 & 1.74e3 & $0.53$ & 4.1e-1 & 4.2e-4 & 3.4e-2 & 1.72e3 & $0.55$ & 4.1e-1 & 4.2e-4 & 3.4e-2 \\ 
2e7 & 2.30e3 & $0.64$ & 4.2e-1 & 4.5e-4 & 4.3e-2 & 2.28e3 & $0.67$ & 4.1e-1 & 4.5e-4 & 4.3e-2 \\ 
5e7 & 3.28e3 & $0.85$ & 4.2e-1 & 4.8e-4 & 5.5e-2 & 3.27e3 & $0.87$ & 4.2e-1 & 4.8e-4 & 5.5e-2 \\ 
1e8 & 4.26e3 & $1.0$ & 4.3e-1 & 5.0e-4 & 6.4e-2 & 4.25e3 & $1.1$ & 4.2e-1 & 5.1e-4 & 6.4e-2 \\ 
2e8 & 5.48e3 & $1.2$ & 4.3e-1 & 5.2e-4 & 7.2e-2 & 5.47e3 & $1.3$ & 4.3e-1 & 5.2e-4 & 7.2e-2 \\ 
5e8 & 7.58e3 & $1.5$ & 4.3e-1 & 5.4e-4 & 8.3e-2 & 7.57e3 & $1.5$ & 4.3e-1 & 5.5e-4 & 8.2e-2 \\ 
1e9 & 9.61e3 & $1.7$ & 4.3e-1 & 5.6e-4 & 9.0e-2 & 9.61e3 & $1.7$ & 4.3e-1 & 5.6e-4 & 9.0e-2 \\ 
2e9 & 1.21e4 & $1.9$ & 4.3e-1 & 5.7e-4 & 9.7e-2 & 1.21e4 & $1.9$ & 4.3e-1 & 5.7e-4 & 9.7e-2 \\ 
5e9 & 1.64e4 & $2.2$ & 4.3e-1 & 5.8e-4 & 1.1e-1 & 1.64e4 & $2.2$ & 4.3e-1 & 5.8e-4 & 1.1e-1 \\ 
1e10 & 2.04e4 & $2.4$ & 4.3e-1 & 5.9e-4 & 1.1e-1 & 2.04e4 & $2.4$ & 4.3e-1 & 5.9e-4 & 1.1e-1 \\ 
2e10 & 2.54e4 & $2.6$ & 4.3e-1 & 6.0e-4 & 1.2e-1 & 2.54e4 & $2.6$ & 4.3e-1 & 6.0e-4 & 1.2e-1 \\ 
5e10 & 3.36e4 & $2.8$ & 4.3e-1 & 6.1e-4 & 1.2e-1 & 3.36e4 & $2.8$ & 4.3e-1 & 6.1e-4 & 1.2e-1 \\ 
1e11 & 4.14e4 & $2.9$ & 4.3e-1 & 6.1e-4 & 1.3e-1 & 4.14e4 & $2.9$ & 4.3e-1 & 6.2e-4 & 1.3e-1 \\ 
2e11 & 5.08e4 & $3.1$ & 4.2e-1 & 6.2e-4 & 1.3e-1 & 5.08e4 & $3.1$ & 4.2e-1 & 6.2e-4 & 1.3e-1 \\ 
5e11 & 6.63e4 & $3.2$ & 4.2e-1 & 6.3e-4 & 1.4e-1 & 6.63e4 & $3.2$ & 4.2e-1 & 6.3e-4 & 1.4e-1 \\ 
1e12 & 8.08e4 & $3.4$ & 4.2e-1 & 6.3e-4 & 1.4e-1 & 8.08e4 & $3.4$ & 4.2e-1 & 6.3e-4 & 1.4e-1 \\ 
2e12 & 9.83e4 & $3.5$ & 4.2e-1 & 6.3e-4 & 1.4e-1 & 9.83e4 & $3.5$ & 4.2e-1 & 6.3e-4 & 1.4e-1 \\ 
5e12 & 1.27e5 & $3.6$ & 4.2e-1 & 6.4e-4 & 1.5e-1 & 1.27e5 & $3.6$ & 4.2e-1 & 6.4e-4 & 1.5e-1 \\ 
 \hline
\end{tabular}
\caption{Same as \cref{tab:appendix_ct18annlo_isoscalar_xs} using the NNPDF4 MHOU PDF set.}
\label{tab:appendix_nnpdf4_isoscalar_xs}
\end{table*}

\end{document}